\documentclass[twocolumn]{article}  

\usepackage{amsmath}
\usepackage{txfonts}
\usepackage{graphicx}
\usepackage{abstract}
\usepackage{authblk}
\usepackage{natbib}
\usepackage{geometry}
\begin{document}

\def\abstractname{ABSTRACT} 

\title{Predicted asteroseismic detection yield for solar-like oscillating stars with PLATO}
\author[1]{M.J. Goupil}
\author[1]{C. Catala}
\author[1]{R.  Samadi}
\author[1]{K. Belkacem}
\author[1]{R.M. Ouazzani}
\author[1]{D. R. Reese}
\author[2]{T. Appourchaux}
\author[3,4]{S. Mathur}
\author[5]{J. Cabrera}
\author[6]{A. Börner}
\author[6]{C. Paproth}
\author[7,8]{N. Moedas}
\author[9]{K. Verma}
\author[1,10]{Y. Lebreton}
\author[11]{M. Deal}
\author[12]{J. Ballot}
\author[13]{W. J. Chaplin}
\author[14]{J. Christensen-Dalsgaard}
\author[7]{M. Cunha}
\author[15]{A. F. Lanza}
\author[16,17]{A. Miglio}
\author[18]{T. Morel}
\author[19,20]{A. Serenelli}
\author[1]{B. Mosser}
\author[21]{O. Creevey}
\author[22]{A.  Moya}
\author[23]{R.A. Garcia}
\author[13]{M.B. Nielsen}
\author[13]{E. Hatt}

\affil[1]{LESIA, Observatoire de Paris, PSL, CNRS, Sorbonne Université, Université Paris Cité, 5 place Jules Janssen, 92195 Meudon, France}
\affil[2]{Institut d’Astrophysique Spatiale, UMR 8617, Université Paris-Saclay, Bâtiment 121, 91045, Orsay}
\affil[3]{Instituto de Astrof\'isica de Canarias (IAC), E-38205 La Laguna, Tenerife, Spain}
\affil[4]{Universidad de La Laguna (ULL), Departamento de Astrof\'isica, E-38206 La Laguna, Tenerife, Spain}
\affil[5]{Deutsches Zentrum für Luft- und Raumfahrt e.V. (DLR), Institut für Planetenforschung, Rutherfordstr. 2, 12489 Berlin, Germany}
\affil[6]{Deutsches Zentrum für Luft- und Raumfahrt e.V. (DLR), Institut für Optische Sensorsysteme, Rutherfordstr. 2, 12489 Berlin, Germany}
\affil[7]{Instituto de Astrofísica e Ciências do Espaço, Universidade  do Porto, CAUP, Rua das Estrelas, 4150-762 Porto, Portugal}
\affil[8]{Departamento de Física e Astronomia, Faculdade de Ciências da  Universidade do Porto, Rua do Campo Alegre, s/n, PT4169-007 Porto, Portugal}
\affil[9]{Department of Physics, Indian Institute of Technology (BHU), Varanasi-221005, India}
\affil[10]{Universit\'e de Rennes, CNRS, IPR (Institut de Physique de Rennes) -- UMR 6251, 35000 Rennes, France}
\affil[11]{LUPM, Universit\'e de  Montpellier, CNRS, Place Eug\`ene Bataillon, 34095 Montpellier, France}
\affil[12]{IRAP, Université de Toulouse, CNRS, CNES, UPS, 14 avenue Edouard Belin,  31400 Toulouse, France}
\affil[13]{School of Physics \& Astronomy, University of Birmingham, Edgbaston, Birmingham B15 2TT, UK}
\affil[14]{Stellar Astrophysics Centre, Department of Physics and Astronomy, Aarhus University, Ny Munkegade 120, DK-8000 Aarhus C, Denmark}
\affil[15]{INAF-Osservatorio Astrofisico di Catania, Via S. Sofia, 78 - 95123 Catania, Italy}
\affil[16]{Dipartimento di Fisica e Astronomia, Università degli Studi di Bologna, Via Gobetti 93/2, I-40129 Bologna, Italy}
\affil[17]{INAF - Osservatorio di Astrofisica e Scienza dello Spazio di Bologna, Via Gobetti 93/3, I-40129 Bologna, Italy}
\affil[18]{Space sciences, Technologies and Astrophysics Research (STAR) Institute, Université de Liège, Quartier Agora, Allée du 6 Août 19c, Bât. B5c, B4000 Liège, Belgium}
\affil[19]{Institute of Space Sciences (ICE, CSIC), Carrer de Can Magrans S/N, 08193, Bellaterra, Spain}
\affil[20]{Institut d'Estudis Espacials de Catalunya (IEEC), Carrer Gran Capita 4, 08034, Barcelona, Spain}
\affil[21]{Université Côte d'Azur, Observatoire de la Côte d'Azur, CNRS, Laboratoire Lagrange, Bd de l'Observatoire, CS 34229, 06304 Nice cedex 4, France}
\affil[22]{Departament d'Astronomia i Astrofísica, Universitat de València, C. Dr. Moliner 50, 46100, Burjassot, Spain}
\affil[23]{AIM, CEA, CNRS, Univ. Paris-Saclay, Univ. Paris Diderot, Sorbonne Paris Cité, F-91191}

\date{}                     
\setcounter{Maxaffil}{0}
\renewcommand\Affilfont{\itshape\footnotesize}

\twocolumn[
\maketitle
\begin{onecolabstract}

In this work, we determine the expected yield of detections of solar-like oscillations for the targets of the foreseen PLATO ESA mission. 
Our estimates are based on a study of the  detection probability, which takes into account the
  properties of the target stars, using the information available in the PIC 1.1.0, 
including the current best estimate of the signal-to-noise ratio (S/N).
  The stellar samples, as defined for this mission, include those with the lowest noise level (P1 and P2 samples) and the P5 sample, which has a higher noise level. 
 For the P1 and P2 samples,  the S/N is high   enough (by construction) that we  can assume that the individual mode frequencies can be measured. 
 For these stars, we estimate the expected uncertainties  in  mass, radius, and age  due to statistical errors induced by uncertainties from the observations only. 
We used a formulation from the literature to calculate the detection probability. We validated this formulation and the underlying assumptions with {\it Kepler} data.  
 Once  validated, we applied this approach to the PLATO samples. Using again  {\it Kepler} data as a calibration set, we  also derived relations to estimate the uncertainties of 
seismically inferred  stellar mass, radius, and age. We then  applied those relations to the main sequence stars with masses equal to or below 1.2 $\rm{M}_\odot$ belonging to the 
 PLATO P1 and P2 samples  and for which we predict a positive  seismic detection.
We found that we can expect  positive  detections of solar-like oscillations for  more than 15 000 FGK stars  in one single field after
 a two-year observation run.  Among them,   1131  main sequence stars with masses of $\leq 1.2 \rm{M}_\odot$  satisfy  the PLATO 
 requirements for the  uncertainties of the seismically inferred  stellar masses, radii, and ages. 
The baseline  observation programme  of PLATO consists of observing  two  fields of similar size 
  (one in the southern hemisphere and one in the northern hemisphere)  for two years apiece. Accordingly, the expected seismic  yields of the mission   
  amount to over  30000 FGK  dwarfs and subgiants, with positive  detections of solar-like oscillations. 
 This sample of expected solar-like oscillating stars   is large enough to enable the PLATO mission's stellar objectives to be amply satisfied. 
The PLATO mission is expected to produce a catalog sample of  extremely well  seismically characterized stars of  a quality that is equivalent to  
the {\it Kepler} LEGACY sample,   but containing a number that is about 80 times  greater, when observing two PLATO fields for two years apiece. 
These stars are a gold mine that will make it possible to make significant advances in  stellar modelling. \\ \\ \\ \\

\end{onecolabstract}
]

\section{Introduction}
 
 The PLAnetary Transits and Oscillations of stars (PLATO) mission  is the  ESA Cosmic Vision M3 mission and its launch is scheduled for the end of  2026. 
 Its main objectives are 1) the  detection and  accurate and precise characterisation of exoplanets down to Earth-size planets in the habitable zone of 
 solar-like stars,  2)  the accurate  and precise determinations of the basic parameters of their host stars (mass, radius, age, etc), and 3) careful 
 statistical analyses of the above characteristics in order 
to better understand the formation and evolution of  planetary and  stellar systems (hereafter, stellar systems). Sufficiently precise determinations of the 
characteristics of these  stellar  systems require very high-quality photometry carried out continuously over long periods of time, hence, the need for a space mission. 
Furthermore, the required accuracy calls for  improvements of  the stellar models  used to estimate the age of the star-planet system. Improving the processes of stellar 
modelling is thus an intrinsic main objective of the PLATO mission \citep[][]{rauer2014,rauer2023}. 
The science operation phase of PLATO is  planned to last for four years with a possible extension of  4.5 years. The baseline is two  long-pointings (LOPs)  
observing one field for two years apiece. PLATO  will  collect high-precision photometric lightcurves of thousands of stars, which will be of
  particular interest for asteroseismological studies. 
To reach its objectives, the PLATO mission has defined a core programme with  several  types of stellar samples. Here, we focus  on the  P1, P2, and P5 samples. 
 The P1 and P2 samples   (hereafter, P1P2)  consist in the brightest  PLATO targets which will be observed with a 25 s cadence. The  P1 sample (resp., P2) 
includes at least 15 000  (resp., 1000) dwarf and subgiant stars (types F5 to K7), with $V \leq  11$ mag (resp. $V\leq 8.5$) observed over the mission 
and a noise level of $<50 ~{\rm  ppm ~h^{1/2}}$. The noise level of those samples has been adapted to enable precise seismological studies. 
The P5 sample  contains at least 245 000 dwarf and subgiant stars (F5-K7), with $V \leq 13$ mag cumulative over two target fields. 
Sampling of these light curves will be 600 s, but it is planned to acquire light curves with a shorter cadence of 50 s for the brightest targets in P5 
or for targets of particular interest.
For that reason, we also consider the stars of the P5 sample. Input information 
about the stars in those samples has been gathered in the Input PLATO Catalogue (PIC, \cite{montalto2021,nascimbeni2022} ). Here, we use   version PICv1.0.0. 
For more details about the PLATO 
project, we refer to \cite{rauer2014} and \cite{rauer2023}.

Asteroseismology, namely, the detection, measurement, and analysis of stellar oscillations, is one of the main tools that will be used in the framework of the PLATO mission to achieve these objectives. The mission is indeed designed  to detect stellar oscillation modes 
for different classes of stars, including solar-type ones which are its prime targets for exoplanet detection. Asteroseismology  can be used to infer 
the stellar properties, either by measuring global seismic parameters such as $\nu_{\rm max}$, the frequency at which the oscillation modes have their maximum amplitude, and/or $\Delta \nu$, the large frequency separation, which characterizes the pattern  of power spectra of solar-like oscillations. In ideal conditions, we can  precisely measure 
 the frequencies, amplitudes, and widths of individual oscillation modes, which then provide tighter constraints on the stellar mass, radius, and age.    

 Starting with the Sun, two decades of observations have demonstrated that solar-like oscillations offer a powerful way to derive precise stellar masses, radii, 
 densities, and ages, provided that high-quality seismic parameters are available \citep[see for instance the reviews by][]{chaplin2013,jcd2016,jcd2018b,garcia2019,jackiewicz2021,serenelli2021}. 
 However, this requires short-cadence (less than 1 mn for dwarfs and subgiants), ultra-high photometric precision (at the level of parts-per-million or ppm), 
 and nearly-uninterrupted long-duration (from months to years)  monitoring.

Here, we investigate the seismic performances of the core programme of the PLATO mission; more precisely, our goal is twofold: 
1) to obtain an estimate of the number of stars of the PLATO catalogue for which solar-like oscillations can be detected 
and 2) to obtain an estimate of the uncertainties on the stellar mass, radius, and age inferences in cases where the data are of 
high enough quality that individual modes can be detected and their frequencies be measured.  By uncertainties here we mean those statistical errors 
which  can only be decreased  with higher quality observations (i.e. higher S/N values and longer observation times). We then estimate the (statistical) uncertainties on
 stellar mass, radius, and age, which result from the propagation of observational errors on the seismic data.
We stress that systematic errors and/or biases  must be added to the statistical errors to obtain the final error budget. In the present case, systematic errors and/or biases  
mostly depend on our ability to improve our stellar modelling. This is only briefly discussed at the end of the paper.

The paper is organized as follows. In Sect.~\ref{global}, we detail the theoretical approach we used to derive the probability of detection of solar-like oscillations. 
The calculation is based on  \cite[][hereafter C11]{chaplin2011} 's methodology which was developed for studying the asteroseismic potential of the
 {\it Kepler} mission and  later on used for the TESS (\cite{campante2016},\cite{schofield2019}) and CHEOPS \citep{moya2018} missions. 
 In order to validate our own computations, we use several samples of  stars observed 
by the NASA {\it Kepler} mission \citep{borucki2010,koch2010} with and without detected solar-like oscillations. These test samples are presented in 
Sect.~\ref{validation}, together with the results about the reliability of the detection probability.  In Sect.~\ref{sect4}, we present our approach to 
estimate the uncertainties in the seismic inferences of stellar masses, radii, and ages (hereafter, MRA) when assuming that frequencies of  individual modes 
can be measured. In Sect.~\ref{PLATO}, we present the computations of the detection probabilities for stars in the P1P2 and P5 samples. 
Those calculations predict the number of stars in each sample for whichvwe expect to detect solar-like oscillations in the core programme of PLATO.
As assumed by the PLATO 
consortium, we consider two observational conditions: 1) we take the adopted noise level arising  from observations  from nominal 24 cameras at the beginning of 
life (BOL), so there is no degradation of the instrument; and 2) we take the noise level arising from observations by 22 cameras only at the end of of life (EOL), 
allowing for some degradation of the instrument as usually taken as reference by ESA.  
We use the noise level given in the  PIC that takes into account the fact that each target is observed by either 6, 12, 18, or 24 cameras 
\citep{montalto2021,nascimbeni2022} .
 The calculations are carried out for one LOP which will 
continuously observe the same field for at least two years. In Sect.~\ref{sect6}, for stars in the P1P2 samples with positive seismic detection, 
we compute the expected uncertainties of the individual frequencies  and deduce the resulting MRA uncertainties using the approach described in 
Sect.~\ref{sect4}. A summary and some discussion are provided in Sect.~\ref{summary}. Finally,
  we  give our  conclusions   in Sect.~\ref{conclusion}.

\section{Global solar-like oscillation detection level}
\label{global}

\subsection{Detection probability}
\label{proba}
In this section, we derive the formalism for computing the detection probability of solar-like oscillations, $P_{\rm det}$, in the photometric power density spectra. 
It is the probability that such oscillations are detected globally in the power spectrum, not to be confused with the probability to detect and measure the 
properties of individual oscillation modes. $P_{\rm det}$ is calculated according to Eqs. 26 and 28 of  C11. The statistics of the power spectrum is a $\chi^2$ with $2N_{\rm b}$ degrees of 
freedom, where $N_{\rm b}$ is the number of independent bins in the envelope band, $\delta \nu_{\rm env}$, of the power spectrum. The probability of having a peak above 
a given level $s$, due to noise only, in the binned power spectrum is then given by Eq.~1 of \cite{appourchaux2004}, namely, 
\begin{align} 
P(s^\prime>s,N_b) = \int_{s}^{\infty}  \frac{1}{\Gamma(N_{\rm b})}  ~ \frac{u^{N_{\rm b}-1}}{S^{N_{\rm b}}} ~e^{-u/S} {\rm d}u \;,
\end{align}
where $S$ is the mean of the power spectrum  and $\Gamma(n)$ is the Gamma function. 

Here we  seek the probability for any value  $s^\prime$ to be larger than a given level  $s$  for a binned power spectrum normalised to the noise level, $S/N$,
\begin{align} 
P(s^\prime>s,N_{\rm b}) = \int_{s}^{\infty }  \frac{1}{\Gamma(N_{\rm b})}  ~ \frac{u^{N_{\rm b}-1}}{(S/N)^{N_{\rm b}}} ~e^{-u/(S/N)} {\rm d}u \; .
\end{align}
With the change of variable  $u'= u/(S/N)$, the probability is given by 
\begin{align} 
\label{threshold}
P(x^\prime>x,N_{\rm b}) = \int_{x}^{\infty }  \frac{1}{\Gamma(N_{\rm b})}  ~ u^{\prime {N_{\rm b}-1}} ~e^{-u^\prime} {\rm d}u^\prime \; ,
\end{align}
which is the probability that the normalised power take any value  $x^\prime = (S/N)^\prime$ larger than a given level  $x= S/N$. 
For $x \rightarrow 0$, that-is 
 $S \rightarrow 0$, corresponding to no signal in the power spectrum, the  above probability is $P=1$ as expected. We note that for convenience, 
 C11 considered a slightly 
 different formulation and calculated the probability that any value  $x^\prime= [(S+N)/N]^\prime$ is larger than a given level  $x= (S+N)/N$, so that in absence of 
 signal $S$, $x=1$, an approach which we use in the following. 
 
As in C11, using the above equation, we first calculate a signal threshold $S_{\rm thres}$ such that the probability for any value $S \geq S_{\rm thres}$ to be due to 
pure noise is smaller than a predefined value $p_{\rm fa}$ (false alarm probability). In practice, we remain very conservative and choose a very small value for this 
false alarm probability, $p_{\rm fa} = 0.1\%$. 
 
In a second step, following C11, we now consider that there is some seismic signal  $S_{\rm mod}$ in the power spectrum. We obtain the probability of detection of this 
signal within the false alarm probability $p_{\rm fa}$ as the probability of being above the threshold defined previously (Eq.~\ref{threshold}),
 but with a new normalisation by $S_{\rm mod}+N$, 
\begin{align}
&P_{\rm det} \equiv P(s^\prime>S_{\rm thres}+N) \nonumber \\
&= \int_{S_{\rm thres}+N}^{\infty }  \frac{1}{\Gamma(N_{\rm b})}  ~ \frac{u^{N_{\rm b}-1}}{(S_{\rm mod}+N)^{N_{\rm b}}} \, {\rm e}^{-u/(S_{mod}+N)} \, {\rm d}u \, .
\end{align}

Choosing as before $u^\prime=u/(S_{\rm mod}+N)$ as the new variable, we obtain 
\begin{align} 
P_{\rm det}   =  \int_{u_{0}}^{\infty }  \frac{1}{\Gamma(N_{\rm b})}  ~ \frac{[(S_{\rm mod}+N)u^\prime]^{N_{\rm b}-1}}{(S_{\rm mod}+N)^{N_{\rm b}}}   \, {\rm e}^{-u^\prime} (S_{\rm mod}+N) \, {\rm d}u^\prime \, ,
\end{align}
which finally becomes
\begin{align}
\label{pdet}
P_{\rm det} = \int_{ u_{0}}^{\infty}  \frac{1}{\Gamma(N_{\rm b})}  ~ u^{\prime N_{\rm b}-1} \, {\rm e}^{-u^\prime} \, {\rm d}u^\prime  \, ,
\end{align}
where 
\begin{align}
\label{uinf}
u_{0} &= \frac{S_{\rm thres}+N}{S_{\rm mod}+N} = \frac{1+(S/N)_{\rm thres}}{1+(S/N)_{\rm mod}},
\end{align}
and $(S/N)_{\rm thres}$ is given by the solution of Eq.~(\ref{threshold}) for $P=p_{\rm fa}$. We note that Eqs.~(\ref{pdet})-(\ref{uinf}) are equivalent to Eqs.~(28) and (29) of C11.

In practice, the probability $P_{\rm det}$ can be computed for any particular target of PLATO or any other mission, 
such as {\it Kepler}, as soon as we have determined: 
(i) the signal-to-noise ratio (S/N) in the power spectrum $(S/N)_{\rm mod}$ and (ii) the number $N_{\rm b}$ of bins to consider in the envelope band of the oscillation modes. 
The derivation of these two quantities is detailed below. The last technical difficulty is then the computation of the $\Gamma$ function for the expected very large 
values of $N_{\rm b}$, which we perform using the classical asymptotic approximation, 
$$ \ln(\Gamma (n)) \approx  (n-1/2) \ln(n) -n +1/2 \ln(2\pi)               ,$$  which is valid for large values of $n$.

We consider that there is detection of power excess due to solar-like oscillations when the probability $P_{\rm det} >0.99$. 
Power spectra showing peaks above $S_{\rm thres}$ as defined by Eq.~(\ref{threshold}) for $p_{\rm fa}=0.1 \%$, but with $P_{\rm det} \leq 0.99 $ 
are considered as potentially indicating solar-like oscillations, but with too little confidence to derive any global seismic parameters, 
let alone properties of individual modes. This is a very conservative position, so that the resulting detection probability
 can be considered as a lower limit of what can be expected with PLATO.
 
 
\paragraph{2.1.1 Global S/N in the power spectrum:}
\label{globalsnr}

The global S/N entering Eq.~(\ref{uinf}) is calculated as
\begin{align}\label{snr}
   (S/N)_{\rm mod}  =  \frac{P_{\rm tot}} {N_{\rm tot}}   \, ,
\end{align}
where $P_{\rm tot}$ is the total power density in the oscillations and $N_{\rm tot}$ the total power density in the noise, both quantities being estimated near
 $\nu_{\rm max}$, the frequency where the oscillations reach their maximum amplitude.
In the following, we detail how these two quantities were calculated.

\paragraph{2.1.2 Oscillation power density  $P_{\rm tot}$:}

The calculation of the oscillation power density $P_{\rm tot}$ (in ppm$^2/\mu$Hz) can be performed using a formulation as established by C11 restricted to 
the calculation of the probability of a global detection of power excess due to solar-like oscillations.
The power density (in ppm$^2/\mu$Hz)   is  given by  (Eq. 19 in C11):  
 \begin{align}
 \label{ptotC11}
 P_{\rm tot}  \simeq \frac{1}{2}  \,  \frac{V_{\rm mod}^2 {\rm A}_{\rm max}^2}{\Delta \nu} \, , 
 \end{align}
 where  $\Delta \nu$ is in $\mu$Hz , $V_{\rm mod}^2 \simeq 3.1$ is the average visibility calculated for a set of 4 modes $\ell=0-3$ \citep{ballot2011}, 
and  $\rm{A}_{\rm max}$ (in ppm) is  the maximum oscillation amplitude of modes, reached at frequency $\nu \simeq \nu_{\rm max}$. 
  The  scaling laws for the determination of $\Delta \nu$ are given in Appendix~\ref{appendixA}.
 The estimate of $\rm{A}_{\rm max}$ is discussed in Sect.~\ref{sec:adoptedinputs} and Appendix~\ref{appendixA}.  

\paragraph{2.1.3 Noise power density $N_{\rm tot}$:}
\label{N_tot}   

We must also estimate the total noise power density$, N_{\rm tot}$. It is composed on the one hand of the   instrumental noise,  $N_{\rm inst}$, 
 which includes photon noise and all other instrumental contributors to the noise, and on the other hand of the stellar intrinsic noise, $N_{\rm gran}$, 
 which we assume is dominated by granulation noise at the relevant frequencies. The total noise power density is then given by:  
\begin{align} 
\label{btot}
N_{\rm tot}   = N_{\rm inst}+N_{\rm gran} \, .
\end{align}

Following \cite[][hereafter S19]{samadi2019}, the granulation noise power density, $N_{\rm gran}$, is calculated as a scaling power law of $\nu_{\rm max}$   (i.e. Eq.~36 in S19)
, as in \cite{kallinger2014}. The instrumental noise,  $N_{\rm inst}$, including photon noise as well as all other sources of noise from the instrument or from 
the background (satellite jitter, readout noise, digitisation noise, stellar background, zodiacal light, etc), will be a major contributor to the total noise entering
 the calculation of the $(S/N)_{\rm mod}$ term involved in Eq.~(\ref{uinf}). It depends on which sample of stars is considered:
 {\it Kepler} stars or PLATO targets and will discussed in Sect.~\ref{sec:adoptedinputs} below. 

\paragraph{2.1.4 Number of bins in the oscillation mode envelope:}
\label{sectNb}

Once $P_{\rm tot}$ and $N_{\rm tot}$, and thus $(S/N)_{\rm mod}$ are determined, we only need to determine the number of frequency bins $N_{\rm b}$ 
in the oscillation envelope in order to apply Eqs.~(\ref{threshold})-(\ref{pdet})-(\ref{uinf}). This number is given by
\begin{equation}
\label{Nb}
N_{\rm b} = Int\left[T_{\rm obs} ~\delta \nu_{\rm env} \times 10^{-6}\right] \, , 
\end{equation}
where $T_{\rm obs}$ is the total time interval of the photometric monitoring, in seconds, $\delta \nu_{\rm env}$ 
is the frequency range over which the oscillations are present in the power spectrum, in $\mu$Hz, and $Int$ denotes the integer part.

The parameter $\delta \nu_{\rm env}$ entering Eq.~(\ref{Nb}) is essential, because with the observing time, it  controls the number of degrees of freedom of the $\chi^2$ statistics followed by the power spectrum.
  
\subsection{Adopted inputs for the calculation of the detection probability $P_{\rm det}$} 
\label{sec:adoptedinputs}

In order to compute the global detection probability $P_{\rm det}$, we must provide as input the S/N, namely, $(S/N)_{\rm mod}$, 
the width of the envelope of the oscillations,  $\delta \nu_{\rm env}$, and the observation time, $T_{\rm obs}$, for each target. 
The S/N, here, $(S/N)_{\rm mod}$ in Eq.~(\ref{snr}) involves the power $P_{\rm tot}$. We estimate  $P_{\rm tot}$ 
at its maximum namely, at $\nu_{\rm max}$ leading $(S/N)_{\rm mod}$ at its maximum which we  denote hereafter $(S/N)_{\rm max}$ for clarity. 
We then use Eq.~(\ref{ptotC11}) assuming the existence of  a regular pattern in a power spectrum every $\Delta \nu$ and not individual modes.  
 
\subsubsection{Amplitude at maximum power}
   
The  amplitude of maximum power, $\rm{A}_{\rm max}$,  involved in $P_{\rm tot}$ is obtained in the literature under the form of an empirical relation 
depending on some combination of the global parameters among stellar mass, stellar radius, effective temperature and/or equivalently the seismic global 
parameters $\nu_{\rm max}$, and $\Delta \nu$.  
We consider two  relations, respectively given by C11 and S19. Both relations are scaled to the solar values.
In Appendix~\ref{appendixA}, we show that our adopted recalibrated theoretical values for $\rm{A}_{\rm max}$, $\rm{A}_{\rm max,scal}$ as given in 
Table~\ref{table:amax1}, are in good agreement with the observed amplitudes at maximum power $A_{\rm max, obs}$, as derived by \cite{lund2017} 
(their Table~3) for the {\it Kepler} LEGACY sample. In Table \ref{table:amax1}, we show how the stellar mass is derived from the scaling relation 
Eq.~(\ref{MRscal3}) of Appendix~\ref{appendixA}, relating the mass $M$ to the effective temperature $T_{\rm eff}$ and $\nu_{\rm max}$, namely,  
\begin{align}
\label{mass}
 \frac{\rm{M}}{\rm{M}_\odot} = \left(\frac{\nu_{\rm max}}{\nu_{\rm max,\odot}}\right)^{-0.28}  \left(\frac{\rm{T}_{\rm eff}}{\rm{T}_{\rm eff,\odot}}\right)^{3/2}  \, .
\end{align}
For the {\it Kepler} samples, $\nu_{\rm max}$ values are taken from the measurements by \cite{lund2017}. For the PLATO targets,  in Sect.~\ref{PLATO} and 
\ref{sect6}, we will take the  stellar radii and effective temperatures from PICv1.1.0    to derive  $\nu_{\rm max}$. For  the solar values, we adopt 
 $\rm{A}_{\rm max,bol,\odot}  = 2.53~$ppm \citep[rms value, see][]{michel2009}, $\nu_{\rm max,\odot}    = 3090 \, \mu$Hz,  $\Delta\nu_\odot  = 135.1~ \mu$Hz, and $\rm{T} _{\rm eff,\odot}  = 5777$ K throughout.

\begin{table} [h]
\centering 
\caption{Adopted amplitude formulation for $\rm{A}_{\rm max,scal}$ ($\rm{A}_{\rm max,C11}$ and 
$\rm{A}_{\rm max,S19}$ are obtained using the scaling relations as in C11 and  S19, respectively  in Appendix.~\ref{appendixA}. 
As described in Sect.~\ref{sec:adoptedinputs} and detailed in
 Appendix~\ref{appendixA},
 the proportionality factors are calibrated to the observations inferred by \cite{lund2017} ($\nu_{\rm max} $ in $\mu$Hz, the mass, $M$, in solar units). } 
\label{table:amax1}
\begin{tabular}{ |l | c | c| c|}
\hline
\hline
 $\nu_{\rm max}  $                      &   $ \leq 2500 $  &  $ > 2500 $   \\
\hline
  $ M \leq 1.15  $               &      $  \rm{A}_{\rm max, C11}*1.31  $                    &    $ \rm{A}_{\rm max, C11} *1.19 $                     \\
\hline 
  $ M > 1.15 $             &            $ \rm{A}_{\rm max, S19} *0.95$      $ \rm{A}_{\rm max, C11}*0.95 $            &                       \\
 \hline
\hline 
\end{tabular}
\end{table} 

\subsubsection{Instrumental noise}   

As discussed in Sect.~\ref{N_tot}, we need to estimate the  instrumental noise,  $N_{\rm inst}$, in the power spectrum. The {\it Kepler}  instrumental noise  used by C11
 is taken from \cite{gilliland2010}, namely, 
\begin{align}
\label{EqKepnoise}
 N_{\rm inst}  = 2 ~ N_{\rm rand}^2 ~10^{-6}~dt \, , 
\end{align}
where $N_{\rm rand}$, the total random noise per time interval of the data series (in ppm$^2$) is given by
\begin{align}
N_{\rm rand}^2  =   \alpha  \left(  1  + 0.1604  \left(\frac{12}{K_p}\right)^5   10^{-0.4~(12-Kp)} \right)  \, , 
\end{align}
with 
\begin{align}
\alpha  =   \frac{10^{5}}{1.28 } 10^{-0.4~(12-Kp) }  \, .
\end{align}  
In the above equations, $K_p$ is the {\it Kepler} magnitude of the star, ${\rm d}t$ is the cadence of the photometric series,  namely,  ${\rm d}t =58.85$ seconds for the {\it Kepler} short cadence mode. The remaining factor of 2 in Eq.~(\ref{EqKepnoise}) accounts for the choice
 of a single-sided power spectrum in C11, a convention that we   adopt for the remainder of the paper. 
 
For the PLATO calculations in Sects.~5 and~6, we use the noise level, $N_{\rm inst}$, that is available for each target in the PLATO catalogue, taking into account 
all instrumental sources of noise according to the most up-to-date understanding of the instrument (\cite{rauer2023} and Sect.~5). 
 
\subsubsection{Width of the envelope of the oscillations, $\delta \nu_{\rm env}$} 
 \label{deltanuenv}

 The parameter $\delta \nu_{\rm env}$ can be measured by the width of the assumed Gaussian-like shape envelope of the oscillations in the power spectrum. 
 Several formulations have been suggested for $\delta \nu_{\rm env}$ in the literature.
 They generally take the form of a scaling relation of the type $\delta \nu_{\rm env} =a \nu_{\rm max}^b$. 
 The coefficients $a$ and $b$ are obtained by fitting the {\it Kepler} data. 
 Their values differ according to whether one considers MS stars of spectral type G, K or hotter stars of spectral type F 
 or  subgiants, or  red giants (\cite{kim2021} and references therein).  
 For instance, \cite{kim2021} found slightly different values   depending on the formulation assumed for the granulation noise background. 
We find that these different relationships remain within the upper and lower limits  $\delta \nu_{\rm env}=\nu_{\rm max}$ and  $\delta \nu_{\rm env}=\nu_{\rm max}/2$, 
 respectively. Specificlly at low $\nu_{\rm max}$ (i.e. for more evolved stars), the curves are close 
 to $\nu_{\rm max}/2$, while the curves approach $\nu_{\rm max}$ at higher $\nu_{\rm max}$ (i.e. for younger stars). 
   In the following, we therefore consider both cases  $\delta \nu_{\rm env}=\nu_{\rm max}$ and  $\delta \nu_{\rm env}=\nu_{\rm max}/2$ , but keep the conservative case $\delta \nu_{\rm env} = \nu_{\rm max} /2$ for our baseline and estimate the changes in the detection 
   predictions when using $\delta \nu_{\rm env}=\nu_{\rm max}$.
  
  \medskip
\subsubsection{Main sequence versus subgiant stars} 
\label{classification}

For purpose of discussion presented later on in this work,  we  distinguish the cases of main sequence stars (MS stars) and subgiants. 
From a stellar evolutionary point of view, the subgiant phase starts when there is not enough hydrogen left at the centre to produce nuclear energy and kinetic 
pressure to sustain gravity.  We use the central hydrogen  mass fraction, $X_c$, to define a threshold.  
 The  MS stars  are then  defined with $X_c \geq 10^{-6}$. 
According to  our stellar models for the range of mass of interest here and the  adopted solar chemical composition, the subgiants satisfy 
 \begin{equation}\label{transition}
\log \left(\frac{L}{L_\odot}\right) > 10\;  \left(\log T_{\rm eff}-3.7532\right)+0.25 \, .
 \end{equation}
 The transition between MS tars and subgiants is located in a HR diagram in Fig.\ref{HR} in Appendix~\ref{appendixC}.

 {\it Mass subsamples:} 
 In the present work, we consider that only  MS stars  with  $M \leq 1.6 ~ \rm{M}_\odot$ can show solar-like oscillations 
  whereas subgiant  being evolved and therefore cooler, can still oscillate  with solar-like oscillations while being more massive. 
  We therefore study more carefully the sample of MS stars with seismic masses below 1.6 $ \rm{M}_\odot$  
 while no mass restriction is made for the subgiants. The mass threshold corresponds approximatively to the transition between stars with no 
 convective core like the Sun and stars with a convective core in the MS. We also draw a specific attention to the  subsample of main-sequence (MS) stars with 
  seismic masses $M \leq 1.2 ~ \rm{M}_\odot$. The reason is that the stellar requirements of the PLATO mission are established 
 for a  star like the Sun (in mass and age).  
  
\section{Validation of the calculation of the detection probability with  {\it Kepler} stars}
\label{validation}
 
The formalism described in Sect.~\ref{global} must be validated before being applied to the stars of the PLATO Input catalogue. Here we use several samples of 
stars observed by {\it Kepler}, in order to verify the performance of this formalism 
 in predicting detectability of solar-like oscillations. In other words, we checked with which confidence level Eqs.~(\ref{threshold})-(\ref{pdet})-(\ref{uinf}), 
 with the above prescriptions for $P_{\rm tot}$, $N_{\rm tot}$ and $N_{\rm b}$, can predict whether or not solar-like oscillations can be detected. 
 For that purpose, we 
 use   two types of {\it Kepler} stars, those for which such oscillations were or were not detected. The confidence level will be measured 
 in terms of false positive and false negative 
 predicted detections. 

\subsection{{\it Kepler} data sets for calibration and validation of theoretical calculations}\label{Keplersamples}
   
 \subsubsection{Sample 1: large sample of stars with solar-like oscillations detected by  {\it Kepler}}
 
The first data set used to construct this sample is a compilation of {\it Kepler} short cadence stars by     \citep[][hereafter S17]{serenelli2017}. 
It includes 415 stars with known detected oscillations, as already reported by C11.
 Since 2011, these stars were further observed over time intervals ranging from 40 days up to 1055 days. For most of these stars, only the global seismic parameters $\nu_{\rm max}$ and $\Delta \nu$ are available. 

A second {\it Kepler} data set used to construct our sample 1  is an updated compilation by  \cite[][hereafter M22]{mathur2022} of the {\it Kepler} 
short-cadence stars with detected solar-like oscillations, derived on the basis of samples from C11, \cite{chaplin2014} and S17. 
It provides a homogeneous catalog of global seismic parameters for 624 stars. 
 
 In order to build a final sample of stars with detected oscillations with all the necessary parameters available,  
 we considered the  set  of 413 stars common to  M22 (updated $\nu_{\rm max}$ and $\Delta \nu$) and S17 (observing intervals and grid-based inferred  stellar 
 mass and radius).  In the following we use the updated $\nu_{\rm max}$ and $\Delta \nu$ from M22. 
 In Appendix~\ref{prediction}, we look at the impact  of choosing the values of  $\nu_{\rm max}$, $\Delta \nu$,  and 
 $T_{\rm eff}$ from S17 instead of M22.
 
\subsubsection{Sample 2: large sample of stars with no oscillations detected by  {\it Kepler}}

As a second sample, we consider the list of 990 {\it Kepler} short-cadence main-sequence solar-like stars for which analyses revealed   no  detected
 oscillations, as published by  \cite[][hereafter M19]{mathur2019}. For each star in sample 2, the value of $\nu_{\rm max}$ is computed according to 
\begin{align}
\label{numaxlogg}
\frac{\nu_{\rm max}}{\nu_{\rm max,\odot}} =\frac{\rm g}{\rm{g}_\odot} \left(\frac{\rm{T}_{\rm eff}}{\rm{T}_{\rm eff,\odot}}\right)^{-1/2}  ,
\end{align}
where the surface gravity $g=GM/R^2$ and $T_{\rm eff}$ are taken from M19.
\footnote{For the star KIC 4464952, we rather use the LAMOST value $T_{\rm eff}=6161\pm 214$K.}. Stellar masses are obtained from the seismic scaling law 
Eq.~(\ref{mass}).  This  provides $\delta \nu_{\rm env}$.The observation time, $T_{\rm obs}$, is taken from KASOC ({\it Kepler} 
 Asteroseismic Science Operations Center provides asteroseismological data).
 
Here again, we focused on stars with masses $M <1.6 ~\rm{M}_\odot$. The resulting set of 833 stars constitutes our final sample of {\it Kepler} 
non-oscillating stars (sample 2).

\subsubsection {{\it Kepler} LEGACY sample}  
  
  Finally, we need  to validate and calibrate the calculation of the oscillation maximum amplitude $\rm{A}_{\rm max}$, as detailed in Appendix~\ref{appendixA},
    as well as \cite{libbrecht1992}'s relation between
   individual mode frequencies, linewidths, and S/Ns (see Appendix~\ref{appendixD}), We then used the {\it Kepler} Legacy sample, which is composed of 66 main-sequence stars with the highest quality of 
    seismic data \citep{lund2017} (in the following). For those stars,  individual modes are identified. Indeed solar-like oscillation modes can be described 
    by  spherical harmonics with spherical degree $\ell$ and  azimuthal order $m$  for their surface geometry and by the radial order $n$ labelling 
    the overtones of a given $\ell,m$ mode. 
     When rotation is not taken into account or cannot be detected seismically, the modes are $m$-degenerate and the mode frequencies do not depend on $m$. This is the case here, so for each individual mode $\ell,n$, the  frequency, amplitude and line width are measured with the highest precision. For those stars, the observed values of $\nu_{\rm max}$ are taken from  L17.   

\subsection{Results of the validation: Performance of the detection probability approach}

Our approach for calculating the probability to globally detect solar-like oscillations was tested against the above  {\it Kepler} samples. 
Using the formalism described in Sect.~\ref{proba} and the various needed inputs as explained in Sect.~\ref{sec:adoptedinputs}, we assessed on the one hand 
the fraction of {\it Kepler} targets with detected oscillations for which we predict no detection (false negatives) and on the other hand the fraction of 
{\it Kepler} targets with no detected oscillations for which we predict detection (false positives). 

In prevision of the investigation for the PLATO case, we made as our baseline the conservative choices of a positive detection when $P_{\rm det} > 0.99$ 
and $\delta \nu_{\rm env} =\nu_{\rm max}/2$. As summarized in Table~\ref{table:table16} in Appendix~\ref{prediction},  considering the total population of
 1349 {\it Kepler} stars (MS stars with masses $M< 1.6 \rm{M}_\odot$  and subgiants of all masses,  hereafter $R_1$ sample) 
 with both predicted false seismic positive  (186 stars) and negative (40 stars) detections of oscillations in the baseline conditions leads to  
 an underestimate of the number of real detections (false negative) by $\sim 3\%$  
 for the PLATO samples. On the other hand, we can see in Table~\ref{table:table16} that one  overestimates the number of real detection (false positive) by 
  14\%. If one considers only MS stars with  masses 
 $M \leq 1.2 \rm{M}_\odot$, we overestimate the  number of real detection by 7\%. These tests using {\it Kepler} results confirm that our approach is valid 
 within the quoted uncertainties and will be used for the PLATO targets in Sect.5.
The detailed results of the calculations of the above results, as well as justifications of the choices made for defining our baseline, 
are given in Appendix~\ref{prediction}.

 We must stress here that  the high percentage  (14\%) of  false positive detection  for the $R_1$ sample 
 is mostly due to subgiant  stars with  masses larger than 1.6 $M_\odot$.   
 If we consider a subsample of  stars including both MS stars  and subgiants with masses less than $1.6 M_\odot$,
   the percentage of false positive detection of    14\%  decreases to $\sim$ 9\%. 
 
False positive detections can be due to actual amplitudes being lower than predicted. 
Several reasons have been put forward to explain lower-than-expected amplitudes for {\it Kepler} stars, including  significant magnetic 
activity (\cite{chaplin2011b}; M19) 
and low metallicity (\cite{samadi2010}; M19). This likely depends on the properties of the stars themselves 
(mass, luminosity, temperature, rotation, magnetism, etc.). 
 M19 provided the iron-to-hydrogen mass fraction [Fe/H] and 
the photometric proxy for magnetic activity $S_{ph}$  \citep{garcia2010, mathur2014, santos2023} which  measures   the amplitude of the
spot modulation in the light curves and must be considered as a lower limit of the stellar activity \citep{salabert2016, salabert2017}. 
M19 then found  that the probability of non detection  of solar-type oscillations  is $\sim 98.7\%$ when $S_{ph} >2000 $ ppm 
(for reference, M19 gave $S_{\rm ph,min}=67.4$ ppm and $S_{\rm ph, max}=314.5$ ppm at the minimum and maximum of activity for the Sun and  that 
the solar oscillation amplitudes decrease by 12.5 \% from minimum to maximum of activity).
Concerning the impact of metallicity, the magnitude of the amplitude decrease due to a low metallicity    remains uncertain and   solar-like oscillations have been 
detected for  some metal poor stars. So there is no clear one-to-one correspondence between metallicity and non detection of solar-like oscillation. 
 Nevertheless it is still of interest to look at  the   146  stars  with false positive detection that have values for  $Sph$ and [Fe/H] in the M19 sample. This subsample 
  includes  100 subgiants among which 72 with masses larger than 1.6 $M_\odot$.
We then note  that none of those 72 subgiants have high stellar activity  and only 38 of them are metal poor compared to the 
Sun ([Fe/H]$<-0.1$). We also considered the whole sample of 146 stars with false positive detection  and available
  values for  $Sph$ and [Fe/H]  and  found  82  stars with a high activity level ($S_{ph} > 2000$) or [Fe/H] $ < -0.1$.  
Removing those 82 stars from the original  sample of 186  stars with false positive detection leaves 104 stars.  
Using 104 stars instead of the original 186 stars with false positive detections, we find that 
the percentage of false positive for the $R_1$ sample  drops to $\sim 7.7$\%. 

In any case,  taking into account stellar activity and metallicity,  and/or  additional specific properties of the stars to explain the whole sample of false detections  
  would   deserve further investigation but  is out of scope of the 
present paper.   Since such detailed  information are not yet available for the PIC 1.1.0 stars  we will therefore keep a conservative  value of   
  14\% for the false positive uncertainty.  Accordingly we  later give  the values of $X$ together with their uncertainties under the form   $X^{+3\%}_{-14\%}$ for the PLATO subsample of 
  MS stars with masses $M< 1.6 \rm{M}_\odot$  and subgiants of all masses. 
  However, as an optimist remark,   let us stress that stars in the M19 sample were observed over one month only.
   As the S/N increases over time,
   we expect to reach smaller amplitudes, everything else equal, with the PLATO mission and therefore a smaller percentage
    of false positive detection due to too small oscillation amplitudes.
    
\section {Uncertainties on stellar properties in cases of individual frequency measurements}
\label{sect4}

For estimating the seismic MRA inference  performances, we go on to consider the case when  the  mode frequencies  can be measured  individually. 
We  derived empirical relations giving  the mass and radius relative uncertainties  as a function of the uncertainty $\delta \nu_{\ell=1,max}$ of  
the $\ell=1$ mode  closest to $\nu_{\rm max}$, the frequency at maximum power (those modes have the smallest uncertainties). 
For that purpose, we used the stellar evolution code CESTAM \citep{morel2008,marques2013} to build a set of stellar models of  MS stars with masses of 
$M\leq 1.2 \rm{M}_\odot$, for which we numerically computed  the individual frequencies using the ADIPLS code \citep{jcd2008}. 
The frequency uncertainties  are obtained from a rescaling of the frequency uncertainties derived for a 'degraded Sun' \citep{lund2017}. We then  used the above frequency set  for each 
synthetic star in the  inference code AIMS \citep{rendle2019,lund2018} updated for the present purpose by one of the co-authors (D. Reese) in order to infer the MRA and their statistical 
uncertainties. It is known that non seismic constraints play only a minor role when the inference includes a large number (e.g. a few dozen) of highly precise individual frequencies. 
We nevertheless include  uncertainties  for non-seismic constraints: a generic 70 K as an uncertainty for 
 the effective temperature and 0.05 dex for the metallicity expected from individual spectroscopic study 
 (e.g see  the PASTEL catalogue, \cite{soubiran2022}).
 For Sun-like stars, differential studies with respect to the Sun are even more precise  and accurate  \citep{morel2021}. 
 On purpose, we  did not introduce any systematic  errors, so that we can estimate the seismic performances specifically due to the quality of the data. 
 We then established a correlation between the MRA uncertainties and the frequency uncertainty $\delta \nu_{\ell=1,max}$ which was then fitted.  We found the following fitted 
 relations (see Eq.~(\ref{MR1c}) in Appendix~\ref{appendixC}):
\begin{align}\label{MR1c}
 \left\{
\begin{array}{ll}
\delta M / M &=   2.083 ~ \delta \nu_{\ell=1,max} + 0.046    \\ 
\delta R / R  &= 0.707  ~ \delta \nu_{\ell=1,max} + 0.149 \, , 
 \end{array}
\right.
\end{align}
where $\delta \nu_{\ell=1,max}$ is in $\mu$Hz. All details of these calculations are presented in Appendix~\ref{appendixC}.
 The  MRA relative  uncertainties (Eq.~\ref{MR1c}) can then be seen as a lower limit 
of what can be achieved given the observational constraints  (which depend on the observing conditions).
Realistic uncertainties require to ad d systematic  errors to obtain the final error budget. This is out of scope of the present work 
(but see  the discussion in Sect.\ref{summary}).     

For the PLATO targets,  we determine the (theoretical) uncertainties on the individual frequencies using the \cite{libbrecht1992} formula
 (Eq.~\ref{sigma_libb} in Appendix~\ref{appendixD}), $\sigma_{\rm Libb}$, which depends on  the S/N for that particular mode   
 and on  the duration of the observation and has proven to yield the right order of magnitude. 
 The S/N  in such a case  is given by the 
 the power per resolved mode -instead of the global power density in the oscillation envelope as before- over the background noise.
 The power per resolved mode is related to the  height of the mode. Accordingly, 
  the  power density per resolved mode  is derived from (S19) and \cite{lochard2003} for a single-sided spectrum \citep[see also][]{appourchaux2004}:
\begin{align}
\label{ptotS19}
P_{\rm mod} = 2 V_1 \, \frac{A^2_{\rm max}}{\pi \Gamma_{\rm max}} \, ,
\end{align}  where $\Gamma_{\rm max}$ is the mode linewidth at $\nu\approx\nu_{\rm max}$, expressed in $\mu$Hz. 
For $\ell=1$ modes,   the square visibility is $V_1=1.5$ \citep{ballot2011}.
The computation of $\rm{A}_{\rm max}$ is described in Appendix~\ref{appendixA} (see in particular Table \ref{table:amax}). Estimates of the mode linewidths are obtained by a fit of 
measurements in \cite{lund2017} as a function of effective temperature (see Appendix~\ref{appendixD}, and in particular Table \ref{table:gamma} for details).
 The \cite{libbrecht1992} formula  predicts the uncertainty  for a single, isolated mode peak such as the  $\ell=0$ modes. For  higher $\ell$ degree modes, 
 one should take into account the fact that they are comprised of multiple components that might not be resolved; and moreover, 
 the exact appearance of the non-radial modes 
 will depend on the angle of inclination presented by the star, currently  unknown for the PLATO targets. 
 We therefore rather adopt an empirical approach
   using  real data from the {\it Kepler} mission: we  establish in Appendix~\ref{appendixD} a relation between the theoretical Libbrecht uncertainty  of 
    a $\ell=1$ mode at $\nu_{max}$,  $\sigma_{\rm Libb,\ell=1,max}$,  and 
      the measured frequency uncertainty $\delta \nu_{l=1,max}$ for the same mode  for  stars of the {\it Kepler} Legacy sample.  
   As found in Appendix~\ref{appendixD} (Fig.\ref{figure:ratsig2}),  the ratio 
 $\delta \nu_{\rm \ell=1,max}/\sigma_{\rm Libb,\ell=1,max}$
  tends to decrease with the effective temperature of the star and the decrease is significant over the effective temperature interval found for the PLATO targets.
 The fit of the ratio $\delta \nu_{\rm \ell=1,max} / \sigma_{\rm Libb,\ell=1,max}$ as a function of effective temperature for 
 the stars of {\it Kepler} LEGACY sample is shown  in Appendix~\ref{appendixD} (see Eq.~\ref{ratsig}) and yields 
 \begin{align}
\label{ratsig4}
\delta\nu_{\rm \ell=1,max}= \sigma_{\rm Libb,\ell=1,max} ~\left(4.89-4.18~\frac{T_{\rm eff}}{6000}\right) \, , 
\end{align}
valid for $5000 < T_{\rm eff} \leq 6200$ K. This roughly corresponds to the effective temperature  range of the PLATO targets 
for which we will compute those uncertainties later on.  
 For each PLATO target, we will  compute  $\sigma_{\rm Libb,\ell=1,max}$ then derive $\delta \nu_{l=1,max}$ using Eq. \ref{ratsig4}, 
 before using Eq.~(\ref{MR1c}) for the stellar mass 
and radius uncertainties. 
 
\section{Expected solar-like oscillations with PLATO}
\label{PLATO}

In this section, we estimate the number of stars for which solar-like oscillations are expected to be detected with the PLATO mission. For this purpose and as already mentioned, we took for each PLATO target star  the  stellar radius and effective temperature from  PICv1.1.0 \citep{montalto2021, nascimbeni2022}. The luminosity is then derived  as: 
\begin{align}
\label{luminosity}
\log \left(\frac{L}{L_\odot}\right) = 2 \log \left(\frac{R}{R_\odot}\right) + 4 ~   \log \left(\frac{T_{\rm eff}}{T_{\rm eff, \odot}} \right) \, .
\end{align}
 
We consider, on one hand, the P1P2 sample and, on the other, the P5 sample, both  for one LOP. We  removed  the hot stars that appear in the instability  strip 
using the criterion from C11
\begin{align} 
\log T_{\rm eff}  > \log(8907)-0.093 \log(L/L_\odot). 
\end{align}
This  eliminated only   a few  stars because the criterion on the temperature of the hot side adopted to construct the PIC is much more severe. We also removed the early red giants 
from the PIC sample, based on their location in a theoretical Hertzsprung-Russell (HR) diagram.  These stars
  are included in a specific scientific validation sub-catalogue of the PIC \citep{aerts2023}. 
  PLATO performances for such evolved  stars were assessed by \cite{miglio2017}. 
  These stars are located at the base of the red giant branch in a HR  diagram which we use as 
   an empirical criterion to remove them. We define the  criterion in terms of luminosity 
   and effective temperature by computing and plotting  a set of evolutionary tracks  with different masses and locating the onset of the red giant branch in the HR 
   diagram. This leads us to remove stars when they satisfy:
\begin{align}
\log T_{\rm eff} \le 3.66+0.05 \log \frac{L}{L_\odot}.
\end{align}
After the removal of hot stars and evolved ones, we are left with a set of 7,009 stars in our P1P2 sample and 130 140 stars in our P5 sample for one LOP.
The stellar mass was derived from the seismic scaling  relation Eq.~(\ref{mass}) where $\nu_{\rm max}$ is evaluated here according to Eq.~(\ref{invertedradius}):
\begin{align}
  \frac{\nu_{\rm max}}{\nu_{\rm max,\odot}} = \left(\frac{R}{R_\odot}\right)^{-1.5625} \ \left(\frac{T_{\rm eff}}{T_{\rm eff,\odot}}\right)^{0.78}. 
\end{align}
Appendix~\ref{appendixA} offers more details. The mass is  used only to consider various subsamples of stars  when analysing the results of the calculations. 
 MS stars with masses larger than 1.6 $\rm{M}_\odot$   are too hot  and therefore  unlikely to show solar-like oscillations, 
except perhaps for the stars with high metallicity. In absence of information about metallicity  at the present time, hereafter we exclude  MS stars with 
predicted seismic masses $M> 1.6 \rm{M}_\odot$.

The detection probability, $P_{\rm det}$, is obtained  using Eq.~(\ref{pdet})  which involves the (S/N)$_{\rm max}$ (Eq.~\ref{snr}), 
the observing time $T_{\rm obs}$ and the  width of Gaussian-like envelope of the oscillation power spectrum  $\delta \nu_{\rm env}$. The amplitudes, $\rm{A}_{\rm max}$,  
used to compute $(S/N)_{\rm max}$  are taken according to Table \ref{table:amax1}. In the calculation  of $N_{\rm tot}$  (Eq.~\ref{btot}), we  used for $N_{\rm inst}$  
the PLATO (random and systematic residuals) noise level included in the PIC1.1.0 , $N_{\rm PIC}$, for EOL conditions, which was then converted in ppm$^2/\mu$Hz.  For the  BOL 
conditions, we used the data provided by one of the co-authors \citep{borner2023} .  Because the convention in our calculation is a single-sided spectrum as for the power density, 
we take $N_{\rm inst} =2 N_{\rm PIC}$. We then added the single-sided stellar granulation background noise as used in S19 (see Sect.~\ref{global}).
 
To remain conservative, we kept only those stars for which the probability of the signal be due to noise is 0.1\% or less and of those stars we kept only stars for which 
the probability of the signal being due to solar-like oscillation is larger than  99 \%.

\subsection{Expected solar-like oscillations within the PLATO P1P2 sample }
 
\begin{figure}[!]
        \begin{center}
                \includegraphics[width=9.cm]{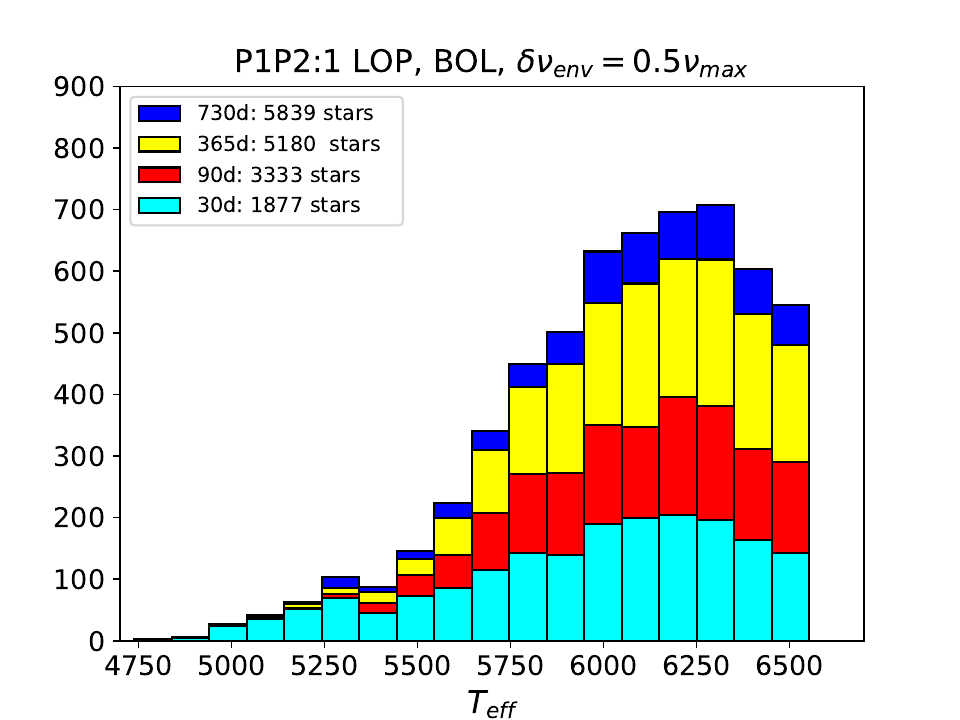}

                \includegraphics[width=9.cm]{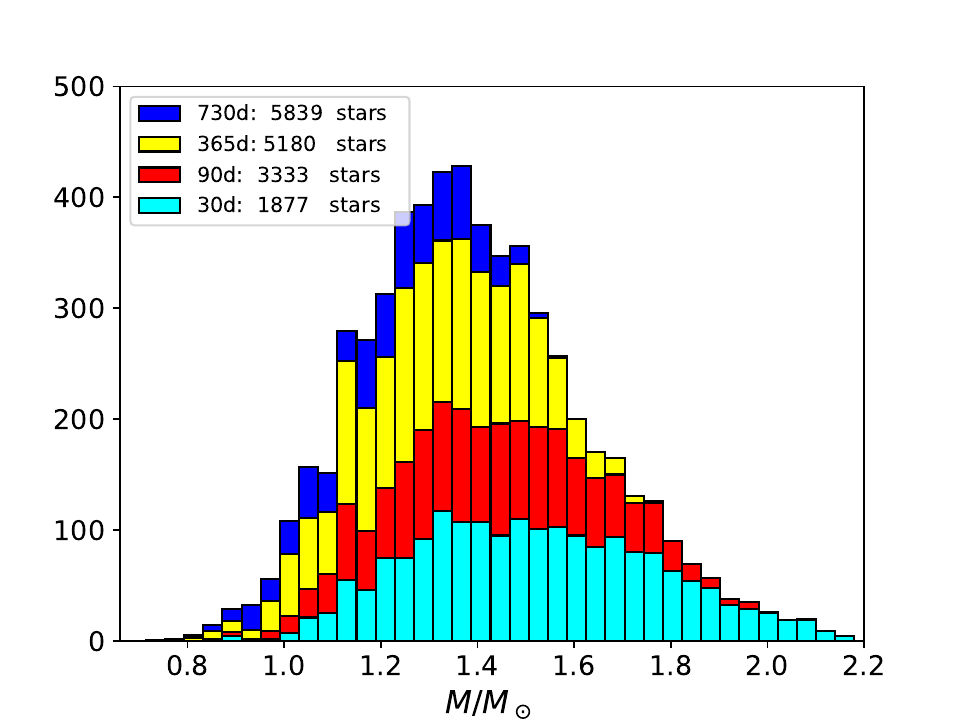}

                \includegraphics[width=9.cm]{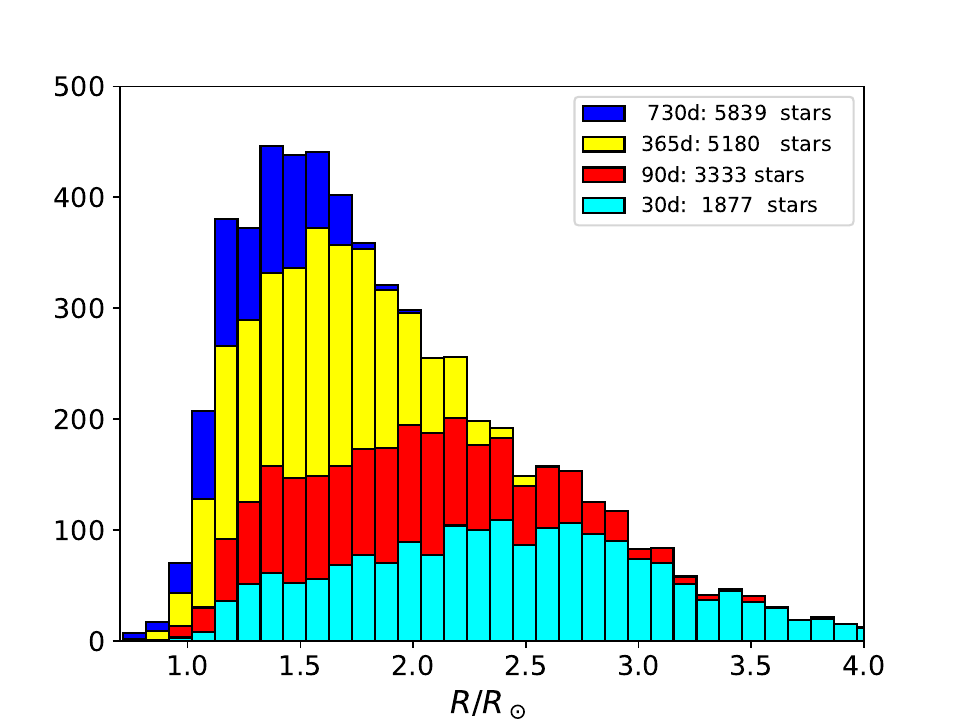}              
                \caption{Histograms  of the number of  stars from the P1P2 sample (subgiants of all masses and MS stars with  masses  
                $M/\rm{M}_\odot <1.6 $) with a 
                probability of $>$ 99 \% of positive detection of solar-like oscillation  in the case of  (1 LOP, BOL,$\nu_{\rm env}=0.5 \nu_{\rm max}$). 
                Top: Distribution in $T_{\rm eff}$, middle in stellar mass, bottom in stellar radius. The color code represents the assumed observing duration.}
                \label{histoP1P2BOL}
        \end{center}
\end{figure}

\begin{figure}[!]
        \begin{center}
                \includegraphics[width=9.cm]{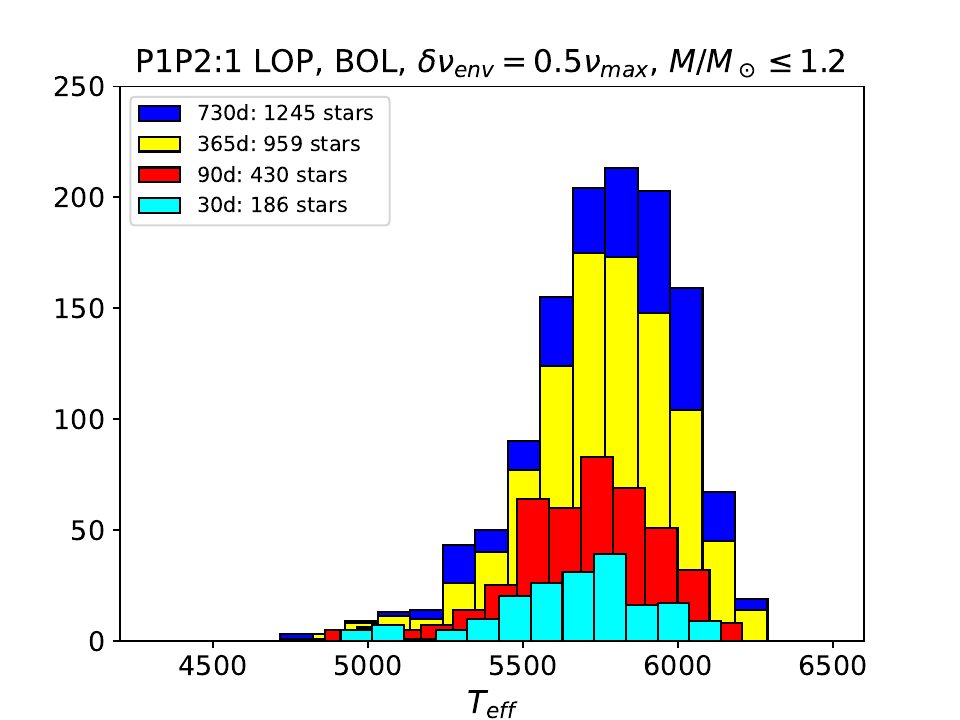}

                \includegraphics[width=9.cm]{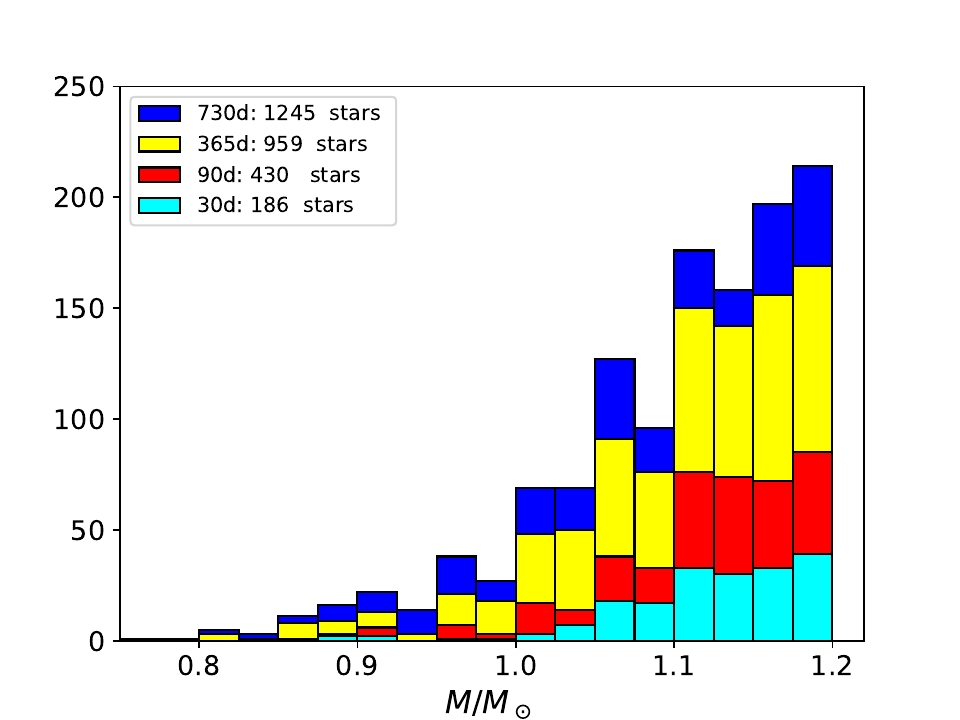}

                \includegraphics[width=9.cm]{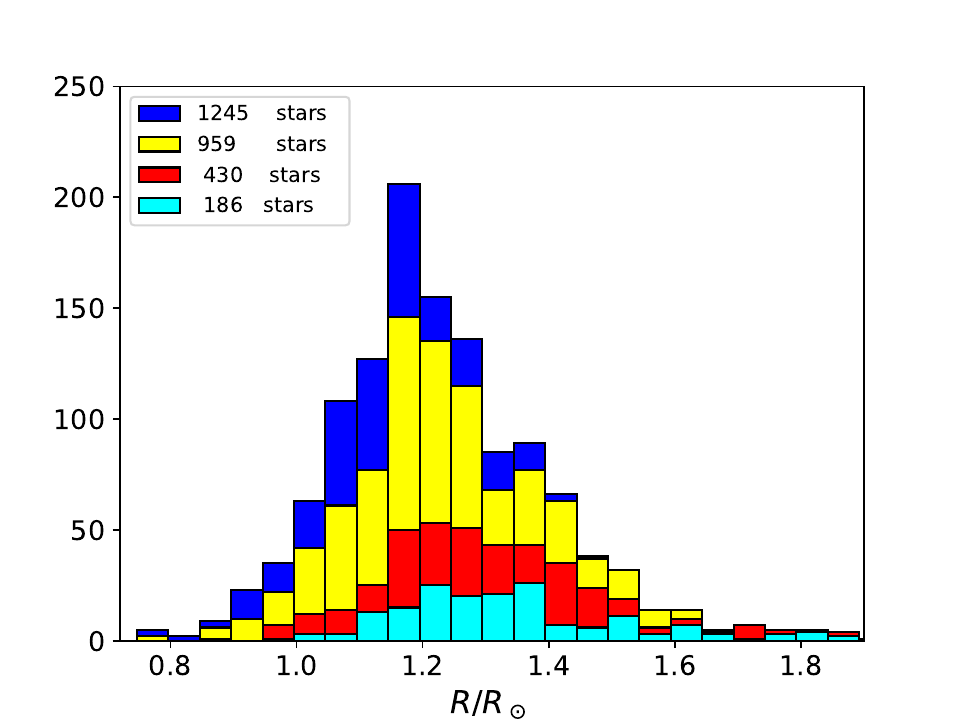}              
                \caption{Histograms  of the number of stars
   from the P1P2 sample (with masses  $M/\rm{M}_\odot \leq  1.2 $)   with a probability $>$ 99\% 
   of  positive detection of solar-like oscillation  in (1 LOP, BOL, $\delta \nu_{\rm env}=0.5 \nu_{\rm max}$) conditions.
    Top: Distribution in $T_{\rm eff}$; middle in stellar mass; bottom: in stellar radius. 
    The color code represents the assumed observing duration.}
                \label{histoP1P2BOL2}
        \end{center}
\end{figure}

 The predicted numbers of P1P2  target stars with positive  oscillation detection obtained in  different mass subsamples are collected in Tables~\ref{table:EOLBOL} 
 and~\ref{table:detection}.  When we apply the  (1 LOP, BOL, $\delta \nu_{\rm env}=\nu_{\rm max}/2$) conditions after two years of observation,
 we expect to detect   solar-like oscillations for at least  5839  stars (of which 2732 MS stars with $M \leq 1.6 \rm{M}_\odot$ and  3107 subgiants.). 
 Figure~\ref{histoP1P2BOL} show the distributions 
 of those stars with expected detected solar-like oscillations as a function of $\log T_{\rm eff}$, stellar mass and radius. One expects to detect solar-like 
 oscillations in a   sub-sample of 1245  stars with $M/\rm{M}_\odot \leq 1.2$ after two years of observation in BOL conditions (Fig.~\ref{histoP1P2BOL2}).
 When restricted to  MS stars with masses $\leq 1.2 \rm{M}_\odot$, the subsample counts 1016 stars. Those stars are small and therefore of the utmost interest for detecting small planets. On the stellar side,  more massive stars   are likely prone to large systematic uncertainties
   because they  developed a convective core, the exact extent of which is   unknown.
   Stars with masses $M\sim 1.2\rm{M}_\odot$ might also develop a convective core  but it is small enough 
   that convective overshoot does not contribute significantly to the total error budget on the age.  

\begin{itemize}
\item  {\it Impact of BOL/EOL conditions:} assuming EOL conditions instead of BOL ones,  we would lose a few tens of percent of stars, 
  mostly at small mass and radius (Table~\ref{table:EOLBOL}).
\begin{table} [h]
\centering
\caption{\label{table:EOLBOL}  Numbers of stars in the P1P2 sample in 1 LOP with expected detection of solar-like oscillations 
after 730 days of observation and assuming $\delta \nu_{\rm env}=\nu_{\rm max}/2$. Stellar masses, $M$, and radii $R$ in solar units.}
\begin{tabular}{ ||l | c | c| c||}
\hline
\hline
cases                &      BOL      &   EOL            \\
\hline
all                  &         5858        &    5553         \\
MS stars             &       2751           &   2449   \\
$M<1.6$              &         4744      &        4439        \\
$M<1.6$, MS stars     &       2732      &     2430        \\
$M \leq 1.2$           &    1245       &     1106          \\
$M \leq 1.2$, MS stars &        1016     &     830      \\ 
$R \leq 1.1$           &      269     &             203       \\ 
\hline
\end{tabular}
\end{table}  

 \item {\it Impact of $T_{\rm obs}$:} as can be expected and be seen in Table~\ref{table:detection}, the observing time plays an important role. The increase 
of the number of stars  with predicted positive seismic detection is a factor $\sim$  5-7 greater when increasing the observing time from   30 days to two years. 
After 30 days, one expects  1877  stars (among which  1596 subgiants), number which increases up to  5858  
stars (among which 3107  subgiants) after two years of observation.
Not surprisingly, stars for which we might not typically detect solar-like oscillations for too short an observing time are  low-mass, MS stars because
 their oscillation amplitudes are too small. 
 It is also interesting to consider  the distributions of stars with positive seismic detection of solar-like oscillations  with their magnitudes.
 Solar oscillations for stars  with magnitude $>10.5$ will be detected  only after about one year of observations. 
 
\begin{table} [h]
\centering
\caption{\label{table:detection} Number of stars in the P1P2 sample  in 1 LOP with    expected  detection of 
solar-like oscillations ($P_{dec}>0.99$) assuming $\delta \nu_{\rm env}=\nu_{\rm max}/2$.  The numbers without parenthesis correspond to 
stars with estimated seismic masses $\leq 1.2 \rm{M}_\odot$, whereas the numbers
 in parenthesis correspond to  stars with all masses. } 
\begin{tabular}{ ||l | c | c| c||}
\hline
\hline
                                         &      BOL               &               \\
\hline
$ \delta \nu_{\rm env}         $             &   730 days        &   30 days    \\
\hline
\hline
 $\nu_{\rm max}/2  $                         &      1245 (5858)   &      186 (1877)    \\
\hline 
  $ \nu_{\rm max} $                          &    1541 (6387)     &  329 (2811)     \\
\hline
\hline 
                                        &      EOL            &                \\
\hline
 $\delta \nu_{\rm env}$             &   730 days        &   30 days    \\
 \hline
\hline
$\nu_{\rm max}/2  $                 &    1106 (5553)     &     151(1591)   \\
\hline 
  $ \nu_{\rm max} $                 &    1389 (6131)   &     267 (2399)    \\
\hline
\end{tabular}
\end{table} 

\item {\it Impact of uncertainties in the probability calculations:} 
the uncertainties on the number of stars with positive seismic detection due to the uncertainty on $\delta \nu_{\rm env}$ (see Sect.2.5) can be estimated   
from Table~\ref{table:detection}. Denoting by $D$ the number of stars with positive seismic detection, the impact of  $\delta \nu_{\rm env}$ uncertainty can 
be estimated as  $(D_{\nu_{\rm max}/2} -D_{\nu_{\rm max}})/7009,$ where 7009 is the total number of stars in the initial sample. This yields
 $\sim$ 4\% and 7.5\%  when considering,  respectively the sample of stars with masses $\leq 1.2 \rm{M}_\odot$ and the sample of all mass stars with 
 positive seismic detection over 730 days in BOL conditions .
      
\begin{table} [h]
\centering
\caption{\label{table:detection2}  Uncertainties in the number of P1P2 stars with predicted positive seismic detection after 730 days of observations in (1 LOP, BOL) conditions.} 
\begin{tabular}{ ||l | c | c| c||}
\hline
\hline
 $\delta \nu_{\rm env} $         &   $\nu_{\rm max}/2$                 &      $\nu_{\rm max}$                      \\
\hline
\hline
MS stars ($M<1.6$)  and subgiants       &  $5839^{+175}_{-818}$             &       $ 6997^{+70}_{-1679} $       \\
\hline 
  $ M\leq 1.2 $                &    $1245^{+37}_{-87}$              &     $ 1541^{+14}_{-170} $     \\
\hline
\hline 
 \end{tabular}
\end{table} 
   
\end{itemize}

We obtain an order of magnitude of the  uncertainties on the number of positive detections, say $X$,  by considering the underestimate due to false negative and 
the overestimate due to false positive detections. We then used the false negative and positive detection rates derived for our samples 1 and 2 of 
 {\it Kepler} stars in the option $\delta \nu_{\rm env}=\nu_{\rm max}/2$ (second column of Table~\ref{table:table16} 
 in Appendix~\ref{prediction}).
 This yields $X^{+3\%}_{-14\%}$ for the sample of MS stars with $M<1.6 \rm{M}_\odot$ and subgiants and 
  $X^{+3\%}_{-7\%}$  for the sample of stars with $M \leq 1.2 \rm{M}_\odot$ when $\delta \nu_{\rm env}=\nu_{\rm max}/2$.
  Such uncertainties are provided in Table~\ref{table:detection2}. 
  In percentages, Table~\ref{table:detection2} indicates a predicted  seismic positive detection rate  for stars with $M\leq 1.2 \rm{M}_\odot$  
  between $\sim$  55-61\% (resp.  65-74\%) of the whole sample of 2099 stars with masses $\leq 1.2 {\rm M}_\odot$  in BOL conditions
   after two years of observation 
  taking into account uncertainties due to $\nu_{\rm max}/2$  (resp.  $\delta \nu_{\rm env}=\nu_{\rm max}$). 
 In the same conditions but for the sample of MS stars with masses $M<1.6 M_\odot$  and subgiants   with all masses, 
  the same uncertainties  yield  predicted seismic positive detections  at the level of 71-86\% (resp. 76-100\%) 
   of the whole sample of 7009 stars in the initial sample taking into account uncertainties due to $\nu_{\rm max}/2$ 
    (resp.  $\delta \nu_{\rm env}=\nu_{\rm max}$)

 We note that uncertainties in the probability calculation and the number of stars  with expected solar-like oscillation detection can also come from the fact that we used  PIC1.1.0 radius and effective temperature to compute the global seismic parameters  and derive the seismic mass.   

 \subsection{Expected solar-like oscillations within PLATO P5 sample}
 
 We carried out the same probability calculation as for the P1P2 sample after removing the same types of stars and assuming again  a positive 
 detection  for  $P_{\rm det} > 0.99$  (1 LOP, BOL,  $\delta \nu_{\rm env} =\nu_{\rm max}/2)$ conditions. We found that the number  of 
 expected postive seismic detections amounts to  9 486  for the sample of MS stars with $M/\rm{M}_\odot <1.6$ and subgiants after two years of 
 observation (Table~\ref{table:detectionP5}). 
\begin{table} [h]
\centering
\caption{\label{table:detectionP5} Numbers of stars  in the P5 sample with expected  positive detection ($P_{dec}>0.99$) of  solar-like oscillations in
 (1 LOP, BOL, $\delta \nu_{\rm env}=0.5 \nu_{\rm max}$) conditions. Masses and  radii are given in solar units.}
\begin{tabular}{ ||l | c | c|  c| c||}
\hline
\hline 
cases                          &      730d     &   365d       & 90d   & 30d      \\
\hline
 all             &    9491           & 5718            &  1599     &   380         \\
\hline
MS stars ($M\leq 1.6 \rm{M}_\odot$)   &      9486     &   5716   &  1599     &  380 \\
\& subgiants  of all masses                 &                 &         &           &           \\
\hline
Subgiants  of all masses                    &     8877    &   5657   &    1599    &    380   \\
\hline
$M \leq 1.2$ :                 &   878        &    392        &  81     &  21    \\  
    MS stars                  &  250         &     43       &  0         &  0      \\
    subgiants                 &  628         &    349       &   81       &  21    \\
\hline
\end{tabular}
\end{table} 

\begin{figure}[!]
        \begin{center}
                \includegraphics[width=9.cm]{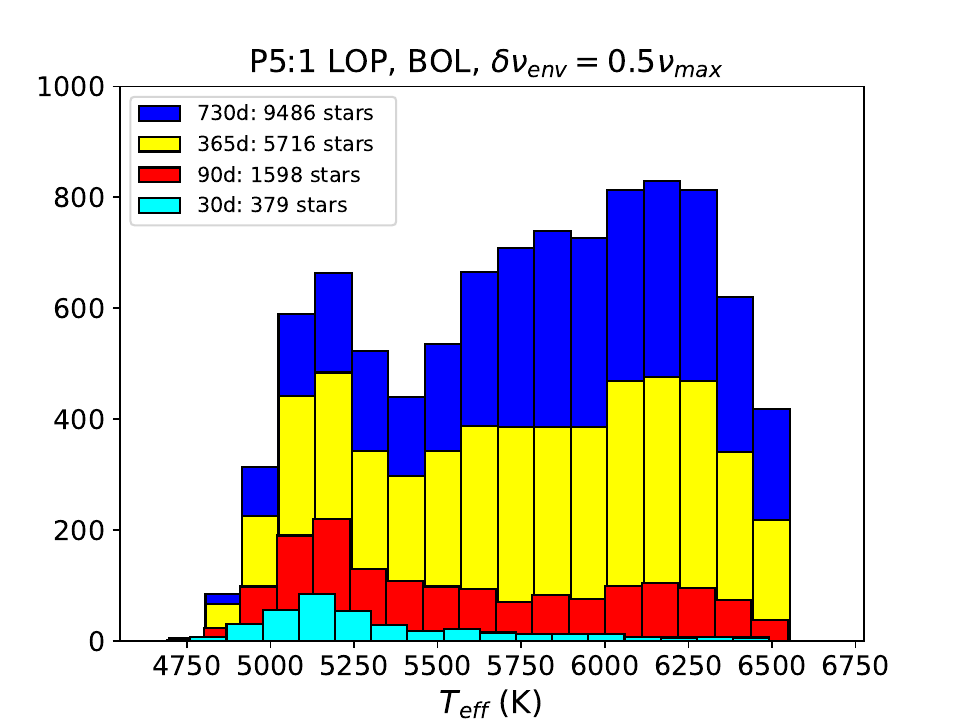}

                \includegraphics[width=9.cm]{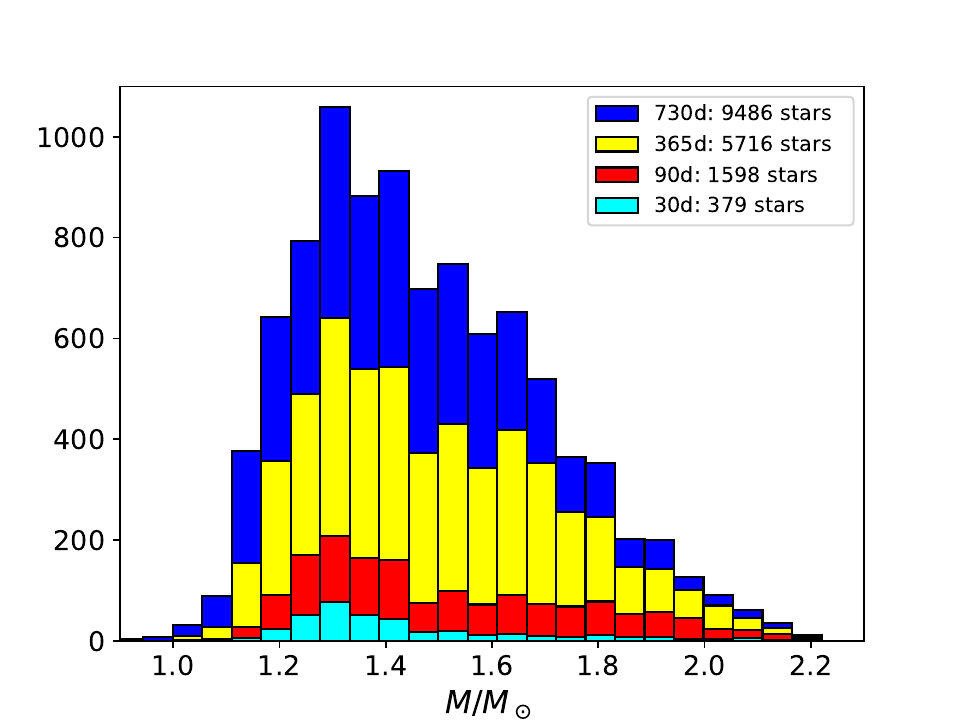}

                \includegraphics[width=9.cm]{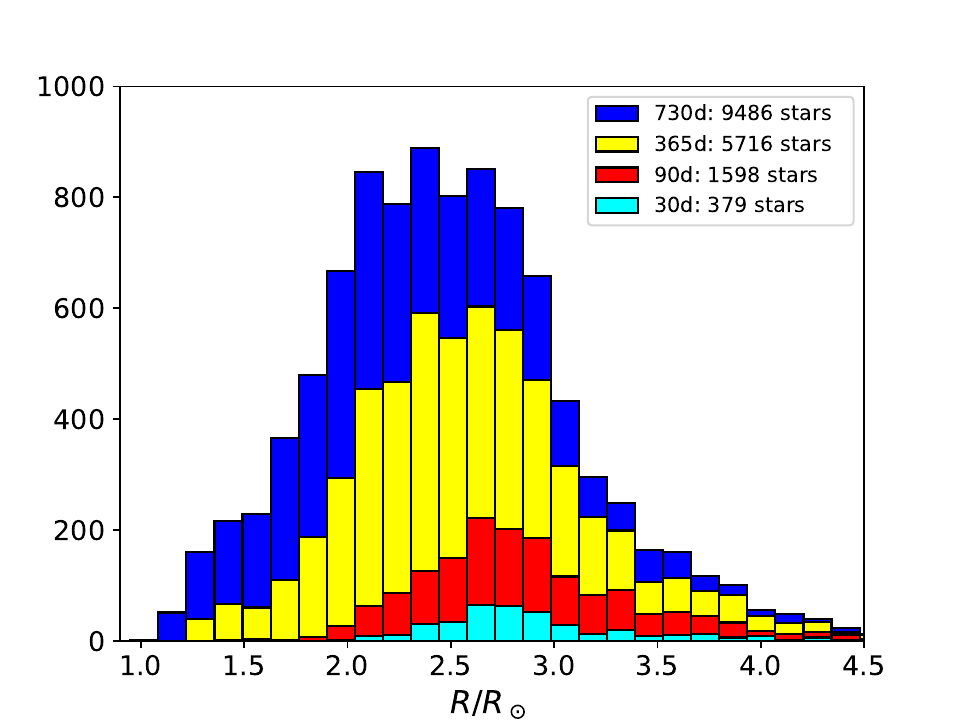}              
                \caption{Histograms of the number of  stars  (MS stars with $M<1.6 \rm{M}_\odot$ and subgiants)  from the P5 
   sample with an expected detection of solar-like oscillations with at least  
   99\% probability assuming 730, 365, 90 and 30 days of  observations. Conditions are (1 LOP, BOL, $\delta \nu_{\rm env}=0.5 \nu_{\rm max}$).}
                \label{histoP5BOL}
        \end{center}
\end{figure}

Here again, we observe a drastic increase in positive seismic detections with the observing time. This is illustrated in Fig.~\ref{histoP5BOL}. 
This figure shows the distributions in $T_{\rm eff}$, mass, and radius of P5  MS stars with masses $M<1.6 \rm{M}_\odot$ and subgiants with all 
masses with expected positive detections for different observing durations. The number of stars significantly increases when $T_{\rm obs}$ increases 
beyond 1 year, in particular toward stars with smaller radii. The subgiants  outnumber significantly the MS stars for the whole sample of stars 
with $M<1.6\rm{M}_\odot$ (Table~\ref{table:detectionP5}). As expected, after only 90 days of observation, 
only 81 stars  with masses $M\leq 1.2 \rm{M}_\odot$ have a seismic positive detection, all subgiants because their amplitudes
 (roughly $\propto L/M$) are  the highest.  

\begin{figure}[!]
        \begin{center}
                \includegraphics[width=9.cm]{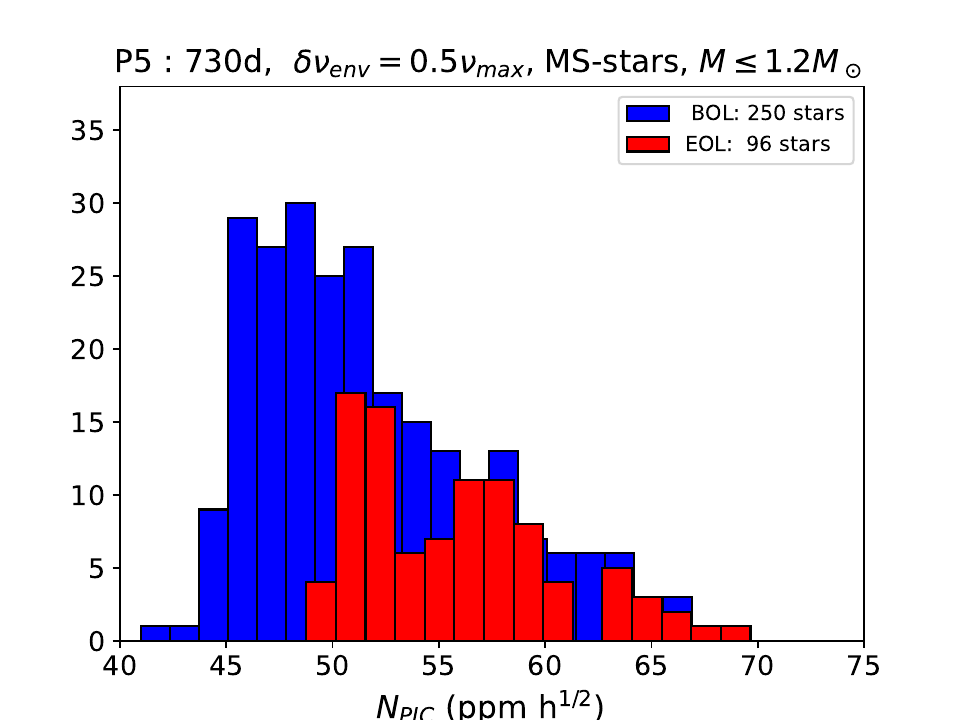}
                \caption{Distributions in $N_{PIC}$ (PLATO noise from the PIC1.1.0, in ppm $\rm{h}^{1/2}$) of MS stars with  $M \leq 1.2 \rm{M}_\odot$ from the P5 
   sample  with an expected positive detection of solar-like oscillations with a   
  probability higher than 99\% assuming 730  days of  observations. Conditions are 
  (1 LOP, $\delta \nu_{\rm env}=\nu_{\rm max}/2$), BOL (blue), EOL(red).}
                \label{histoSNR5BOLEOL}
        \end{center}
\end{figure}     
  
Figure~\ref{histoSNR5BOLEOL}  compares the distributions of the PLATO noise level, $N_{\rm PIC}$,  taken from the PIC1.1.0 between BOL and EOL conditions for stars 
with expected positive detection of solar-like oscillations in the case of (1 LOP,  $\delta \nu_{\rm env}= \nu_{\rm max}/2$) for 730 days of observation.
By construction, following the PLATO ESA requirements, the stars belonging to the  P1P2 sample  have $N_{\rm PIC} \leq 50$ ppm h$^{1/2}$ whereas the stars 
with higher noise  levels constitute the P5 sample. This was based on the EOL conditions. Assuming that the more optimistic  BOL conditions hold, we find
 that 115 MS stars with mass $M \leq 1.2 \rm{M}_\odot$ in the P5 sample have  $N_{\rm PIC} \leq 50 ~$ppm.h$^{1/2}$ and  could be reclassified as P1 stars, 
 increasing the number of P1P2  stars from 1016 (Table 2) to 1131 -that-is an  increase of  the positive detection rate from 15\% to  above 17\%.

\section{PLATO seismic performances for MRA inferences in  the P1P2 sample}
\label{sect6}

For the subset P1P2 stars with expected solar-like oscillations, the detection and highly precise measurement of individual frequencies for a 
significant number of modes is ensured by the selection of a high S/N by construction. This will allow  us to satisfy the requirements
 that must be achieved by the PLATO mission  (PLATO Science Requirements Document PTO-EST-SCI-RS-0150, ESA document,  June 2021) which are: a mass uncertainty better than 15\%, 
 a radius uncertainty lower than 2\% and an age uncertainty as low as 10\% for a star like the Sun or the PLATO reference star defined  as $1 \rm{M}_\odot, 1 R_\odot$, and $T_{\rm eff}=6000K$.
        
\begin{figure}[!]
        \begin{center}
                \includegraphics[width=9.cm]{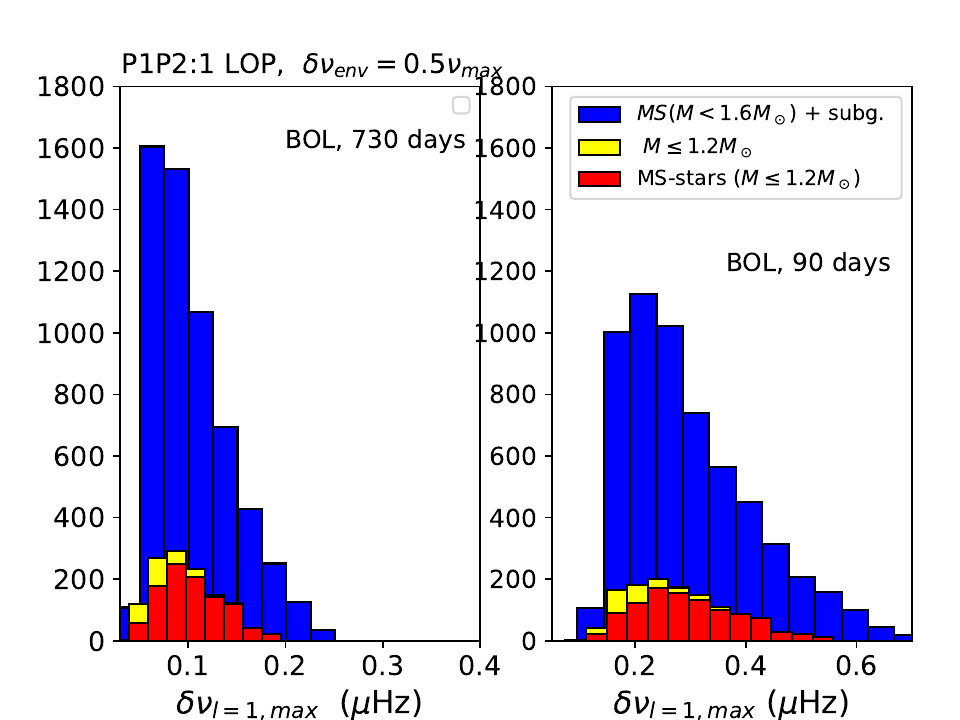}

                \includegraphics[width=9.cm]{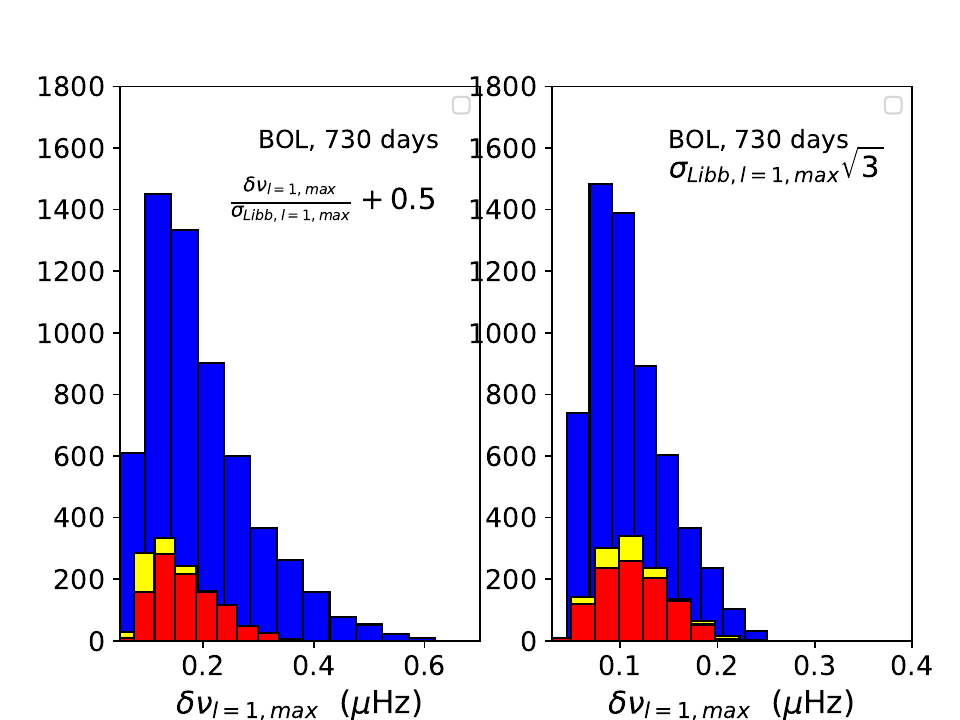}
                \caption{Distribution of frequency uncertainties $\delta\nu_{\ell=1,max}$
    for the sample of P1P2 stars with expected  positive seismic detection. The color code corresponds to different mass samples as indicated in the  top right panel.
     The top panels correspond to the frequency uncertainties computed with Eq.~(\ref{ratsig4}) for observation lengths of 730d (top-left panel) and 
     90 days (top-right panel).
    The bottom panels show the frequency uncertainties distribution when one takes into account the scatter in the relation obtained with Eq.~(\ref{ratsig4})  (left panel) 
     or the uncertainties in the mode linewidths at low effective temperature  width, as per Appendix~\ref{appendixD}  (right panel). }
                \label{histosigma}
        \end{center}
\end{figure}     

 \subsection{PLATO seismic performances for oscillation frequencies for the P1P2 sample}
 \label{sect6.1}

For the PLATO targets, assuming individual  frequencies are available, we can only determine the (theoretical) 
uncertainty on the frequencies using the \cite{libbrecht1992}  formula (Eq.~\ref{sigma_libb}), $\sigma_{\rm Libb,\ell=1,max}$. 
We then use Eq.~(\ref{ratsig4}) to relate 
$\delta \nu_{\rm \ell=1, max}$ to $\sigma_{\rm \ell=1, max}$ for each target. 
We computed $\sigma_{\rm Libb,\ell=1,max}$ then $\delta\nu_{\rm \ell=1,max}$  for each target of our P1P2 sample of stars   with  expected  positive
 seismic detection in BOL condition for 730 days and 90 days.  We also included the case where we take into account the scatter in the 
 relation Eq.~(\ref{ratsig4})  and the measurement uncertainties in the mode linewidths derived from the {\it Kepler} stars which leads to multiplying 
 by $\sqrt{3}$ the Libbrecht's uncertainties  $\sigma_{\rm Libb,\ell=1,max}$ (Appendix~\ref{appendixD}). The corresponding distributions of  
frequency uncertainties $\delta\nu_{\rm \ell=1,max}$ are shown in Fig.~\ref{histosigma}. In the case  (BOL, 730 days, $\delta \nu_{\rm env}=\nu_{\rm max}/2$) 
conditions, the  bulk of uncertainties are concentrated below 0.1 $\mu$Hz. Including the scatter in the relation Eq.~(\ref{ratsig4}) shifts up the maxima of 
the distributions by about 0.07 $\mu$Hz. The uncertainties remain below the PLATO  requirement  of frequency uncertainties  0.3-0.5 $\mu$Hz. In case of 3 months
 observations, the shift is higher, about 0.1-0.13 $\mu$Hz and the bulk of uncertainties  reach 0.2-0.3 $\mu$Hz. When the mode linewidth are increased by  
 a factor $\sqrt{3}$, the uncertainties are only  slightly shifted with the bulk of uncertainties 
concentrating around 0.1 $\mu$Hz. We also computed the frequency uncertainties in EOL conditions for 730 days but the associated degradation of the signal has 
only a small impact and is not shown. 

\subsection{PLATO seismic performances for stellar  MRA inferences for the P1P2 sample}
\label{sect6.2}

We now turn to the MRA uncertainties  from seismic inferences resulting from the error propagation due to $\delta \nu_{\rm \ell=1, max}$.  
We focused on stars with masses  $M\leq 1.2 \rm{M}_\odot$. We used Eq.~(\ref{MR1c}) to estimate the  mass and radius uncertainties  as discussed in 
Sect.~\ref{sect4}. We also used  the constraint   $\delta \nu_{\rm \ell=1, max}$ as a proxy  for the constraint on the age uncertainty. 
The condition $\delta M/M \leq 3\%$ (often used to obtain an age uncertainty at the  level of 10\%) can also be added as an additional constraint.
  
We derived the numbers of P1P2 stars with a positive seismic  detection while adding successive constraints on the uncertainties  giving rise to three cases as follows:
 \begin{itemize}
 \item  case I : $\delta M/M\leq 15\%$ \& $\delta R/R \leq 2 \%$   \\
  \item case II : $\delta M/M\leq 15\%$ \& $\delta R/R \leq 2 \%$ \&   $\delta \nu_{\rm \ell=1, max} \leq 0.2 \mu$Hz \\
  \item case III : $\delta M/M\leq 15\%$ \& $\delta R/R \leq 2 \%$ \&   $\delta \nu_{\rm \ell=1, max} \leq 0.2 \mu$Hz \&  $R/R_\odot \leq 1.1$ 
  \end{itemize}
We consider the frequency uncertainties $\delta \nu_{\rm \ell=1,max}$ as given by Eq.~(\ref{ratsig4}) or Eq.~(\ref{ratsig}) without (case a) and with (case b)  a $+0.5$ shift due
 to the  scatter in the fitted relation between $\delta \nu_{\ell=1,max}$ and $\sigma_{Libb,\ell=1,max}$ (Appendix~\ref{appendixD}). The results are listed  in Table~\ref{generic}.
   
\begin{table} [h]
\centering
\hskip -0.5 truecm
\caption{\label{generic}  Number of MS dwarfs with $M \leq 1.2 \rm{M}_\odot$ stars   in the P1P2 sample with expected  positive seismic detection and
 satisfying different  cases of MRA uncertainties.  Assumed conditions are  (1 LOP, BOL, 730 days, $\delta \nu_{\rm env}= \nu_{\rm max}/2$). 
 Labels a) and b) refer to  frequency uncertainties assumed without and with a $+0.5$ shift due to the  scatter in the fitted relation between 
 $\delta \nu_{\rm \ell=1,max}$ and $\sigma_{\rm Libb,\ell=1,max}$ (Eq.D.3).}
\begin{tabular}{ ||l | c | c|   c||   c  |c  ||}
\hline
\hline 
case           &   BOL      &   EOL \\
\hline
Ia             &   1016     &    880  \\
IIa            &   1016    &    880     \\
IIIa          &     260      &   195        \\
\hline
Ib             &   1016   &    880  \\
IIb            &   729  &   599          \\ 
IIIb          &    206     &   146      \\
\hline
\end{tabular}
\end{table}    

\begin{figure}[t]
        \begin{center}
                \includegraphics[width=9.cm]{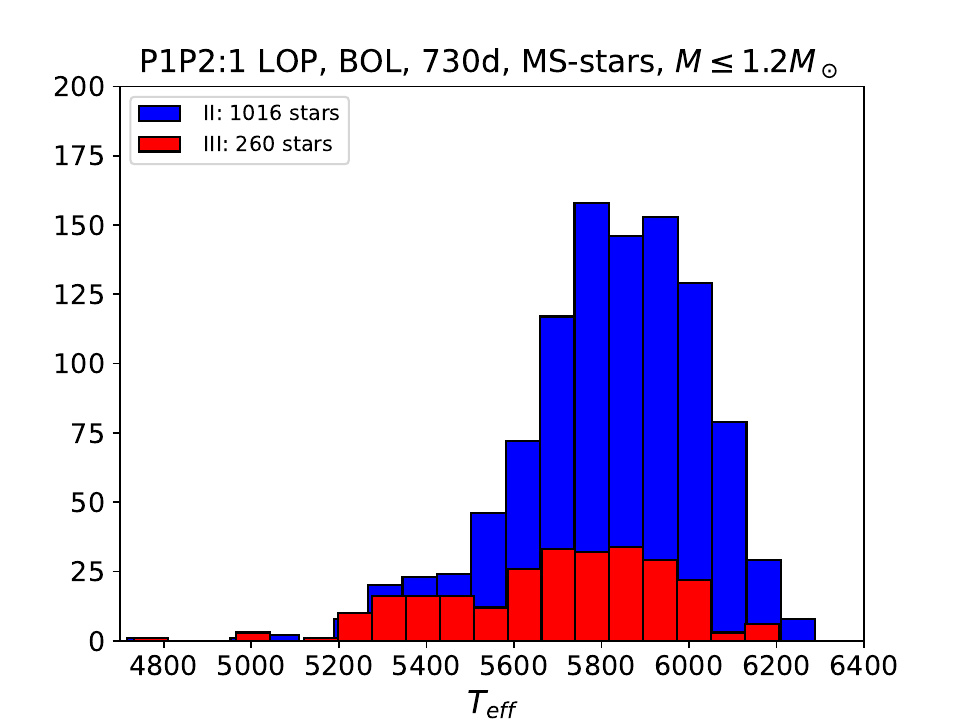}
                \includegraphics[width=9.cm]{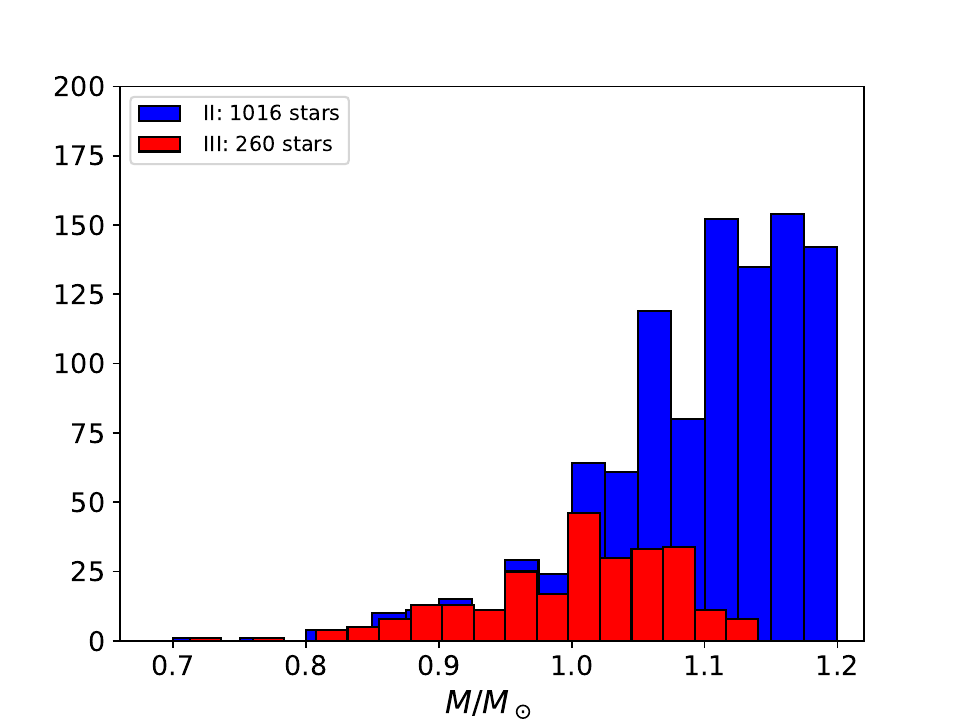}
                \caption{Histograms in $T_{\rm eff}$ (top), and mass   (bottom)   
       for MS stars  with $M/\rm{M}_\odot \leq 1.2$  in the P1P2 sample assuming 
        (1 LOP, BOL, 730 days, $\delta \nu_{\rm env}= \nu_{\rm max}/2$) conditions
 with expected detection of solar-like oscillations and satisfying   cases IIa and IIIa listed in Table~\ref{generic}.}
                \label{histoP1P2sigma3}
        \end{center}
\end{figure}

Figure~\ref{histoP1P2sigma3} shows the distributions of MS stars   with masses $M \leq 1.2 \rm{M}_\odot$ corresponding to  the cases listed in Table~\ref{generic} as a function of  
$T_{\rm eff}$, stellar mass and radius (1 LOP, BOL, $\delta \nu_{\rm env}=0.5 \nu_{\rm max}$ 730 days) conditions. We expect that the number of stars decrease when adding new constraints. Cases I do not reduce the number of stars compared 
with the initial sample of P1P2 MS stars  with masses $M \leq 1.2 \rm{M}_\odot$ with positive seismic detection. This means that the main constraints are the S/Ns imposed by design and the detection probability. The PLATO requirements for the mass, radius  and age uncertainties (case II) of P1P2 stars are automatically satisfied in the PIC, provided the oscillations are detected.

\begin{figure}[t]
        \begin{center}
                \includegraphics[width=9.cm]{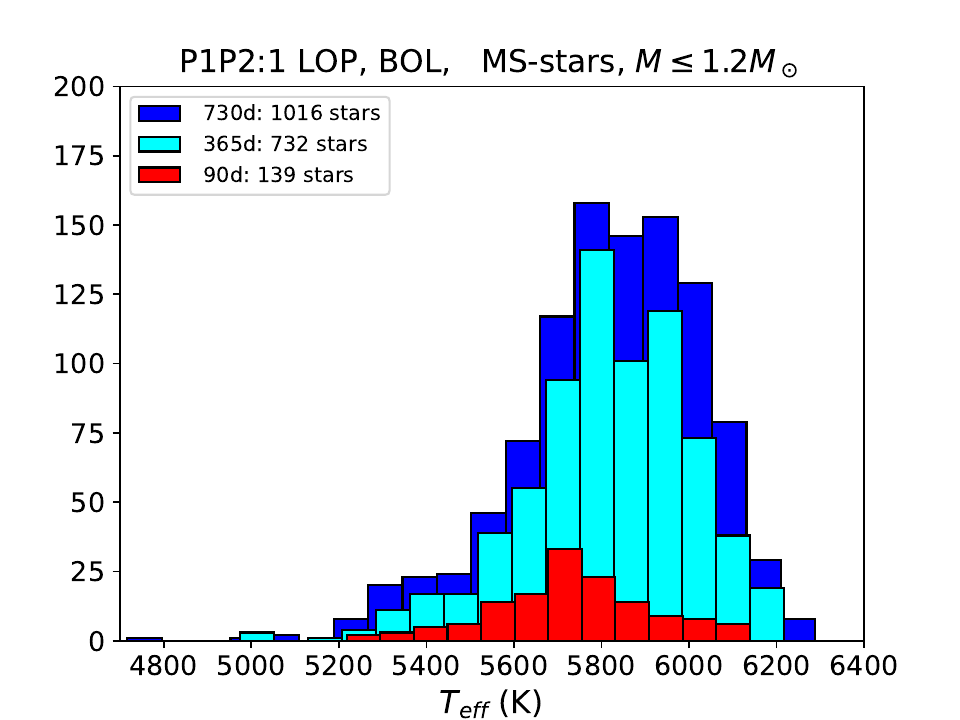}
                \caption{Histogram  in $T_{\rm eff}$  for the P1P2 subsample of 
   MS stars with $M/\rm{M}_\odot \leq 1.2$  assuming (1 LOP, BOL, $\delta \nu_{\rm env}=0.5 \nu_{\rm max}$) conditions
   and case IIa  in Table~\ref{generic}.}
                \label{histoP1P2sigma4}
        \end{center}
\end{figure}

Figure~\ref{histoP1P2sigma4} shows the histogram  of the evolution of the number of these stars when  the observation time increases from 90 days to 730 days, 
assuming (1 LOP, BOL) conditions for MS stars with $M/\rm{M}_\odot \leq 1.2$,  
 $\delta M/M  \leq 15\%;\delta R/R  \leq 2\%  $, and  $\delta \nu_{\rm \ell=1, max} \leq 0.2 \mu$Hz. 
The gain of stars satisfying the PLATO requirements  is particularly significant for stars like the reference star when going from 90 days to a year. 
We note that no such star is  found when observing over a short period of time of 30 days.
  
We also computed the numbers of stars  satisfying other constraints such as the case ($\delta M/M\leq 3\%$ ;   $\delta R/R \leq 2 \%$);   
the case($\delta \nu_{\rm \ell=1, max} \leq 0.5 \mu$Hz) or the case ($\delta M/M\leq 15\%$ ;   $\delta R/R \leq 2 \%$   \&  $\delta \nu_{\rm \ell=1, max} \leq 0.5 \mu$Hz). 
The number of stars remains the  same as in the initial sample of P1P2 MS stars  with masses $M \leq 1.2 
\rm{M}_\odot$ with positive seismic detection. 
 
\paragraph{Taking into account  uncertainties in frequencies and mode linewidths $\Gamma$ values:}
The above  numbers of stars remain unchanged when including    $\sim$ 5-$\sigma$ uncertainties due to scatter in Eq.~(\ref{MR1c}) 
(Fig.~\ref{dMdRdA} in  Appendix~\ref{appendixC}): $\delta M/M (1 \pm  0.5\%)$;  $\delta R/R (1 \pm 0.1\%)$. In contrast, in the extreme case where one takes into account a scatter 
of $+0.5$ in the $\delta \nu_{\rm \ell=1, max}/\sigma_{\rm Libb,\ell=1,max}$ fitted relation for  all stars, the number of MS stars with $M\leq 1.2 \rm{M}_\odot$  
satisfying the PLATO requirement  decreases by about 25\%.

The impact of degrading the frequency uncertainty   $\delta \nu_{\ell=1,max}$ by $\sqrt{3}$    for stars with $T_{\rm eff}\leq 5650$K (due to uncertainties in the values the mode 
linewidths, as per Appendix~\ref{appendixD}), instead of the original values of $\delta \nu_{\rm \ell=1,max}$ -   leads to a decrease of the number of stars by 38\% in case IIb with 
(BOL, 730d, $\delta \nu_{\rm env} =\nu_{\rm max}/2$) conditions.
 
We then conclude that even in the conservative case we considered for the detection probability and in the above worst cases for the frequency uncertainties, 
 the PLATO  mission  should yield a set of MS stars  with a seismic characterisation between $\sim$ 80 and 100  times the {\it Kepler} Legacy sample   
assuming  two observing  fields  (i.e. doubling the number of stars 
obtained for 1 LOP) and  depending on whether we assume $\delta\nu_{\rm env}=\nu_{\rm max}/2$ or $\nu_{\rm max}$. 
   
\section{Summary and discussion}
\label{summary}
 
  The present study is part of the scientific preparation for the  ESA's PLATO mission, which will be launched towards the end of 2026.
In this work, we calculate the theoretical probability of detecting solar-like oscillations for stars belonging to the PLATO Input Catalogue (PIC). 
More specifically, we considered bright stars (magnitude of 11 and brighter) of the FGK spectral type on the main sequence (masses lower than 1.6 $M_\odot$) 
and the subgiant branch.  The calculation takes into account the estimated noise level for each individual star, 
provided by the PIC.  Our results indicate that the proportion of stars with positive detections of solar-like oscillations lies within a range of 
70\% to 100\% for a continuous observation of two years. 
 The lower (upper) value of this range comes from the assumed narrow (wide) bandwidth of the oscillation spectrum in the Fourier domain for each star, 
which is the main uncertainty in the calculation. It also depends on the  beginning-of-life or end-of-life conditions of the PLATO instrument.
  
   The CoRoT and Kepler missions have taught us that individual oscillation modes can be detected for stars with a noise level of 50 ppm in one hour or less. 
  For the stars in our sample  that satisfy this criterion and with positive seismic detection
    we have estimated the uncertainty 
  in the individual frequency measurements at the maximum of the power spectral density  for each star  based on the results of the {\it Kepler} mission.
   This enabled us to assess the
  propagation of this statistical error on the seismic inference of the mass, radius, and age of each star. We found that 
 $\sim$  47- 61\% of the sample of MS stars with masses $M\leq 1.2 \rm{M}_\odot$ with statistical uncertainties 
 below the PLATO requirements of  15\%, 2\%  for the stellar masses and  radii,  respectively and satisfying oscillation frequency uncertainty $\leq 0.2 \mu$Hz 
 at maximum power density amplitude as a proxy for  10\% uncertainty  of the age of a Sun-like star. Those uncertainties are small  enough that they leave 
 margins for including  systematic errors  while still keeping  the  total error budget satisfying the PLATO requirements.
We note that the  masses used to define  various mass samples, especially the sample represented by $M\leq 1.2 \rm{M}_\odot$, are seismic masses derived from scaling relations. 
 As such, they are approximated  as are  the number of targets found in each  mass subsample  but the order of magnitude remains 
 correct. 
   
We  also stress that  for a few  stars, it may be expected that  additional  errors can come from unexpected issues in the data acquisition or in the variable behavior of the star 
\citep[such as magnetic activity,][]{perez2019,karoff2019,thomas2021,santos2023} that can alter the measurements of the frequencies and therefore the MRA uncertainties. From the  {\it Kepler} experience, this could add  an uncertainty up to  0.3 $\mu$Hz to the statistical uncertainties but it is difficult at this stage to foresee for which stars  in the PIC this can happen and this was ignored here. 
 
For the P5 sample, the noise level is higher- again by construction- than for the P1P2 sample in the same (EOL or BOL) conditions. We find a percentage of 
7.3-4.3 \% P5 stars with expected  positive seismic detection after 730 days of observation in BOL and EOL conditions, respectively. Among those, a percentage of 
0.5-0.2\% of P5 MS stars with masses $M\leq 1.2 \rm{M}_\odot$ are expected to show positive seismic detections. Among them,  115 P5  MS stars with masses 
$M\leq 1.2 \rm{M}_\odot$ and with a PIC noise level lower than 50   ppm . h$^{1/2}$  in BOL conditions satisfy the above PLATO requirements, which means that those stars 
could be re-classified as P1 stars.

\begin{figure}[t]
	\begin{center}
		\includegraphics[width=9.cm]{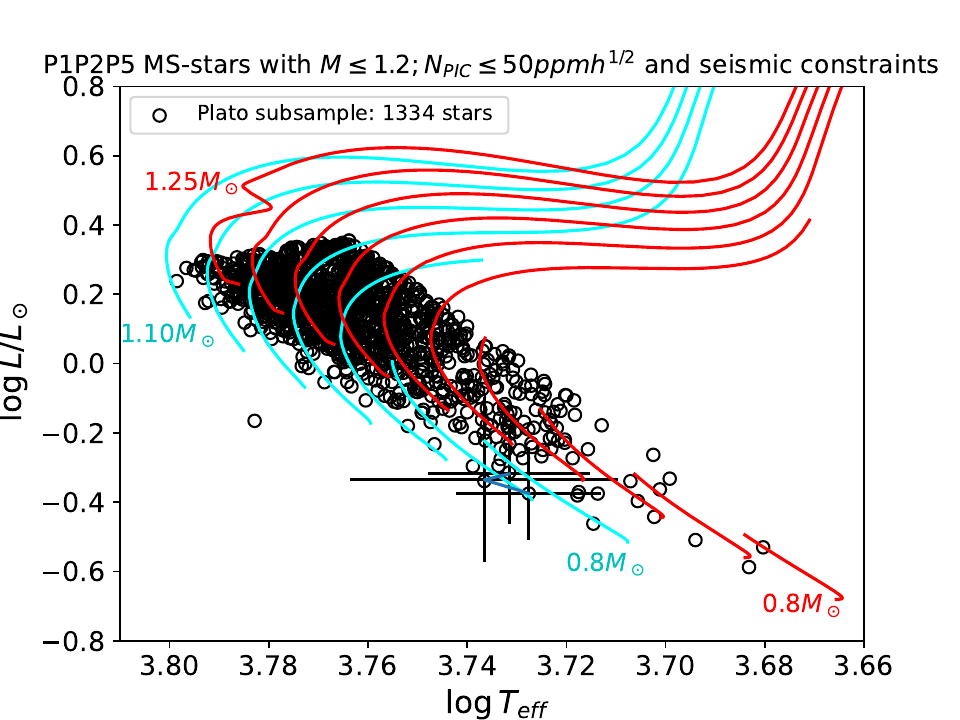}
		\caption{P1,P2, P5 stars with $M\leq 1.2 \rm{M}_\odot$, $N_{PIC} \leq 50 $ppm . h$^{1/2}$ 
		and satisfying constraint of case II in Tab.\ref{generic}.
  Crosses indicate the (representative) PIC
  uncertainties  for three target stars.  Evolutionary tracks of stellar models built with 
  a given initial chemical composition and $\alpha_{MLT}$ value  are represented by the same color: 
  cyan ($\alpha_{MLT}=1.845$, $Y_0=25$, $(Z/X)_0=0.0136$,  no atomic diffusion included, masses in the range [0.8 -1.10]$\rm{M}_\odot$  in steps of 0.05 $\rm{M}_\odot$)
    or red ($\alpha_{MLT}=1.642$, $Y_0=0.25$, $(Z/X)_0=0.0246$, atomic diffusion included, masses in the range [0.8 -1.25]$\rm{M}_\odot$  
    in steps of 0.05 $ \rm{M}_\odot$).}
		\label{fig:HRMStars}
	\end{center}
\end{figure}

Accordingly, and as a whole, the calculations yield  a total of 1131 MS stars with masses $M\leq 1.2 \rm{M}_\odot$ for which one expects a positive seismic
 detection and seismic analyses providing mass, radius, and age satisfying the above PLATO requirements in BOL consitions after  two years of observation for one LOP. 
    The stars of this sample are plotted in a HR diagram in Fig.~\ref{fig:HRMStars}. For each target, the luminosity is derived with Eq.~(\ref{luminosity})  
    and the stellar radius 
and the effective temperature, and their uncertainties taken for the PICv1.1.0. Uncertainties are plotted for three stars as representative of the typical PIC
 uncertainties in the HR locations. Overplotted over the PLATO targets locations, evolutionary tracks of stellar models cover the mass range of 
 $[0.8-1.2] \rm{M}_\odot$.  The stellar models were built with the CESTAM code \citep{morel2008,marques2013} assuming  the solar relative  chemical abundances 
 AGSS09 \citep{asplund2009} with  different  initial values for $X_0,Y_0,Z_0$ (namely, the mass fractions of hydrogen $X_0$, helium  $Y_0$, and metallicity 
 representing all heavier chemical elements collectively counted as $Z_0$). The convection is described with the classical MLT formulation \citep{cox1968} involving the mixing length parameter, $\alpha_{MLT}$ (a free parameter representing the efficient of convective transport in 1D stellar models). Otherwise, the input physical assumptions are  similar to those of the reference model A described in \cite{lebreton2014c}. The evolutionary tracks in Fig.~\ref{fig:HRMStars} were computed until an age of 14 Gyr (on purpose greater than the age of the Universe) for the lowest mass stellar models  or stopped at an arbitrary phase of the red giant branch for the most massive ones. Hence
 assuming one single chemical composition and $\alpha_{MLT}$ values- usually taken as for the  Sun- would clearly not reproduce the whole extended region in the HR 
 occupied by  the PLATO targets with the lowest masses. This remains true even taking into account the observational uncertainties in luminosity and 
 effective temperature and the fact that  several  stars might belong to binary systems.

 The anticipated PLATO sample  of $\sim$ 2793 well characterized main sequence stars with masses of $M <1.6 M_\odot$ in one LOP 
 will contribute to  the PLATO set of best seismically characterized stars  and is roughly 42 times larger than the {\it Kepler} Legacy sample
  and will complement the latter in providing tight constraints on stellar modelling. 
  All these results will be  reviewed and revised when confronted in 2.5 years from now with the real PLATO data after launch.


In the present work, we have purposefully estimated only  the statistical uncertainties on the frequencies and resulting MRA seismic inferences in order to 
appreciate the added value due to the expected high quality of the PLATO photometric data. The total error budget, however,  must include the uncertainties due 
to the various systematic
 errors that can be identified but not fully corrected. Of all stellar properties, the stellar age is by far the most challenging to determine accurately.  
  In most cases indeed, 
 stellar ages of single, field stars can  only be  determined through stellar modelling \citep{soderblom2010, jcd2018}; thus,  their accuracy strongly
  depends on the
 degree of reliability  of the available  stellar models. Systematic errors are indeed  expected to come mostly from insufficiently realistic   
 stellar modelling  \citep[e.g.][]{lebreton2010,jcd2018,lebreton2014a,lebreton2014b,salaris2017,dupret2019,buldgen2019}.
  
Thanks to the high quality of data acquired by space missions such as  CoRoT \citep{baglin2009} and  {\it Kepler}, and currently TESS \citep{ricker2015}, 
theoretical studies  were  initiated in order to identify and  quantify the impact of the main  systematic errors that contribute the most to the error budget. 
Studies have been carried out as  pure theoretical investigations and hare and hounds exercises \citep[e.g.][]{Appourchaux2006,lebreton2014b,reese2016,cunha2021} or
 by modelling specific or sets of  stars with the best seismic observations such as the small set of CoRoT stars or  the larger set of  {\it Kepler} LEGACY stars 
\citep[e.g.][among others]{Appourchaux2008,benomar2010,metcalfe2012,lebreton2014c,SA2015,SA2017,creevey2017,bellinger2017,nsamba2018,valle2020,ong2021,farnir2020,betrisey2022,betrisey2023}. 
At present, due to lack of accuracy,  the age uncertainties for solar-like filed stars can increase roughly by  $\sim$ 5 to 25\% depending on poorly modelled  physical
 processes and unknown initial  chemical composition and to  $\sim $ 15\%- 50\%  for a more massive star with a convective core.    

  We consider below two  illustrative cases : the degraded Sun and the two  best studied   stars of  the Legacy sample.

\subsection{ The `degraded' Sun}  A natural test of the accuracy of seismic modelling and characterisation is to look at the Sun-as-a-star and compare the results of the seismic MRA 
inferences to the independently known values of the mass, radius, and age of the Sun. This is now done routinely when inferring the MRA for various  sets of {\it Kepler} stars.
 We  therefore carried out several MRA inferences using  the  data  of the so-called `degraded Sun' of \cite{lund2017}. The  frequencies 
  of   the  'degraded' Sun  and their  uncertainties were built to match the quality of the {\it Kepler} Legacy sample
   ($\delta \nu_{\rm \ell=1,max} \sim 0.15 \mu$Hz with oscillation modes in the range 
   $\ell=0,n=16-27; \ell=1,n=15-27; \ell=2, n=16-24; \ell=3,n=20$ , \citep{lund2017}). Here again  we used a grid-based approach (GBM) with the AIMS code. 
   The observational constraints besides the oscillation frequencies and their uncertainties  were taken as:    
   $T_{\rm eff} = 5777 \pm 77$K, $[Fe/H] = 0    \pm 0.1$, $\nu_{\rm max}=3090 \pm 3.9 \mu$Hz. We compare the results obtained when using two different grids of stellar models. 
   The \cite[][hereafter C21]{cunha2021}
 for  which we recall that  the stellar models were computed with MESA evolutionary code \citep[][and references therein]{paxton2018} and the frequencies 
 were computed using with the oscillation code ADIPLS \citep{jcd2008}. The second grid  of stellar models (hereafter Mo23) was computed by  one of the
  co-authors (namely, N.~Moedas, for more details see Moedas et al., in prep.) using also the MESA code and the frequencies were computed using 
  the GYRE oscillation code \citep{townsend2013}. The input physics, values of free parameters  and the reference solar relative chemical abundances differ 
  between both grids. This allows to assess  the impact of  the main uncertainties in modelling solar-like stars. Intended to be applied to real stars 
  unlike C21,   Mo23 used a more  updated stellar physics (similar to models D1 in \cite{moedas2022}, namely, it included atomic diffusion of chemical elements 
  that helioseismology taught us is crucial for the modelling of solar-like stars. The reference solar abundances are AGSS09 \citep{asplund2009} in Mo23 whereas 
  it is GS98 \citep{grevesse1998} in C21. For sake of simplicity, Mo23 kept the mixing length value of $\alpha_{MLT}$ fixed to the calibrated solar value 
   whereas C21 let the  convection parameter be adjusted in the fitting process. Because  here we dealt with real stars, we had to include  surface-effect 
   corrections and adopted the \cite{ball2014b}'s correction  in the AIMS inferences with both grids.
  
  Following \cite{reese2016} and C21,  in the particular case of the  Sun, we can measure the biases (or departure from accuracy) with
$d_{\rm X,rel}\equiv (X_{\rm fit}-X_{\rm true})/X_{\rm true}$ for $X=M,R,A$. We wish then to compare these values to the relative statistical uncertainties 
 $\delta_X\equiv |\delta X|/X_{\rm fit}$ (\%)  (where  $|\delta X|$ corresponds to one standard deviation) 
 provided by the MRA inferences with the AIMS' code. Finally it is also informative to estimate how large are the departures from accuracy compared to   statistical uncertainties 
 derived from the GBM approach $d_{\rm X,norm} \equiv  |(X_{\rm fit}-X_{\rm true})|/\delta X$  since  we will have only access to the last ones in most PLATO 
 stars.   The departure between the seismically inferred  values for the solar MRA and the known solar values (taken here as $4.6 \pm 0.4$ Gyr for the solar age
  \citep{houdek2007} as measured by  $d_{\rm X,rel}$  are given in Table~\ref{table:sun}  for the two grids. As expected the accuracy is much higher in the case 
  of the Mo23 grid than for the C21 grid mainly because the C21 grid does not included atomic diffusion. These figures are  similar to the values derived in  
  other works which all adopt various different  input assumptions and which fall in the range  ($\sim 0.3-4$\%) for the mass, ($\sim 0.1-2$\%) for the radius, 
   and for the age $\sim 2-9$\% when atomic diffusion is included and  $\sim 15-16$\% when atomic diffusion is not included
    \citep[e.g.][]{SA2017,creevey2017,rendle2019,jiang2021,SA2022,metcalfe2023}. 

On the other hand, the 1$\sigma$ uncertainties given by the inference calculations as measured by $\delta_X$  are comparable  between the two grids
 (in the typical range 0.5-4\% and 3-8\% for the mass and age), relatively independently of the accuracy. Accordingly the  departures from accuracy  
as measured in terms of $\delta X$ uncertainties ($d_{\rm X,norm}$,  last column of Table~\ref{table:sun}) significantly differ between 
the two calculations and between the fitted parameters MRA. In addition in both cases, the inaccuracies of the derived  luminosities, which were not included as 
input  constraints   amount  to 5\% in both cases. As we want to decrease the inaccuracies at the level of or below  the statistical uncertainties, this 
shows that  
there is room for improvements in the inference process or  in the present solar modelling.  This also emphasized the importance of building
 a set of  stars with determination of mass and/or radius, and or age (benchmark stars) independently 
of stellar modelling as is done for the  GAIA mission (\cite{heiter2015b}
\cite{heiter2015a})  and in preparation  for the PLATO mission  \citep{maxted2023}.

\begin{table} [h]
\centering
\hskip -0.5 truecm
\caption{\label{table:sun}  MRA inference for the 'degraded Sun' of one standard deviation or    68\% credible intervals about the median values. }
\begin{tabular}{ ||l | c | c|   c||   c  |c  ||}
\hline
\hline 
grid              &   $\delta_ X$ (\%)  & $ d_{X, rel}$ (\%)  &    $d_{\rm X,norm}$ \\
\hline
  mass             &                       &                             &                                        \\
\hline
Mo23        &   0.20   & +0.74     &   3.7       \\
C21         &   0.21   & -2.51     &   13            \\
\hline
\hline
radius             &      &                                  &                                       \\
\hline
Mo23   &     0.07    &  +0.02     &   0.32     \\
C21    &     0.08    &  -1.03    &    13         \\
\hline
\hline
age                &     &        &                              \\
\hline
Mo23   &    1.37  &    + 2.1      &  0.97       \\
C21    &    1.53  &    - 11      &   6.5                  \\
\hline
\hline
Validation     &            &         &           \\
luminosity     &            &         &          \\
Mo23    &   0.71   &  5.5    &   7.8     \\
C21    &    0.88   & 5.3     &   6.35         \\
\hline
\hline 
\end{tabular}
\end{table}    
 
In the case of the `degraded Sun',  the net error budget for the age remains close to $\sim $ 10 \% accuracy. One must nevertheless keep in mind that free parameters
 entering the solar and stellar modelling (namely, the initial helium abundance, convective efficiency  parameter, $\alpha_{MLT}$) are calibrated  for the Sun so that  
 the solar model reproduces the radius, luminosity at the age of the Sun.  On the seismic side, the surface corrections of the frequency are  designed for the 
 theoretical frequencies of the stellar models  to match the  observed ones. Inaccuracy in the solar modelling are then either compensated or minimized by such
  procedures. This cannot be done for other solar-like stars and one must either attribute arbitrarily the solar values to the free parameters or adjust them during 
  the fitting process for other stars. This can lead  to hidden inaccuracies.  What can then be and is done  is rather studying the sensitivity of the fitted results 
  to changes in the physical description or the values of the free parameters.  This is what was carried out for the two brightest ($V\sim 6$) solar-like stars 
 from the {\it Kepler} LEGACY, which we discuss below.

\subsection{Best studied stars of the Kepler Legacy}

The two  best studied   stars of  the Legacy sample, the stars 16 Cyg A (KIC12069424) and B (KIC12069449) belong to a multiple system and 
show solar-like oscillations \citep{metcalfe2012,lund2017}. They are  bright stars for which  interferometric radii are available. We can also assume that they were born with the same chemical 
composition and have the same age. Unlike the Sun, we have no independent measurements of the masses and independent determinations of their ages. 
On the other hand,   the  information of a common age and interferometric radii can act as validation of the inferred results and assessment of  the accuracy of 
the MRA inferences.   The most recent studies dedicated to 16 Cyg A,B were those of  \cite{bazot2020}, \cite{farnir2020}, \cite{nsamba2022},  \cite{buldgen2022}   who
 provided  references to former works.   The uncertainties are found  of the order of 4\%  and 15\% for the masses and ages, respectively.  
 The interferometric radii are well reproduced with uncertainties of 2\%. As an illustration,  we carried  out  seismic inferences for both stars 
 with the two already mentioned grids C21 and Mo23. 
  We  inferred the MRA for  each  {\it Kepler} star independently   using again the  \cite{ball2014a} surface-effect correction for the frequencies. 
  The  observational constraints are listed in  Table~\ref{table:16Cyg1}. The  sample of frequencies are those provided by \cite{lund2017}. As can be seen in 
  Table~\ref{table:16Cyg2}, the uncertainties  derived from the MRA inference processes are  small, namely,
 $\sim$ 1\% or below in all MRA cases and with both grids. They are smaller for the  C21 grid than the Mo23 grid.  At that level of relative uncertainties, 
 this is  likely due to a difference in the properties of the grids  such  as the number of adjusted  free parameters or to a difference in density of stellar 
 models in the parameter space around the  studied stars.  This would deserve further investigation but is out of scope in the present study. As for the accuracy, 
 the relative departure of the inferred radius of each star from its respective  interferometric radius is slightly above $1\%$ for both grids (Table~\ref{table:16Cyg3}), showing that the radius is well constrained by seismology, rather independently of the physical description of the stellar models in the grids. Table~\ref{table:16Cyg3} also gives $\Delta A=  (A_X-<A>)/\delta A$,  the relative differences between the age of each star and their average age, $(A_X-<A>)$ for $X=16CygA$ and $16CygB$ 
  and $<A>=(A_{16CygA}+A_{16Cyg_B})/2$ compared to the relative uncertainties of the inferred ages, $\delta A$. This shows that the departure from a common age is of the order of the  inference uncertainties.  

\begin{table} [h]
\centering
\hskip -0.5 truecm
\caption{\label{table:16Cyg1}  Observational constraints for the MRA inferences for 16 Cyg A and B. References:
(a)  \cite{morel2021} (uncertainties increased arbitrarily); (b) \cite{lund2017}; (c) \cite{white2013} ; (d)  \cite{metcalfe2012}. }
\begin{tabular}{ ||l | c | c|   c||     c||c|c||  ||}
\hline
\hline               
            &   16Cyg A       &   16 Cyg B     \\    
\hline 
$T_{\rm eff}^{(a]}$  (K)      &  5800  (50)     &  5750.0  (50)         \\ 
$[Fe/H]^{(a]}$       (dex)   &  0.11  (0.05)    &   0.08 (0.05)       \\ 
$\nu_{\rm max}^{(b)}$ ($\mu$Hz)      &  2188.5 (4.6)   &   2561.3 (5.6)          \\
 \hline               
 Validation             &                                &                \\
 $R/R_\odot^{(c)}$      &      1.22 (0.02)               &     1.12 (0.02)       \\
 $L/L_\odot^{(d)}$      &      1.56 (0.05)               &      1.27 (0.04)         \\
\hline                 
\end{tabular}                   
\end{table}

\begin{table} [h]
\centering
\hskip -0.5 truecm
\caption{\label{table:16Cyg2}  Results of MRA inferences (\%)  for 16 Cyg A and B: 68\% credible intervals about the median. 
16 Cyg A $ \delta \nu_{\rm \ell=1,max}=0.10 \mu$Hz (53 modes); 16 Cyg B : $ \delta \nu_{\rm \ell=1,max}=0.04 \mu$Hz  (52 modes)}
\begin{tabular}{ ||l || c |    c|    c   ||}
\hline
\hline 
                 &  $\delta  M/M$  &       $\delta R/R$   &       $\delta A/A$          \\
\hline
16Cyg A          &                   &                     &                             \\
Mo23             &   0.33            &    0.12             & 1.01                                  \\
C21              &   0.10           &     0.03            &  0.77                        \\    
\hline       
16Cyg B         &                   &                      &                                  \\
Mo23            &   0.43             &     0.14             &   0.85                          \\           
C21             &   0.14            &      0.06             &   0.95                               \\
\hline
\hline 
\end{tabular}
\end{table}

\begin{table} [h]
\centering
\hskip -0.5 truecm
\caption{\label{table:16Cyg3} Relative differences (\%)
between  the interferometric radius and the corresponding inferred radius 
 $\Delta R = (R_{int}-R_X)/R_{int}$ for each star. Relative differences  between ages of both stars compared to the inference uncertainties  
$\Delta A\equiv (|A_{16CygX}-<A_{16Cyg}>|/\delta A$ where $<A_{16Cyg}>$ is the mean of ages of 16 Cyg A and 16Cyg B. }
\begin{tabular}{ ||l || c |    c|    c   ||}
\hline
\hline 
                   &   Mo23                      &    C21                                     \\
  $\Delta R$       &    1.27-1.24               &  1.27-1.32                                        \\
   $\Delta A$       &  0.65-0.78                 &    1.19-1.00                                 \\
 \hline
\hline 
\end{tabular}
\end{table}    
 
Along with  the Sun,  the 16Cyg system  is often used  to test the sensitivity of the inferred results to the use of new/updated inference approaches or new and updated physical processes implemented in stellar modelling (e.g. \cite{bellinger2016}, \cite{morel2021},  \cite{nsamba2021}, \cite{rendle2019}, \cite{SA2022}, \cite{ong2021},  \cite{verma2022}, \cite{farnir2023}, \cite{betrisey2023}, \cite{metcalfe2023}). For instance, \cite{farnir2020} carried out a comprehensive study of the 16 Cyg A,B binary  system by estimating the sensitivity of 
several uncertainties of the MRA inferences- each at a time- in the physical description of both stars.  
Taking the extremum values about their centroid values of the full set of calculations, the authors found  relative differences of $\pm$ 3.7\% and $\pm$ 7\% for the 
mass and age of 16 CygA and for CygB. The centroid values fall in the same ranges as found by  previous authors.  However they were not able to find stellar  
models of both stars with a common age and the same chemical composition while assuming the same physical description. They had to give up either the assumption of
the  same chemical composition or assume that the stars undergo different efficiency of the atomic diffusion, probably counteracted by additional transport 
processes yet to be identified. In our illustrative case,  we give in Table~\ref{table:16Cyg2} the relative differences  of the median values  for the mass, radius, 
and age of each star resulting from the inferences  using the two  grids Mo23 and C21. As is well-known and already seen above with the `degraded Sun' discussion, the age is 
the most affected by differences between the two grids. Here again the main reason is  the inclusion or not of atomic diffusion. 

\begin{table} [h]
\hskip -0.5 truecm
\caption{\label{table:16Cyg4}  Sensitivity to different input physics, number and values of free parameters and reference solar chemical composition
$DM\equiv (M_{Mo23}-M_{C21})/M_{Mo23}; DR \equiv  (R_{Mo23}-R_{C21})/R_{Mo23} ;  DA \equiv (A_{Mo23}-A_{C21})/A_{Mo23}$ }
\begin{tabular}{ ||l || c |    c|    c   ||}
\hline
\hline 
                        &   16 CygA          &  16 CygB   \\
 \hline
 DM  (\%)              &          3.26      &   2.73     \\
 DR   (\%)            &          1.14    &    0.97     \\
 DA   (\%)             &         -13.00      &   -13.43  \\
  \hline
\hline 
\end{tabular}
\end{table}    

This illustrates the lessons we can learn from the study of  seismically well characterized stars.

\section{Conclusion}
\label{conclusion}

 We estimated the detection probability of solar-like oscillations for the target stars of the ESA project  PLATO as provided by the version 1.1.0 of
the PLATO input catalogue. The targets  belong to  different samples: stars  with the lowest expected noise level constitute
 the P1P2 sample  (main sequence and subgiant FGK stars with magnitude less or equal to 11) and 
the P5 sample contains similar types of  stars but with  a higher noise level.
 A positive  detection was assumed whenever the probability that the signal is due to noise is less or equal to 0.1\% and
   the probability of the signal be due to solar like oscillation is larger than 99\%.
  We then found that we can expect positive detections of solar-like oscillations for $5858^{+176}_{-879}$ stars in the P1P2 sample
      in one single field after a two-year run of observation assuming the instrument remains nominal over the two years. 
  The given uncertainties are due to false negative and false positive detections as calibrated with Kepler data and
  likely mostly due the fact that we could not take into account the stellar activity or a non solar chemical composition.   
For the P5 sample, we find a positive detection of $ 9491^{+285}_{-1424}$  stars in the same observing conditions and 
assuming the same relative  uncertainty percentages.   As a whole,  we can  expect more than 15000 stars with solar-like oscillations 
to be compared to the Kepler solar-like oscillating  (main sequence and subgiant stars)
sample of   624 stars (M22).

   The S/Ns of  the targets in the  P1P2 sample is (by construction of the sample) 
    high  enough  that  individual mode frequencies can be measured  with high precision. For the  P1P2 targets for which we predicted 
    a positive seismic detection, we computed the  expected frequency uncertainties.
    We used the error propagation due to those 
     frequency uncertainties to estimate the relative uncertainties that we must expect for the seismically inferred masses, radii, and ages of those targets.

         Focusing on main sequence stars with masses of $\leq 1.2M_\odot$, we found that about 1131 stars
satisfy the PLATO requirements for the uncertainties of the seismically inferred stellar masses, radii, and ages in one single field after
a two-year run of observation.
Those stars   will constitute 
  an enlarged set of well characterized stars, compared to the {\it Kepler} LEGACY sample,  which contains 66 stars,      out of which about 31 main sequence stars
  with mass    $\leq 1.2M_\odot$. 
   
    We note that the PLATO mission is expected to operate for four years, with  possible extensions over 4.5 more years. 
   This will  make possible  to  more than double the number of detection of
   solar-like oscillators or to increase signicantly the precision of the measurements (of frequencies and then of MRA inferences), depending on whether we 
   observe several fields   or remain longer on  one field. 
   
  We must  stress that to the above  uncertainties, we must add uncertainties due to systematic errors that mostly arise 
  from imperfect physical description  of our stellar models.  Those can contribute up to 5 to 10\% to the age uncertainties depending on the type of stars. 
  Ongoing  theoretical works are therefore currently
  addressing  the main problems of inaccuracy. Tests and validations of improvements in the physical description of stellar models must use the best 
  seismically characterized stars at our disposal.  While 
  the well-characterized stars of the {\it Kepler} Legacy sample helped us to identify  such stellar modelling biases and  offering a path to solving them, 
  further advances are currently limited by the small number of stars able to bring tight constraints on the various modelling issues of solar-like oscillating 
stars. It is therefore one of the key  goals of the PLATO mission to collect a sufficiently large  number of  stars with the highest quality  data that can serve  as 
benchmark stars or calibration stars to improve stellar modelling.  
   The  expected sample of PLATO solar-like oscillators will provide a much larger diversity of well-characterized 
stars than that available today.  This will result in a larger and denser parameter space in terms of mass, age, chemical composition, and rotation rate. 
This will then make it possible to reduce the uncertainties in stellar modelling, particularly with regard to the internal transport processes that mainly affect 
the determination of stellar ages.
 
 {\footnotesize
 \paragraph*{acknowledgements}
 We would like to thank the anonymous referee for his pertinent comments, which helped to improve the manuscript.
This work presents results from the European Space Agency (ESA) space mission PLATO. The PLATO payload, the PLATO Ground Segment, and PLATO data processing are 
 joint developments of ESA and the PLATO Mission Consortium (PMC). Funding for the PMC is provided at national levels, in particular by countries participating 
 in the PLATO Multilateral Agreement (Austria, Belgium, Czech Republic, Denmark, France, Germany, Italy, Netherlands, Portugal, Spain, Sweden, Switzerland, Norway, 
 and United Kingdom) and institutions from Brazil. Members of the PLATO Consortium can be found at https://platomission.com/. The ESA PLATO mission website is 
 https://www.cosmos.esa.int/plato. We thank the teams working for PLATO for all their work.
M.C. acknowledges the support of Fundação para a Ciência e Tecnologia FCT/MCTES, Portugal, throughnational funds by these grants UIDB/04434/2020, UIDP/04434/2020.FCT, 2022.03993.PTDC (DOI:10.54499/2022.03993.PTDC) and CEECIND/02619/2017. 
T.M. acknowledges financial support from Belspo for contract PLATO mission development.  
Funding for the Stellar Astrophysics Centre was provided by The Danish National Research Foundation (Grant DNRF106). 
M.J.G.,  C.C., R.S., K.B., R.M.O, D.R., Y.L., B.M.,  and J.B. acknowledge support from the Centre National d’Etudes Spatiales  (CNES).
A.S. acknowledges grants PID2019-108709GB-I00 from Ministry of Science and  Innovation (MICINN, Spain), Spanish program Unidad de 
Excelencia Mar\'{i}a de Maeztu CEX2020-001058-M, 2021-SGR-1526 (Generalitat de  Catalunya), and support from ChETEC-INFRA (EU project no. 101008324) and the Plan de Recuperación, Transformación y  Resiliencia (PRTR-C17.I1)
A.M. acknowledges support from the ERC Consolidator Grant funding scheme (project ASTEROCHRONOMETRY, G.A. n. 772293.
S.M.~acknowledges support from the Spanish Ministry of Science and Innovation (MICINN) with the Ram\'on y Cajal fellowship no.~RYC-2015-17697, the grant no. PID2019-107061GB-C66, and the grant no.~PID2019-107187GB-I00, and through AEI under the Severo Ochoa Centres of Excellence Program 2020--2023 (CEX2019-000920-S).
}
\bibliographystyle{aa}
\bibliography{biblio.bib}       

\appendix

\section{Validation of adopted input for our theoretical calculations}
\label{appendixA}

We determine and validate our theoretical relations and calculations  by comparing them  with the appropriate {\it Kepler} data sets described in Sect.\ref{Keplersamples}. The goal is to choose the proper recipes for the input parameters that enter the calculation of either the detection probability or the MRA uncertainties for the PLATO target stars. 
 
\subsection{Global seismic parameters and validation of the scaling relation for the seismic mass}
 \label{scalings}

 The scaling properties of the global seismic parameters are discussed in \cite{chaplin2013,hekker2020} and references therein;
 $\Delta \nu \propto \bar \rho^{1/2} \propto (M/R^3)^{1/2}$ and $\nu_{\rm max}\propto g/T_{\rm eff}^{1/2}$  where $\bar \rho$ and $g$ 
 are the mean density and the  gravity,  respectively. When scaled with the solar values, this gives
 \begin{align}
 & \frac{\nu_{\rm max}}{\nu_{\rm max,\odot}} = \left(\frac{M}{\rm{M}_\odot}\right) \left(\frac{R}{R_\odot}\right)^{-2} \left(\frac{T_{\rm eff}}{T_{\rm eff,\odot}}\right)^{-1/2}   \label{numax1a}\\
 & \frac{\Delta \nu}{ \Delta \nu_{\odot}} = \left(\frac{M}{\rm{M}_\odot}\right)^{1/2} \left(\frac{R}{R_\odot}\right)^{-3/2} \,   \label{numax1b}.
 \end{align}
 For  the solar values, we adopt  $\rm{A}_{\rm max,bol,\odot}  = 2.53~ ppm$ \citep[rms value, see][]{michel2009}, $\nu_{\rm max,\odot}    = 3090\mu $Hz,  $\Delta\nu_\odot  = 135.1 \mu$Hz, and $T_{\rm eff,\odot}  = 5777$ K.
 
These relations can be inverted to provide the seismic mass and radius provided that $\nu_{\rm max}$, $\Delta \nu$, 
and $\rm{T}_{\rm eff}$ are known. The inverted relationships read as follows: 
 
 \begin{align}
 &  \frac{M}{\rm{M}_\odot}  =   \left(\frac{\nu_{\rm max}}{\nu_{\rm max,\odot}}\right)^3   \left(\frac{\Delta \nu}{ \Delta \nu_{\odot}}\right)^{-4} \left(\frac{T_{\rm eff}}{T_{\rm eff,\odot}}\right)^{3/2}   \label{numax2a}\\
 &  \frac{R}{R_\odot}       =   \left(\frac{\nu_{\rm max}}{\nu_{\rm max,\odot}}\right)   \left(\frac{\Delta \nu}{ \Delta \nu_{\odot}}\right)^{-2} \left(\frac{T_{\rm eff}}{T_{\rm eff,\odot}}\right)^{1/2}   \,   \label{numax2b}.
 \end{align}


Further, it is also well accepted that $\Delta \nu$ and $\nu_{\rm max}$ are tightly 
correlated 
\citep[][]{stello2009,huber2011,serenelli2017}. Indeed a fit to the {\it Kepler} data yields
\begin{align}
\label{numaxDeltanu}
\frac{\Delta \nu}{\Delta \nu_\odot} = a \left(\frac{\nu_{\rm max}}{\nu_{\rm max,\odot}}\right)^s.   
\end{align}
We then  use this scaling relation between $\nu_{\rm max}$ and $\Delta \nu$ into Eq.~(\ref{numax1a})  above and  inverting the resulting equations  yields 
the seismic mass as 
 \begin{align}
 \label{MRscal2a}
 \frac{M}{\rm{M}_\odot} = a^{-4}  \left(\frac{\nu_{\rm max}}{\nu_{\rm max,\odot}}\right)^{3-4s}  \left(\frac{T_{\rm eff}}{T_{\rm eff,\odot}}\right)^{3/2}. 
\end{align}
Similarly, the seismic radius is obtained as
 \begin{align}
 \label{MRscal2b}
\frac{R}{R_\odot} = a^{-2}  \left(\frac{\nu_{\rm max}}{\nu_{\rm max,\odot}}\right)^{1-2s}  \left(\frac{T_{\rm eff}}{T_{\rm eff,\odot}}\right)^{1/2}. 
\end{align}

\begin{figure}[t]
        \begin{center}
                \includegraphics[width=9.cm]{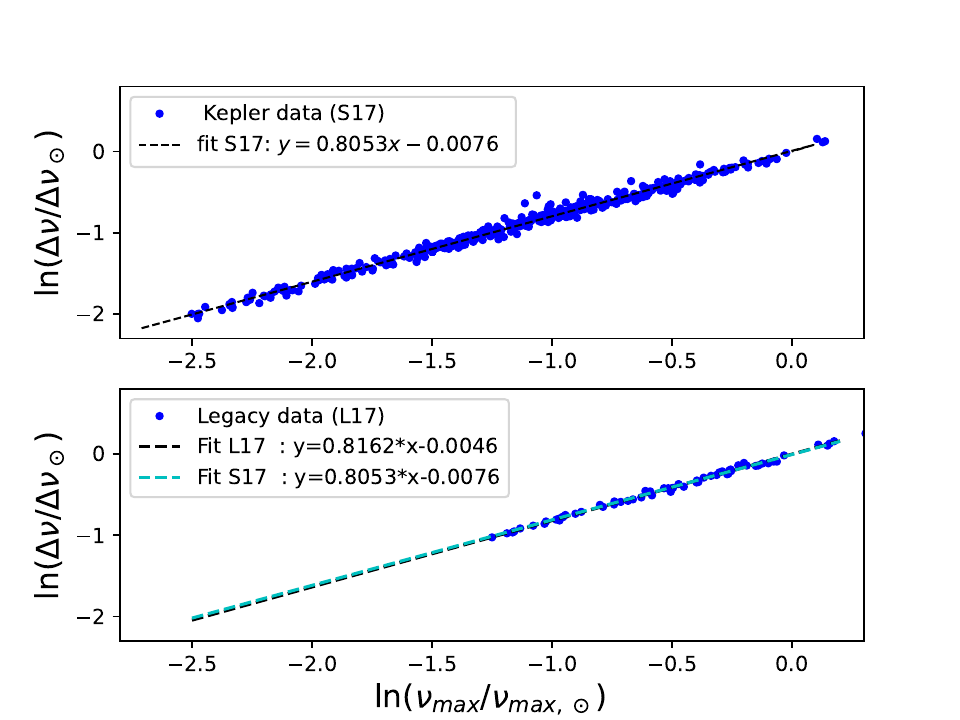}
                \caption{Scaled $\Delta \nu$ as a function of scaled $\nu_{\rm max}$ in logarithmic scales with
    $\Delta \nu_\odot=135 \mu$Hz and $\nu_{\rm max} =3090 \mu$Hz. 
  Blue dots: {\it Kepler} short cadence data.  Top: S17's catalogue (S17's table 3). Bottom: Legacy data (L17). Dashed lines are linear fits.} 
                \label{lnumaxlDeltanu}
        \end{center}
\end{figure}

For consistency, we rederived  the relation Eq.(\ref{numaxDeltanu}) by fitting our {\it Kepler} data set of solar-like oscillating stars. Figure~\ref{lnumaxlDeltanu}
shows the variation of  $\Delta \nu$ as a function of   $\nu_{\rm max}$ for the  short cadence {\it Kepler} data  on one hand from the Legacy sample from 
\cite{lund2017} (L17) and on the other hand from the S17 catalogue (their Table~3). A fit of the S17's data gives $a= 0.992$ and
 $s=0.805$ (assuming $\Delta \nu_\odot=135 \mu$Hz  and $\nu_{\rm max,\odot} = 3090\mu$Hz) while a fit to L17's data gives $a=0.995$ and $s=0.816$,
  both in satisfactory agreement with relations found in the literature. 
  
 We then adopt  $s=0.82$ and $a=1$ in the rest of the paper. We then obtain, from Eq.~(\ref{MRscal2a}), the  scaling for the seismic mass as:
\begin{align}
\label{MRscal3}
 \frac{M}{\rm{M}_\odot} =   \left(\frac{\nu_{\rm max}}{\nu_{\rm max,\odot}}\right)^{-0.28}  \left(\frac{T_{\rm eff}}{T_{\rm eff,\odot}}\right)^{3/2}. \end{align}
For the radius, we obtain, from Eq.~(\ref{MRscal2b}):
\begin{align}
\label{seismicradius}
\frac{R}{R_\odot} = \left(\frac{\nu_{\rm max}}{\nu_{\rm max,\odot}}\right)^{-0.64}  \left(\frac{T_{\rm eff}}{T_{\rm eff,\odot}}\right)^{1/2}. 
\end{align}
It is known that the global seismic scaling relations Eqs.~(\ref{numax1a}) and Eqs.~(\ref{numax1b}) are accurate at the level of a few percents 
(\cite{huber2017}, S17 and  references therein). This is enough for our purpose since masses 
and radii are only used here to delimitate some sub-samples of stars of interest.
 
 When $\nu_{\rm max}$ is not known, it is derived by inverting Eq.~(\ref{seismicradius}) 
 \begin{align}
 \label{invertedradius}
&  \frac{\nu_{\rm max}}{\nu_{\rm max,\odot}}    = \left(\frac{R}{R_\odot}\right)^{-1.5625} \ \left(\frac{T_{\rm eff}}{T_{\rm eff,\odot}}\right)^{0.78} \\
&  \frac{\Delta \nu}{ \Delta \nu_\odot }  = \left(\frac{\nu_{\rm max}}{\nu_{\rm max,\odot}}\right)^{0.82} 
\end{align}

The seismic parameter  $\nu_{\rm max}$  can be obtained from a seismic analysis 
(e.g. from {\it Kepler} data)  and $T_{\rm eff}$ is obtained from a spectroscopic study. Because for the PLATO targets, $\nu_{\rm max}$   are not yet 
known, we use the stellar radius and $T_{\rm eff}$ from the PIC to derive  $\nu_{\rm max}$. 

\subsection{Oscillation power amplitude $\rm{A}_{\rm max}$ }
\label{amp}

Both formulations, Eqs.~(\ref{ptotC11}) and (\ref{ptotS19}), for the power, $P_{\rm tot}$, involve the oscillation maximum amplitude $A_{\rm max}$.
In order to adopt a theoretical relation for the amplitudes, we compare the amplitudes given by  two semi-empirical relations derived in C11 and S19  
with   the measured amplitudes, $\rm{A}_{\rm max,obs}$, of  the {\it Kepler} Legacy sample by \cite{lund2017} (their Table~3). 
We recall that  this sample is composed of 66 main-sequence stars with the highest quality of  seismic data, 
and for which individual oscillation modes are identified and their  frequencies, amplitudes, and line widths are measured with the highest level of precision.
   
On the theoretical side, C11  derived  the relation (Eq.~9 in C11): 
\begin{align}
\label{AmpC11}
\frac{\rm{A}_{\rm max}}{ \rm{A}_{\rm max,bol,\odot}}= \beta_{is} ~  \left( \frac{R}{R_\odot}\right)^{2}   \left(\frac{T_{\rm eff}}{T_{\rm eff,\odot}}\right)^{1/2},   
\end{align} 
where 
\begin{align}
\beta_{is} = 1-{\rm e}^{(T_{\rm eff}-T_{\rm red})/1550 {\rm K}}      
\end{align}
and 
\begin{align}
 & T_{\rm red} (K)=8907 \left(\frac{L}{L_\odot}\right)^{-0.093}  ,\\
 & \frac{L}{L_\odot}  = \left(\frac{R}{R_\odot}\right)^2 \left(\frac{T_{\rm eff}}{T_{\rm eff,\odot}}\right)^4  \, .
\end{align}
 
S19 used the relation between $\rm{A}_{\rm max}, \nu_{\rm max}, \Delta\nu_\odot$, and $T_{\rm eff}$ established by \cite{Corsaro2013} based on their model 4$\beta$  (Eq.~19 , Table 3 in \cite{Corsaro2013}) 
\begin{align}
\label{AmpS19}
  \frac{\rm{A}_{\rm max}}{\rm{A}_{\rm max,bol,\odot}}  =  1.41   ~\left(\frac{\nu_{\rm max}}{\nu_{\rm max,\odot}}\right)^{-2.314} 
     \left(\frac{\Delta \nu}{\Delta \nu_\odot}\right)^{2.088}  \left(\frac{T_{\rm eff}}{T_{\rm eff,\odot}}\right)^{-2.235}  \, , 
\end{align}
valid for $4000 \mu$Hz $ >\nu_{\rm max}>150\mu$Hz. We eliminate $\Delta \nu/\Delta \nu_\odot$ with Eq.~(\ref{numaxDeltanu}). With $s=0.82$, $a=1$, this yields 
the second  relation that we will consider (S19): 
\begin{equation}\label{AmpS19}
\frac{\rm{A}_{\rm max}}{\rm{A}_{\rm max,bol,\odot}}=   1.41 \left( \frac{\nu_{\rm max}}{\nu_{\rm max,\odot}}  \right)^{-0.610} \left(\frac{T_{\rm eff}}{T_{\rm eff,\odot}}\right)^{-2.235}  \, . 
\end{equation}
Both relations involve the effective temperature $T_{\rm eff}$, C11 also involves the stellar radius and S19 involves $\nu_{\rm max}$. We also need the 
stellar mass   to discriminate between  low and high mass cases. The stellar radius and the stellar  mass  are  given by the   scaling relation  Eq.~(\ref{MRscal3}). 

\begin{table} 
\centering
\caption{Adopted amplitude formulation for $\rm{A}_{\rm max,scal}$. Masses, $M,$ in solar units and $\nu_{\rm max}$ in $\mu$Hz. } 
\label{table:amax}
\begin{tabular}{ |l | c | c| c|}
\hline
\hline
 $\nu_{\rm max}$                      &   $ \leq 2500 $  &  $ > 2500 $   \\
\hline
  $ M \leq 1.15$               &      $  \rm{A}_{\rm max, C11}*1.31  $                    &    $ \rm{A}_{\rm max, C11} *1.19 $                     \\
\hline 
  $ M > 1.15 $             &      $ \rm{A}_{\rm max, S19}*0.95 $                    &    $ \rm{A}_{\rm max, C11}*0.95 $                     \\
 \hline
\hline 
\end{tabular}
\end{table} 

\begin{figure}[t]
        \begin{center}
                \includegraphics[width=9.cm]{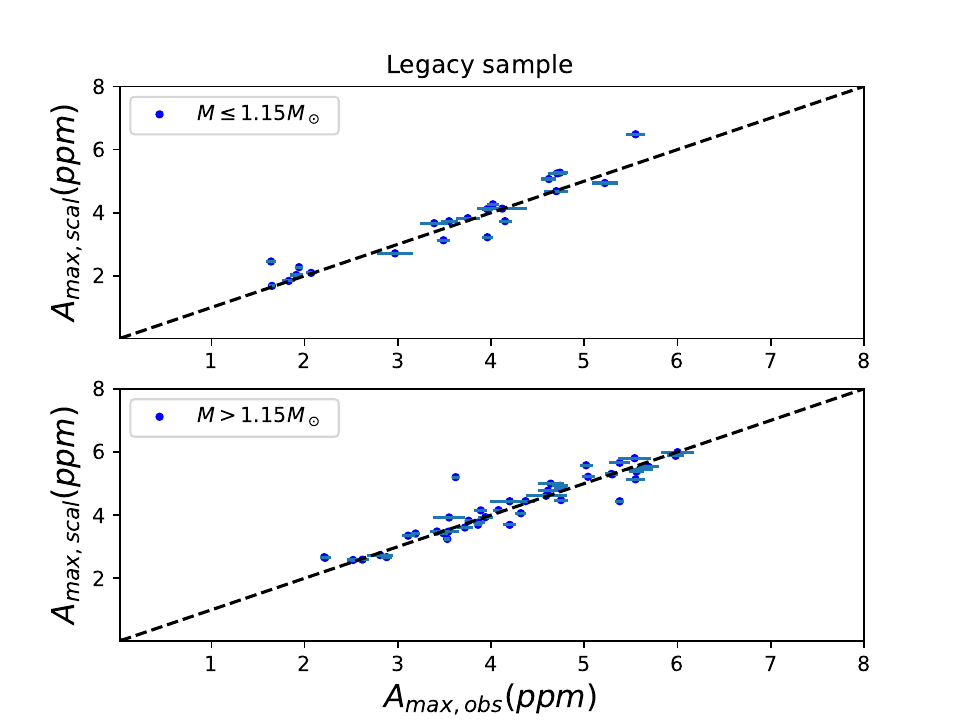}
                \caption{Comparison of  the amplitudes $\rm{A}_{\rm max, scal}$ computed  as 
      given in Table~\ref{table:amax} with the observed amplitudes $\rm{A}_{\rm max, obs}$ for the LEGACY sample  from \citep{lund2017}.} 
                \label{kepAmp4}
        \end{center}
\end{figure}     

Figure~\ref{kepAmp4} compares  the  amplitudes $\rm{A}_{\rm max, scal}$ as given in Table~\ref{table:amax} to the observed amplitudes  $\rm{A}_{\rm max,obs}$ for 
the Legacy sample. The comparison leads us to adopt as best estimate the purely empirical amplitude relations as given in Table~\ref{table:amax}. As can be seen
 in  this figure, our empirical calibration of $\rm{A}_{\rm max}$ is quite satisfactory. The rescaling factors 1.19 and 1.31 in Table~\ref{table:amax}  can be explained 
 as arising from uncertainties in the estimate of the {\it Kepler} noise. The threshold $M=1.15 \rm{M}_\odot$ is arbitrary but reflects  the fact that the scatter in $\rm{A}_{\rm max}$ is larger 
for stars in the high mass regime corresponding to higher effective temperature for which measurements of frequencies, linewidths, and amplitudes 
are more difficult \citep{appourchaux2012}. On the theoretical side, the $1.15 \rm{M}_\odot$ threshold corresponds to the discrimination between main sequence
 stars with and without  a convective core.

\section{Validation for the probability of  detection } 
\label{prediction}

In order to validate  our calculations, we computed the  detection probability for the samples of {\it Kepler} stars with known detection (sample 1) 
 on one hand and the sample of stars for which 
no solar-like oscillations were detected (sample 2) on the other hand. For sample 1, we estimate the percentages of false negative 
detections (i.e.  the number of stars which we predict no seismic detection). For sample 2,  we estimate the percentages of false 
positive detections (i.e. the number of stars from which we predict a seismic detection). 
For the results, we consider 3 types of mass regimes: R1 (MS stars with  $M<1.6 M_\odot$ and subgiants all masses), R2 (stars with $M \leq 1.2$),  R3  (MS stars with  $M \leq 1.2)$

\subsection{Probability of detection for the  sample 1 and sample 2}

 When computing the detection probability for  sample 1  (413 stars), we assumed as our baseline the following input assumptions:   
 $\nu_{\rm max}$, $\Delta\nu$ and $T_{\rm eff}$ are from M22, masses 
are derived with Eq.\ref{numax2a} and   $\delta \nu_{\rm env}= \nu_{\rm max}/2$.  The yields are listed in Table~\ref{table:table9}.
 We find, in the R1 mass regime when $P_{\rm det} \leq 0.99$, 
 a percentage of  about $\sim 10 \%$ of false negative detections when the detection probability is  $P_{\rm det} \leq 0.99$ and $\sim 5\%$ when $P_{\rm det} \leq 0.90$.

\begin{table}[h]
\centering
\caption{ Numbers of false negative detections for  sample 1   with known detected oscillations assuming the  baseline conditions. 
Number of stars predicted with  NO seismic detection. Masses in solar unit.} 
\label{table:table9}
\begin{tabular}{ |l | c | c| c|}
 \hline 
\hline
mass  regime                    &     R1              &        R2                    &  R3   \\
nb of stars                     &   413 stars         &       276 stars             &   52 stars \\ 
 \hline 
\hline 
$P_{\rm det} \leq 0.90$        &       20             &         9                    &     7 \\ 
                               &     $ \sim$   5\%     &       $ \sim$    3\%         &    $ \sim$  13\%     \\ 
\hline 
$P_{\rm det} \leq 0.99$       &       40               &        16                   &     11     \\ 
                              &  $ \sim$   10\%        &   $ \sim$  6\%             &   $\sim$  21\%    \\
\hline
\hline
\end{tabular}
\end{table}

The choice of  $P_{\rm det} \leq 0.99$ obviously  predicts a  larger  number  of false negative detections compared to the choice of  $P_{\rm det} \leq 0.90$ and is therefore more conservative for estimating
 the percentage of positive detections. As a sanity check, we verified that all stars in the Legacy sample are found with a detection probability equal to  1 as expected.

\medskip 
We next  considered the 990  {\it Kepler}  stars  in sample 2. For that sample, we assumed as our baseline the following input assumptions:   
 $\nu_{\rm max}$ derived from  $\log g$ and $T_{\rm eff}$ given by  M19, $\Delta \nu$ from the scaling relation Eq.\ref{numaxDeltanu} with $a=1,s=0.82$, masses 
are derived with   Eq.\ref{MRscal3} and   $\delta \nu_{\rm env}= \nu_{\rm max}/2$.  Table~\ref{table:table10} provides the yields of the calculations of the   
detection probability for those stars. Assuming a seismic  positive detection when $P_{\rm det} > 0.99$, 
we find that the percentage of false positive detections 
is $\sim$ 20\%, 11\%, and  7\% for the R1, R2, and R3 mass regimes, respectively .  
 
\begin{table}[h]
\centering
\caption{Numbers of false positive detections for the stars of sample 2 assuming the baseline case for that sample. Numbers of stars  predicted 
with seismic positive detection.
 Masses are in solar units.} 
\label{table:table10}
\begin{tabular}{ |l | c | c| c|}
\hline 
\hline
mass  regime         &       R1  &  R2   &  R3   \\
nb of stars          &      936   & 396   & 361  \\
\hline 
\hline 
$P_{\rm det} > 0.90$         &  279  & 64  &  44    \\ 
                             &$ \sim$   30\%  &$\sim$ 16\%  & $\sim$ 12\%   \\ 
\hline 
$P_{\rm det} > 0.99$         &    186     &44  &   27  \\    
                             &    $\sim$  20\%    &  $\sim$ 11\%  &   $\sim$   7\%   \\ 
\hline 
\hline
\end{tabular}
\end{table} 
 
When  smoothing the PSD in order to decrease the noise level in case of a low S/N, it is found that the  measured amplitudes are decreased by about  6\% compared 
to the true amplitudes \citep{lund2017}. Assuming detection when $P_{\rm det} > 0.99$, the impact of decreasing  our theoretical amplitudes by 6\%  leads to a percentage of false positive 
detections  $\sim$ 17\%, 10\%, 6\%    for the R1, R2 and R3 mass regimes, respectively, which is slightly lower that for the baseline case
 ($\sim$ 21\%, 11\% and  7\% in Table \ref{table:table9}).   
 The impact is not important and we will  no longer consider the case of decreasing the amplitudes.  Here  
 the choice of $P_{\rm det} > 0.99$ rather than 0.90 provides a lower number of false positive detections.

 \medskip
 Hence, in  both cases above, $P_{\rm det} > 0.99$ is the conservative  choice,
  namely, lowering  the number of positive detections and maximising the number of false detections. 
 We therefore adopted a  detection threshold at 99\% when considering the PLATO targets.
 
  There are a number of assumptions behind the above calculations which lead to some uncertainties in the results. 
  In the next sections, we estimate the impact of  the main sources of uncertainties in predicting the detection probability 
  (assuming the conservative case of a  detection threshold at 99\%). 

 \subsection{Impact of uncertainties in the input quantities}
   
 We estimated  the impact on the detection probability of the uncertainties in the various  input quantities. We carried out the calculations,  changing 
  one or several input  quantities at a time  and the results must be  compared  to the results obtained with the baseline cases. 
     
     We assessed first the impact on the number of false negative detections with sample 1 
      when  using the values of $\nu_{\rm max}, \Delta \nu$, and $T_{\rm eff}$ from the  S17 
 instead of using  those of M22.  The results are listed in Table~\ref{table:table12}.
 We note that  the number of stars-  in each mass regime- can change depending on the adopted assumptions, specifically because the seismic mass 
 is computed using a scaling relation which involves
 $\nu_{\rm max},\Delta\nu$, and $T_{\rm eff}$.
The false negative detections in case of the S17 input parameters are  $\sim$ 11\%, 14\%,  16\% for R1, R2, and R3, respectively (Table~\ref{table:table12}), namely,   slightly greater than $\sim$ 10\%, 6\%, and 21\%,  respectively,  when using the set $\nu_{\rm max}, 
\Delta \nu$ and $T_{\rm eff}$ from M22 in Table~\ref{table:table10}.
   The impact  is higher when considering  the smaller sample of  
 MS stars with  $ M \leq 1.2$  (R3) than the whole sample (R1).   
 
 \begin{table}[h]
\centering
\caption{ False negative detections for  sample 1.  Each row corresponds to one change compared to the baseline case.
  Figures in parenthesis correspond to the total number of stars in each mass regimes and considered case. Masses, $M$,  are in solar units. } 
\label{table:table12}
\begin{tabular}{ |l | c | c| c|  }
 \hline 
\hline
mass regime                            &   R1                     & R2                       & R3                    \\
 \hline 
\hline 
 $\nu_{\rm max,S17}$                  &   44  (413)               &     19  (276)            &      17  (64)                 \\
                                      &    $\sim$    11\%          &      $\sim$   7 \%       &      $\sim$    27\%                \\  
\hline 
 $T_{\rm eff, S17}$                    &   59 (413)                 &      17 (224)          &     8 (50)                  \\ 
                                       &    $\sim$    14 \%        &        $\sim$    8\%     &      $\sim$   16\%                                  \\
\hline
$\Bigl(\nu_{\rm max},\Delta \nu$       &   65   (413)              &     22 (224)            &       15 (53)                   \\ 
 \& $T_{\rm eff}\Bigr)_{\rm S17}$     &      $\sim$ 16 \%          &     $\sim$  10\%       &       $\sim$   28 \%                    \\
\hline 
  $\Delta \nu_{\rm scaling}$          & 33 (413) & 12  (276)    &   11  (52)  \\
                                      &    $\sim$    8\%          &     $\sim$  4 \%            &     $\sim$    21\%              \\        
\hline 
$\delta \nu_{\rm env}=\nu_{\rm max}$  &   14  (413)                        &       6  (276)            &         5  (56)           \\
                                      &      $\sim$    4\%                        &       $\sim$   2\%              &    $\sim$    9\%             \\
\hline
\hline
\end{tabular}
\end{table}

\medskip

As a second type of uncertainty, we  considered the changes when one uses the scaling relation Eq.~(\ref{numaxDeltanu}) (with $s=0.82$ and $a=1$)
 for $\Delta \nu$, $\Delta \nu_{\rm scaling}$,
which is what we have to use in absence of an observed value (case of sample 2 and Plato targets)  
in Appendix~\ref{appendixA}.   We computed the detection probability  using our adopted  sample  of oscillating stars, sample 1.
As can be seen from Table~\ref{table:table12}, we find  false negative detections at the level of 8\%, 4\%, and 21\% for the R1, R2, and R3 mass regimes, respectively.
These figures must be compared with 10\%, 6\%, and 21\%, respectively  found with the baseline case (Table~\ref{table:table10}). 
Here the changes in the false negative percentages come  from the fact that changing the values of $\nu_{\rm max}$ and  
$\Delta \nu$ impacts the mass derived with the scaling relation Eq.~(\ref{mass}) 
and, thus, the number of stars in each subsample. The impact remains small. 
 
\medskip
The third type of uncertainty concerns our choice  for the width of the envelope of the oscillations in a power spectrum. 
We then compare the yields obtained assuming  the baseline case $\delta \nu_{\rm env} = \nu_{\rm max} =1/2$  and  the more optimist case   $\delta \nu_{\rm env} = \nu_{\rm max}$
 for sample 1. In Table~\ref{table:table12},  we can see that the false negative detections are   about 2.5 smaller
   than 
   when assuming  $\delta \nu_{\rm env}=\nu_{\rm max}/2$. One then underestimates the detection rates when assuming 
$\delta \nu_{\rm env}=  \nu_{\rm max}$   compared to the  $\delta \nu_{\rm env}=  \nu_{\rm max}/2$ case.
The comparison between the two options for $\delta \nu_{\rm env}$ can also  be made with 
sample 2 for the false positive detection rates.   One finds a 
higher false positive detection rate when assuming $\delta \nu_{\rm env}=  \nu_{\rm max}$ (Table~\ref{table:table14})
  instead of $\delta \nu_{\rm env}=  \nu_{\rm max}/2$ (Table~\ref{table:table10}).   
Therefore, we overestimate   the positive detection rate when assuming $\delta \nu_{\rm env}=  \nu_{\rm max}$  compared  with the  
$\delta \nu_{\rm env}=  \nu_{\rm max}/2$ case.

\begin{table}[h]
\centering
\caption{False positive  ($P_{\rm det} > 0.99$) for  sample 2. 
The baseline is assumed except for  $\delta \nu_{\rm env}=\nu_{\rm max}$. Masses, $M$, in solar unit. } 
\label{table:table14}
\begin{tabular}{ |l | c | c| c|}
\hline 
\hline
mass    regime                             &   R1   &  R2   &  R3                  \\
                                          &   936  &  403    & 370                \\
\hline                    
  $\delta \nu_{\rm env}=\nu_{\rm max}$    &    316       &   76    &    54          \\   
                                          & 34\%          &  19\%        &     16\%             \\
\hline
\hline
 \end{tabular}
\end{table} 
    
As a net result and to remain conservative, we chose the option which gives a lower false positive detection rate to the cost of  a higher false negative rate. 
This justifies taking   $\delta \nu_{\rm env}=  \nu_{\rm max}/2$  for our baseline condition. 
We will nevertheless provide also the results in the more optimistic  case $\delta \nu_{\rm env} =  \nu_{\rm max} $  for the  PLATO targets.
 
\subsection{Global uncertainties for the  predicted percentage of predicted seismic detection}

In anticipation of the investigation for the PLATO case, we considered as a single sample the total sample of 1349 {\it Kepler} stars  
with detected solar-like oscillation (413 stars, sample 1) and with no detection of 
solar-like oscillation  (936 stars from sample 2) in the R1 mass regime and similarly the total sample of 620 stars in the R2 mass regime and 
the total sample of 412 stars in the R3 mass sample. 
   
\begin{table}[h]
\centering
\caption{\label{table:table16} Number of stars and associated percentages of predicted false  negative and positive detections when considering 
 the  whole {\it Kepler} population with and without real detection in each regimes and the baseline assumptions.   
 Masses, $M$, are in solar units. } 
\begin{tabular}{ |l | c | c| c|  c|}
\hline 
\hline
  mass regime           & R1                    &     R2               &   R3                 \\
 nb of stars             &  1349               &         620           &   412                 \\
\hline 
\hline
False negative         &    40    ($\sim$ 3\%)     &    16    ($\sim$  3\%)   &   11  ($\sim$  3\%)    \\
 \hline 
False positive         &     186    ($\sim$ 14\%)    &       44   ($\sim$ 7\%)   &  27   ($\sim$   7\%)                    \\ 
 \hline
\hline
\end{tabular}
\end{table}

According to Table \ref{table:table16}, the   calculations predicting the number of seismic positive detections applied to PLATO targets in Sect. 5
 underestimate it  by   3.0\%  for all mass regimes and 
overestimate it by    14\%, 7\%, 7\%  for the R1, R2, and R3 mass regimes respectively in the baseline conditions. 
 We then used those rates 
   to estimate the uncertainties on the predicted 
 number of positive detections, say $X$, for the PLATO targets in Sect.5, as follows: 
  $ X^{+3\%}_{-14\%}$  for  R1,   $X^{+3\%}_{-7\%}$ for R2, and $X^{+3\%}_{-7\%}$ for R3.

\begin{table}[h]
\centering
\caption{\label{table:table17} Number of stars and associated percentages of predicted false  negative and positive detections when considering 
 the  whole {\it Kepler} population with and without real detection in each regimes and the baseline assumptions except for $\delta \nu_{\rm env} = \nu_{\rm max}$.  
 Masses, $M$, are in solar units. } 
\begin{tabular}{ |l | c | c| c|  c|}
\hline 
\hline
  mass regime           & R1                    &     R2               &   R3                 \\
 nb of stars             &  1349               &         679           &   426                \\
\hline 
\hline
False negative         &    14    ($\sim$ 1\%)     &    6    ($\sim$  0.9\%)   &   5 ($\sim$  1\%)    \\
 \hline 
False positive         &    326   ($\sim$ 24\%)    &       76   ($\sim$ 11\%)   &  54   ($\sim$   13\%)                    \\ 
 \hline
\hline
\end{tabular}
\end{table} 
 
   Similar estimates
   assuming $\delta \nu_{\rm env} = \nu_{\rm max}$ can be found in Table~\ref{table:table17}.

\section {Uncertainties of the inferred mass, radius, and age}
\label{appendixC}

We seek  relative uncertainties for the mass, radius, and age, $\delta M/M, \delta R/R, \delta A/A$, which can be computed for the PIC targets.  
  In accordance with the stellar requirements of the PLATO mission (which is the most challenging goal),  we focus exclusively on the MS stars with  
  masses $\leq 1.2 \rm{M}_\odot$.

We  want to estimate the purely statistical uncertainties for the  seismically inferred masses, radii, and ages
 generated by the propagation of seismic observational uncertainties only.    
In order to  eliminate as much as possible all systematic errors  (which will be briefly discussed in the conclusions/discussion section), 
we  built a set of synthetic stars 
 which  masses and ages covering the ranges of interest here:
 $ (M/\rm{M}_\odot, A(Gyr)) = (0.9,3.),(1.0,2),  (1.0,4.57), (1.0,9.0), (1.03,4.6),$  $(1.08,4.6),(1.15,2.0),(1.15,9.0)$ 
and infer their seismic masses, radii, and ages by means of a routinely used  a grid-based approach, as described in \cite{cunha2021}.

\subsection{A set of synthetic stars and their frequencies} 

\begin{figure}[t]
        \begin{center}
                \includegraphics[width=9.cm]{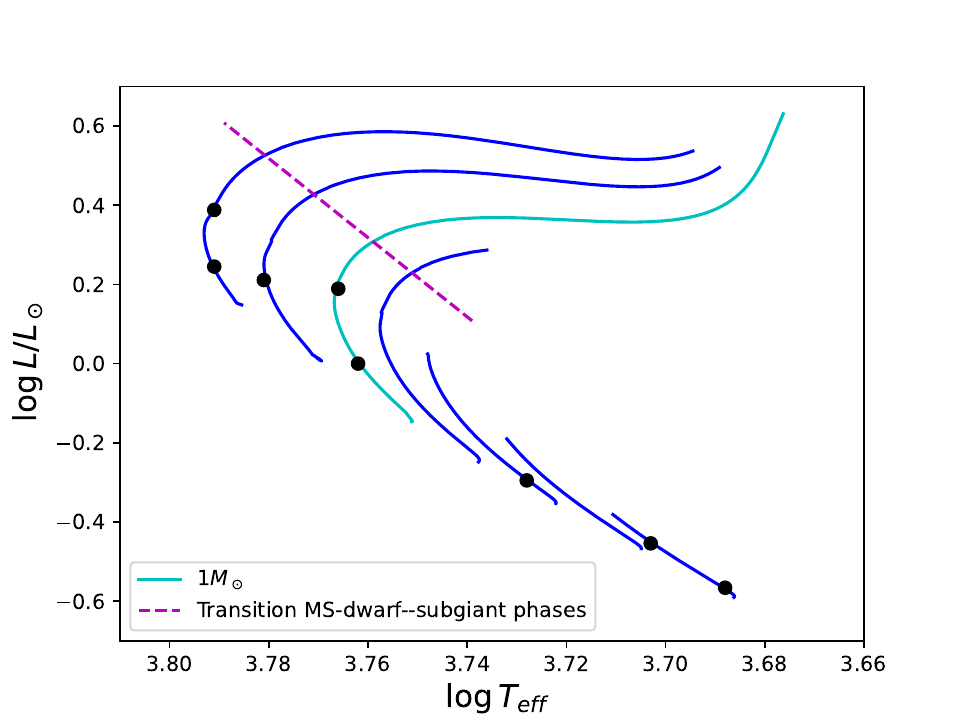}
                \caption{Evolutionary tracks  for masses from $0.8$ to $1.15 \rm{M}_\odot$ in step of $0.05 \rm{M}_\odot$ (blue solid curves) in a HR diagram.
    The black dots indicate the location of
    the synthetic targets stars studied here. 
    The  magenta dashed line shows  the location of models with their central hydrogen content   $X_c \sim 10^{-6}$   which we take here
     as the  transition between MS stars and subgiant phases. The transition follows the empirical relation :
     $\log L/L_\odot = 10~(\log T_{\rm eff} - 3.7532)+ 0.25$  for that range of mass given
       the adopted chemical composition and physics of the stellar models.}
                \label{HR}
        \end{center}
\end{figure}

     We built the stellar models of the synthetic stars 
    with the evolutionary code  CESTAM \citep{morel2008,marques2013} 
 with  the input physics as much as possible similar to that of the stellar models 
 included in the input grid.  The locations of those fictitious stars  in the HR diagram are shown in Fig.~\ref{HR}. 
 
The frequencies for each synthetic star are calculated with ADIPLS \citep{jcd2008}.
 The adopted  uncertainties on the frequencies are scaled with respect to those of the `degraded Sun' (corresponding to the Sun seen as a star)
\citep{lund2017}  which was used as a reference for the studies  of the {\it Kepler} Legacy sample in \cite{SA2017}. 
For the degraded Sun,  the $\ell=1$ mode frequency  closest to $\nu_{\rm max}$ ($\nu_{\rm max,\odot} = 3090 \mu$Hz) 
with the smallest uncertainty ($\delta \nu_{\rm \ell=1,max,\odot}=0.057 \mu$Hz) 
is $\nu_{\rm \ell=1,max,\odot}= 2963.3 \mu$Hz.

  For each synthetic star, the frequency uncertainty for each frequency $\nu_{nl}$ is taken as 
  $\delta \nu_{n,\ell} = x \times \delta \nu_{n,\ell,\odot}$. We  infer the MRA  uncertainties for the cases 
   $x=1, 3, 5, 7,   10,$ which cover the range of uncertainties  for the PLATO  P1P2 sample.  
 
For sake of simplicity, we keep the same number and types of  modes as given above 
 for which frequencies are computed  in all considered cases, although 
 this number decreases when the S/N decreases. Some impact of such degradation is discussed in \cite{cunha2021}.

\subsection{MRA inference  for the set of synthetic stars}

 The stellar MRA   and their uncertainties are obtained with  the inference code AIMS   
  \citep{rendle2019,lund2018}. AIMS  reads as an input a precomputed grid of stellar models. For convenience, we adopt the 
  stellar grid described and used in \cite{cunha2021}.   

  The observational constraints for each star are $T_{\rm eff}, L/L_\odot, [Fe/H],$
 and the frequencies of the individual modes $\ell=0$ ($n=16-27$), 1 ($n=16-27$), 2 ($n=15-27$) 
 for the degree $\ell$ and radial order $n$ similar to  the synthetic star known as Zebedee  (a young Sun) studied in \cite{cunha2021}.
 In order to eliminate systematic errors that are not relevant in this section, we take as central values 
 the exact values of $T_{\rm eff}, L/L_\odot, [Fe/H]$ from the models of the synthetic stars.
  The uncertainties (1$\sigma$) adopted are  $70K$, $0.3$, $0.05$ respectively, as expected at the time of PLATO launch.
  
\begin{figure}[t]
        \begin{center}
                \includegraphics[width=8.cm]{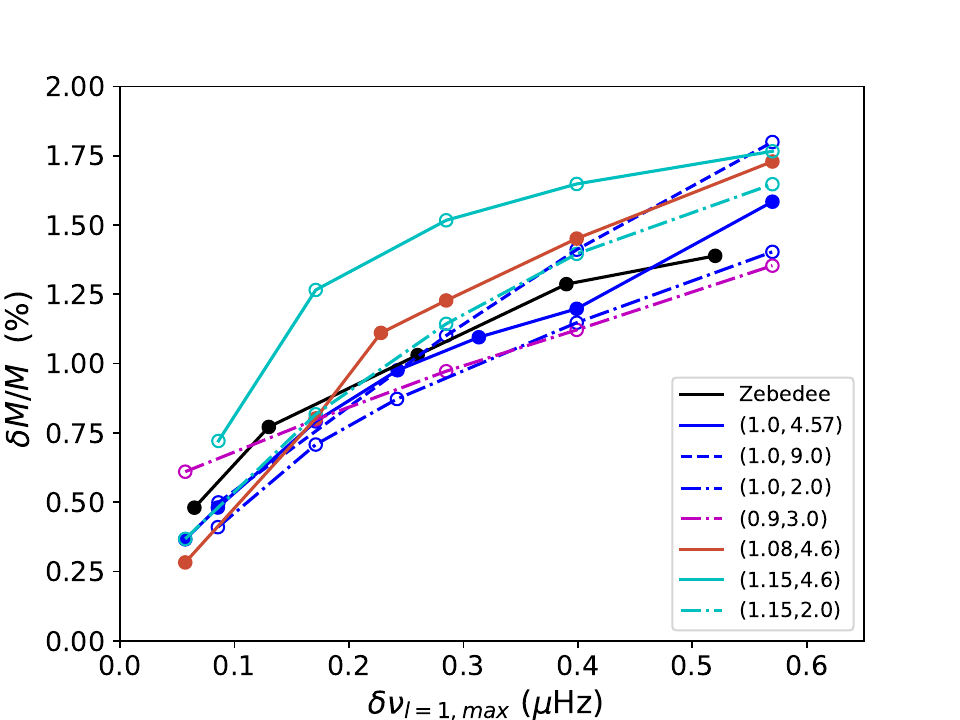}
                \includegraphics[width=8.cm]{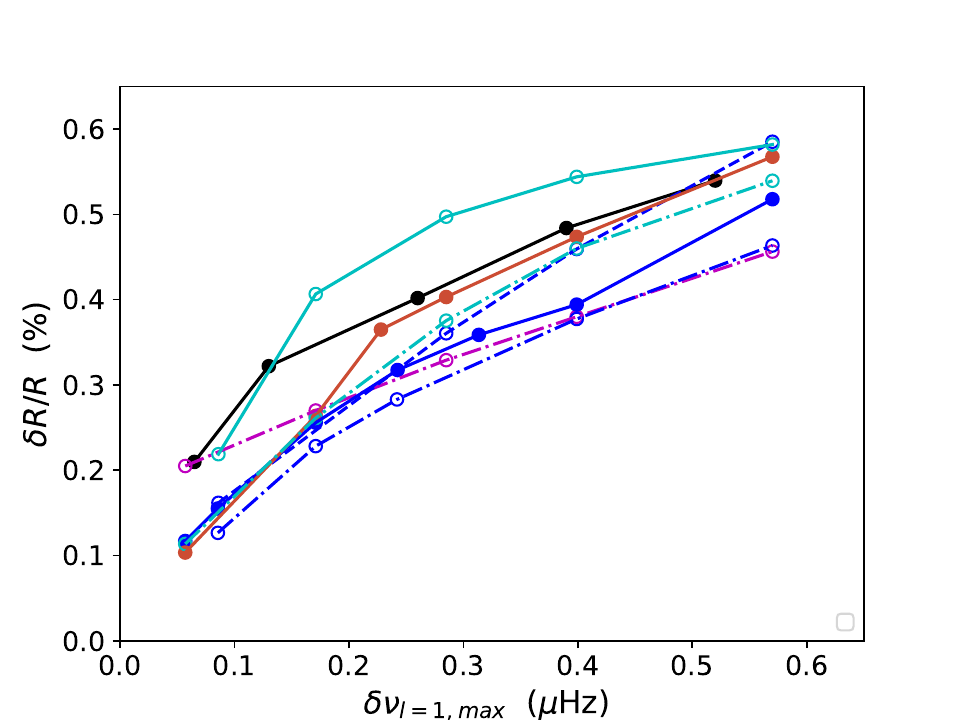}
                \caption{Relative uncertainties for the masses (top) and radii (bottom) with increasing frequency uncertainties
    for all synthetic stars as a function of $\delta\nu_{\rm \ell=1,max}$. 
The color code is given in the panels as ($M/\rm{M}_\odot$, age (Gyr)). The solid blue corresponds to a Sun. 
The solid curve with  black dots represents  a case similar to Zebedee (1.0165 $\rm{M}_\odot$, 3.085 Gyr) in \cite{cunha2021}. }
                \label{dMdRdA1}
        \end{center}
\end{figure}     
For comparison with the results in \cite{cunha2021}, we also inferred  the stellar mass, radius and age, and 
their uncertainties for the Zebedee case  (real values of the corresponding stellar model: $M/\rm{M}_\odot,A(Gyr)= 1.0165,3.085)$. 
In that particular case, 
the uncertainties are taken from \cite{cunha2021}. Since the input frequencies  for that synthetic star included
 surface-effect corrections according to the formulation of \cite{ball2014a,ball2014b}, we also included the  surface effect correction
  according to the same prescription when inferring the stellar parameters for Zebedee with the 
AIMS code.  The results compare well with the results published in \cite{cunha2021} and we do not show them.

\begin{figure}[t]
        \begin{center}
                \includegraphics[width=8.cm]{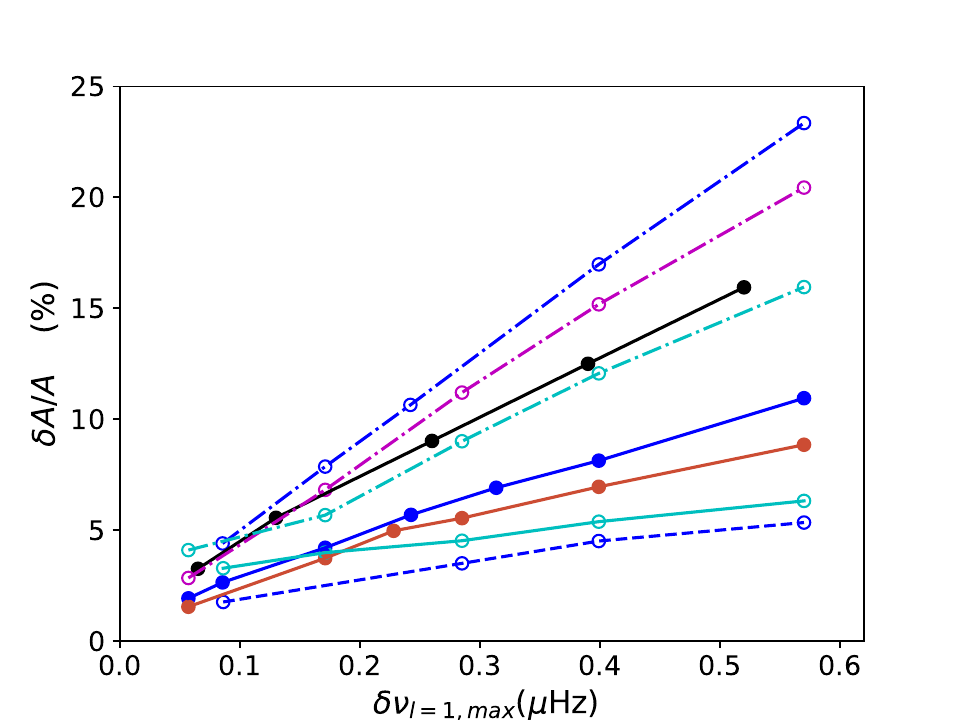}
                \includegraphics[width=8.cm]{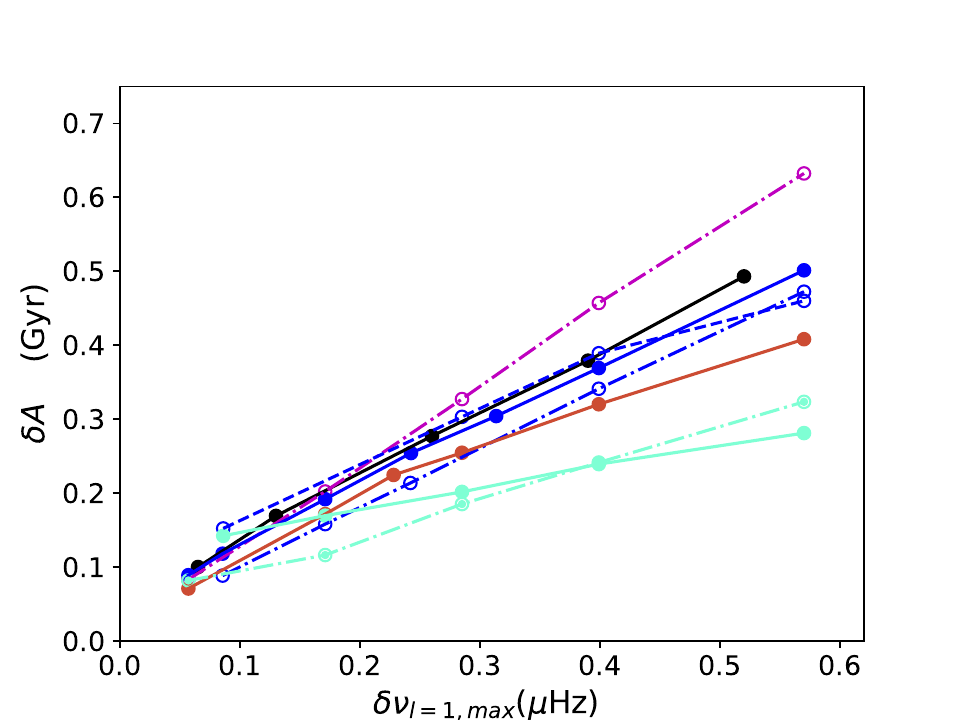}
                \caption{ Same as in Fig.\ref{dMdRdA1} but for the  age uncertainties (top: relative, bottom: absolute). }
                \label{dMdRdA2}
        \end{center}
\end{figure}

 The variations of the relative uncertainties for the inferred  mass, radius, and age 
 are shown  as a function of the uncertainties $\delta \nu_{\rm \ell=1,max}$  in Figs.~\ref{dMdRdA1} and~\ref{dMdRdA2}. As expected the MRA uncertainties increase 
 with 
 increasing  $\delta \nu_{\rm \ell=1,max}$.  
 It can be seen that the precision  for the inferred mass and radius is so high that a comfortable margin is left for systematic errors 
 that must be added quadratically in order to  obtain realistic  mass and radius uncertainties
  while still satisfying the PLATO requirements. This is also true of stars like the Sun (in mass and age).
  Indeed for such a  Sun-like star,   with a frequency uncertainty $\delta \nu_{\rm \ell=1, max}=0.2 \mu$Hz,  the relative age uncertainty is below 5\%.

\begin{figure}[t]
        \begin{center}
                \includegraphics[width=9cm,height=8.5cm]{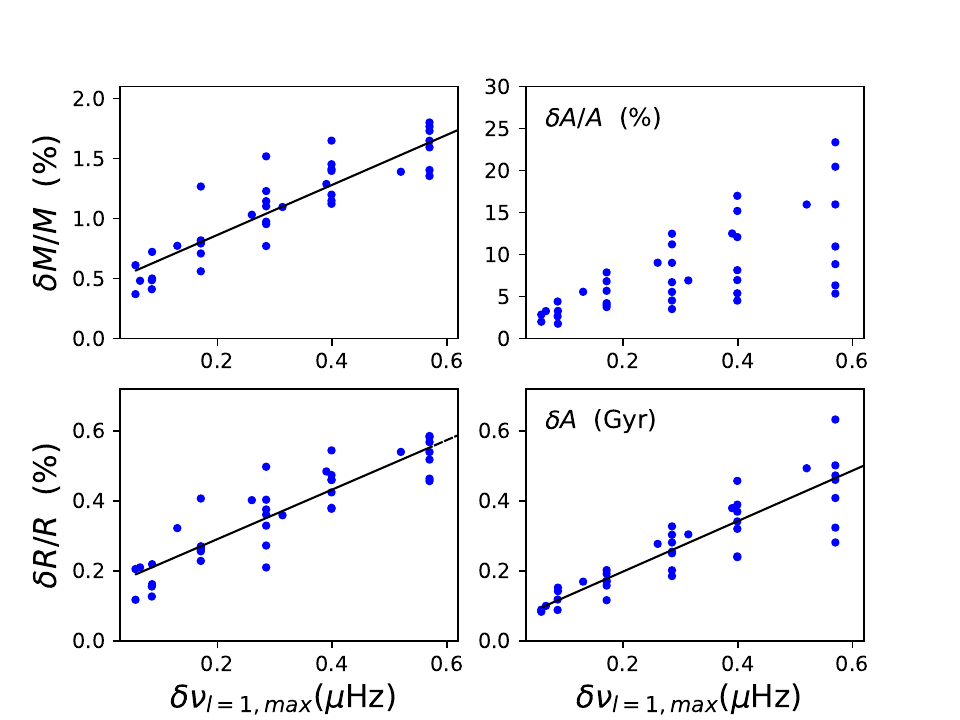}
                \caption{\label{dMdRdA} Uncertainties  for all  synthetic stars (blue dots). Solid black curve:  linear  fits (see text).}
                \label{synt3}
        \end{center}
\end{figure}

In view of  application  to the PLATO samples in Sect. 6, we show  the same results in  Fig.~\ref{synt3} than in Fig.~\ref{dMdRdA1} and Fig.~\ref{dMdRdA2}, respectively, 
  but without 
 the mass and age information.  We then carried out linear fits of the general trends for the mass and radius uncertainties.   
For the mass and radius relative uncertainties (in \%), the fits give:
  \begin{eqnarray}
  \label{MR1c2}
 \left\{
\begin{array}{lll}
\delta M / M  &=    2.083 ~ \delta \nu_{\ell=1,max} + 0.046    \\  
\delta R / R  &=   0.707  ~ \delta \nu_{\ell=1,max} + 0.149 \\
 \end{array}
\right.
,\end{eqnarray}
where $\delta \nu_{\rm \ell=1,max}$ is in $\mu$Hz.  The above relations are used  for the PIC targets in Sect.~\ref{PLATO}.
The  scatter (about 0.5\% and 0.1\% for the the mass and radius relative uncertainties, respectively) due to  different masses and ages 
 remains acceptable for our purposes. 
 
 We also see that there is too large a scatter  for enabling a meaningful fit for $\delta A/A$.  
This would be possible for  the 
 absolute age uncertainty, $\delta A$. We actually  found 
 \begin{align}
 \delta A    =  0.702   ~ \delta \nu_{\rm \ell=1,max}   + 0.054 \, , 
 \end{align}
  but for the PIC targets,  the age is unknown and  we can only use the relative age uncertainty so we will rather adopt 
an alternative criterion.
 As shown in Fig.~\ref{dMdRdA} for $\delta A/A $, the relative statistical uncertainty amounts roughly to 7\% 
 when $\delta \nu_{\rm \ell=1,Libb,\nu_{\rm max}} <0.2 \mu$Hz. 
We then adopt the criterion:
\begin{align}
 \delta \nu_{\rm \ell=1,Libb,\nu_{\rm max}} <0.2 \; \mu{\rm Hz,} 
\end{align} 
 to select the cases for which relative age  uncertainties of $<10$\%  can be expected. 
   This is in accordance with \cite{appourchaux2020}.
 This leaves as much  margin to allow for systematic errors (that must be added in quadrature to obtain the final uncertainties). 
 This is a challenge that has driven -and still drives- many  theoretical studies in the community 
 (see Sect. 7 for a brief discussion).

\section{Relation between $\sigma_{\rm Libb,\ell=1}$ and $\delta \nu_{\rm \ell=1,max}$}
\label{appendixD}

In absence of observations as it is the case today for the PLATO targets, 
 a convenient way of estimating the uncertainties of individual frequencies of 
 solar-like oscillation mode is the generally accepted \cite{libbrecht1992} relation:
\begin{align}
\label{sigma_libb}
\sigma^2_{\rm Libb}= f(\beta) \frac{\Gamma}{4 \pi T_{\rm obs}}
\end{align}
with 
\begin{equation}
f(\beta) = \left(1+\beta\right)^{1/2} \left[ \left(1+\beta\right)^{1/2} +\beta^{1/2}\right]^3 
\end{equation}
where $\beta=1/(S/N)$ is the inverse of the S/N, $\Gamma$ (in $s^{-1}$) is the FWHM linewidth of the mode, 
and $T_{\rm obs}$ (in $s)$  the duration of the observation. It is well accepted that this statistical estimate of the frequency
 uncertainties  of  solar-like oscillation modes represents well the reality. 
 
\begin{figure}[t]
        \begin{center}
                \includegraphics[width=9.cm]{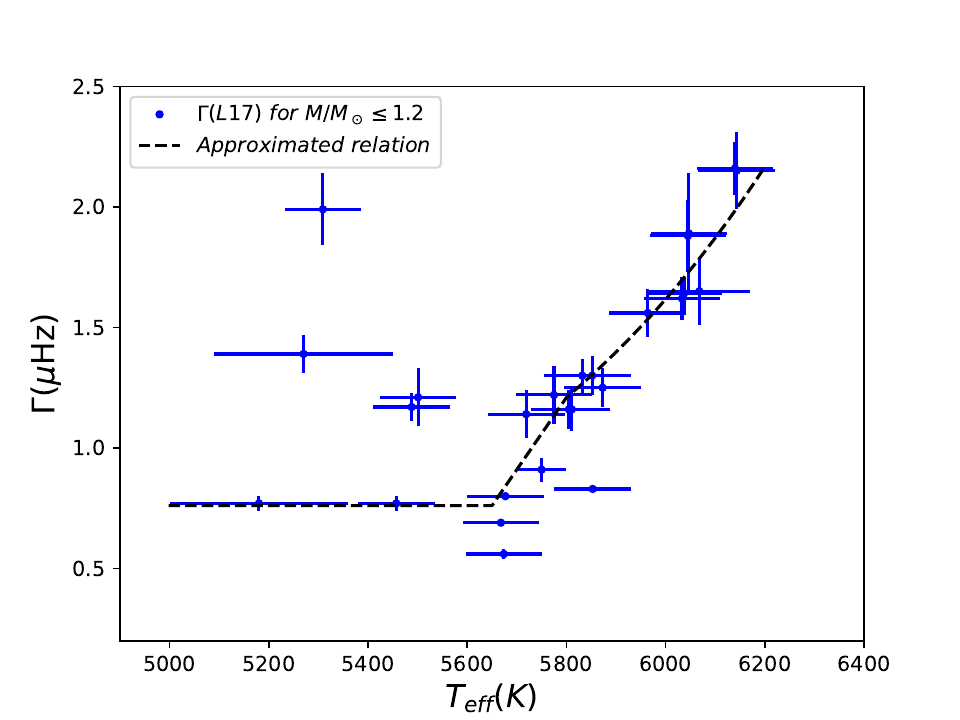}
                \caption{Variation of  $\Gamma$ with the effective temperature. Blue crosses are the observed values
      while the dashed line  represents the computed $\Gamma$ values.}
                \label{gamma}
        \end{center}
\end{figure}     

To proceed further, we  denote $\sigma_{\rm Libb,\ell=1,max}$ the \cite{libbrecht1992}  frequency uncertainty 
on the frequency of a $\ell=1$ mode 
closest to $\nu_{\rm max}$. 
It is evaluated for each star with $(S/N)_{\rm max}$ given by Eq.~(\ref{snr})  and $\Gamma$ in Table~\ref{table:gamma} below. 

  \subsection{Deriving $\Gamma_{\rm \ell=1,max}$ for estimating $\sigma_{\rm Libb, \ell=1,max}$}
  
  For each star, we take from  \cite{lund2017} the values of the linewidth, 
  $\Gamma_{\rm \ell=1,max}$, of the 
  $\ell=1$ mode with the  frequency closest to $\nu_{\rm max}$. Those values are  represented in Fig.~\ref{gamma}
   as function of the effective temperature. We restrict the case to P1P2 stars with mass $\leq 1.2 \rm{M}_\odot$  
   which are the PLATO targets  for which we will estimate the MRA uncertainties in Sect.~\ref{sect6}.
   One clearly notes two regimes: one regime at high effective temperature 
   where $\Gamma_{\rm \ell=1,max}$   increases almost linearly with the effective temperature and 
   one regime at low effective temperature where
    the scatter and the large uncertainties prevent from establishing a trend with $T_{\rm eff}$. In the low regime, 
     we therefore keep $\Gamma_{\rm \ell=1,max}$ constant at  the lowest value $\Gamma_0$ and consider also the case when 
     $3 \Gamma_0$ for $T_{\rm eff} \leq 5650$K  (Fig.~\ref{gamma}).
     We then adopt  for  $\Gamma_{\rm \ell=1,max}$ (in $\mu$ Hz) the  scaling relation adapted from
    \cite{appourchaux2012} (Table 2) and given  in Table~\ref{table:gamma}.
 
\begin{table}[h]
\centering
\caption{Empirical relation for the linewidth $\Gamma_{\rm \ell=1,max}$ derived from a fit of the \cite{lund2017}'s data (Fig.~\ref{gamma}).
\label{table:gamma}
}
\begin{tabular}{ |l | c | c| c|}
\hline 
\hline
$T_{\rm eff}$ (K)    &         $ \Gamma_{\rm \ell=1,max} $        \\
\hline
  ]5800,6400]     &    $0.2+0.97 \left( T_{\rm eff}/T_{\rm eff,\odot}\right)^{10}$  \\
  ]5650,5800]     &    $0.76+ 17.3 \left(T_{\rm eff}-5600\right)/T_{\rm eff,\odot} $      \\
  ]4900,5650]    &    $\Gamma_0=0.76$      \\
  \hline 
 \hline
\end{tabular}
\end{table} 
  
  \subsection{Converting    $\sigma_{\rm Libb, \ell=1,max}$  to  $\delta \nu_{\rm \ell=1,max}$  }
  
Using {\it Kepler} LEGACY data, we have derived in Appendix~\ref{appendixC} the MRA uncertainty as a function of the observed individual frequency
 uncertainty, $\delta \nu_{\rm max}$. We then need to convert $\sigma_{\rm Libb,\ell=1,max}$ into the frequency uncertainty $\delta \nu_{ \rm \ell=1, max}$.
  Again here we use the {\it Kepler} Legacy data to carry out  that calibration.
  
\begin{figure}[t]
        \begin{center}
                \includegraphics[width=9.cm]{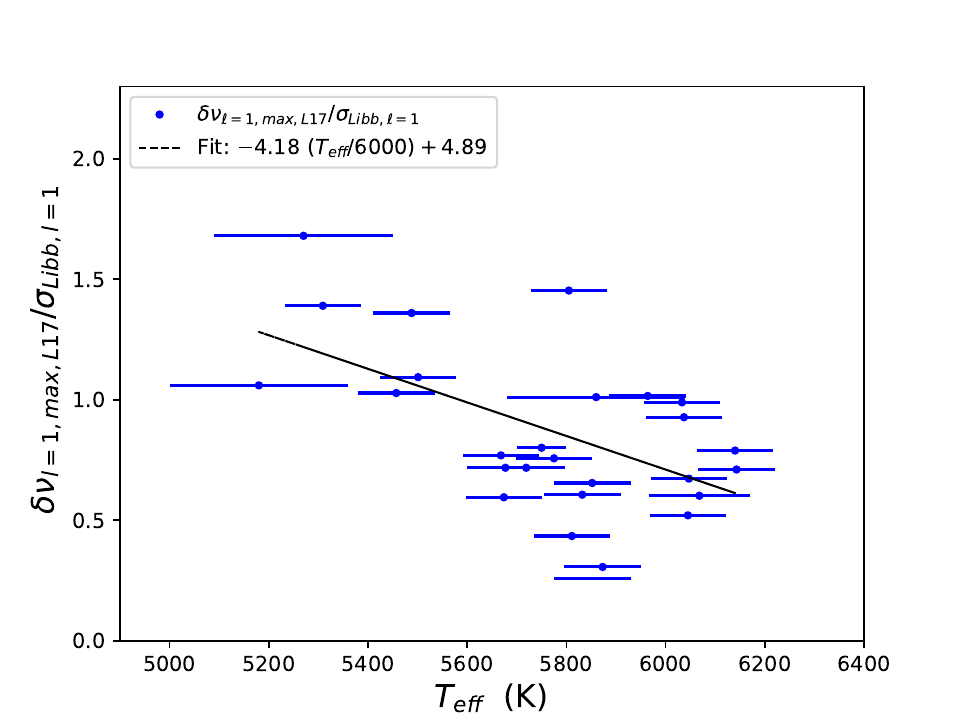}
                \caption{Variation of  the ratio $ \delta \nu_{\rm \ell=1,max}/\sigma_{\rm Libb} $ with the effective temperature (blue crosses). 
        The black curve is a linear fit. }
                \label{figure:ratsig2}
        \end{center}
\end{figure}

For each star, Fig.~\ref{figure:ratsig2} displays the ratio $\delta \nu_{\rm \ell=1,max}/\sigma_{\rm Libb, \ell=1,max}$   as a function
 of the effective temperature. There is a clear trend: the ratio decreases with the effective temperature. 
 For convenience, we derive a linear fit to represent that 
 trend about that fit. The linear fit gives:
\begin{align}
\label{ratsig}
 \frac{\delta\nu_{\rm \ell=1,max}}{\sigma_{\rm Libb,\ell=1,max}}  = 4.89-4.18~\frac{T_{\rm eff}}{6000 K} 
 \end{align}
 valid for $5000 < T_{\rm eff} \leq 6200 ~K$. The scatter about the linear relation is about $\pm 0.5 $. In Sect.6.2, 
 we will therefore consider the effect of adding $\sim  0.5 $  to the above linear relation  on the results for the PLATO targets.  
 
\end{document}